\documentclass[times,final]{elsarticle}

\makeatletter
\def\ps@pprintTitle{%
 \let\@oddhead\@empty
 \let\@evenhead\@empty
 \def\@oddfoot{}%
 \let\@evenfoot\@oddfoot}
\makeatother

\usepackage[utf8]{inputenc}
\usepackage[T1]{fontenc}

\usepackage{geometry}
\usepackage{soul}
\usepackage{color}
\usepackage[usenames,dvipsnames]{xcolor}

\let\OLDthebibliography\thebibliography
\renewcommand\thebibliography[1]{
  \OLDthebibliography{#1}
  \setlength{\parskip}{0pt}
  \setlength{\itemsep}{0pt plus 0.3ex}
}

\usepackage{amsmath}
\usepackage{amssymb}
\usepackage{latexsym}
\usepackage{bm} 
\usepackage{bbm} 

\usepackage{etoolbox}
\robustify\bfseries

\usepackage{caption} 
\usepackage{subcaption} 
\usepackage{booktabs} 
\usepackage{siunitx} 

\newdefinition{remark}{Remark}
\newcommand{\remarkend}{~\hfill$\blacksquare$}

\usepackage{algorithm}
\usepackage{algpseudocode}
\usepackage{float}
\newfloat{algorithm}{t}{lop}

\DeclareMathOperator{\diag}{diag}
\DeclareMathOperator{\sign}{sign}
\DeclareMathOperator{\vecm}{vec}

\providecommand{\abs}[1]{\lvert#1\rvert}
\providecommand{\norm}[1]{\lVert#1\rVert}

\newcommand{\vect}[1]{\boldsymbol{\mathbf{#1}}} 

\usepackage{url}
\usepackage{xcolor}
\definecolor{newcolor}{rgb}{.8,.349,.1}

\usepackage{hyperref}

\begin{document}


\begin{frontmatter}
	\title{An iterative domain decomposition, spectral finite element method on non-conforming meshes suitable for high frequency Helmholtz problems} 
	
	\author[1]{Ryan Galagusz\corref{cor1}}
	\cortext[cor1]{Corresponding author.}
	\ead{ryan.galagusz@mail.mcgill.ca}
	\author[1]{Steve McFee}
	\ead{steve.mcfee@mcgill.ca}
	
	\address[1]{Department of Electrical \& Computer Engineering, McGill University, Montreal, Quebec, H3A 0E9, Canada}
	

	\begin{abstract}
		The purpose of this research is to describe an efficient iterative method suitable for obtaining high accuracy solutions to high frequency time-harmonic scattering problems. The method allows for both refinement of local polynomial degree and non-conforming mesh refinement, including multiple hanging nodes per edge. Rather than globally assemble the finite element system, we describe an iterative domain decomposition method which can use element-wise fast solvers for elements of large degree. Since continuity between elements is enforced through moment equations, the resulting constraint equations are hierarchical. We show that, for high frequency problems, a subset of these constraints should be directly enforced, providing the coarse space in the dual-primal domain decomposition method. The subset of constraints is chosen based on a dispersion criterion involving mesh size and wavenumber. By increasing the size of the coarse space according to this criterion, the number of iterations in the domain decomposition method depends only weakly on the wavenumber. We demonstrate this convergence behaviour on standard domain decomposition test problems and conclude the paper with application of the method to electromagnetic problems in two dimensions. These examples include beam steering by lenses and photonic crystal waveguides, as well as radar cross section computation for dielectric, perfect electric conductor, and electromagnetic cloak scatterers.

	\end{abstract}
	
\end{frontmatter}

\section{Introduction}

The goal of this paper is to describe an efficient iterative method suitable for obtaining high accuracy solutions to high frequency time-harmonic scattering problems. These problems are modelled by the Helmholtz equation and its variants (for example, problems with variable coefficients). To achieve high accuracy, we use a spectral finite element method \cite{Sherwin1999,Canuto2007}. We do so because experimental and theoretical results regarding dispersion errors (sometimes called pollution errors) for finite element methods applied to the Helmholtz equation \cite{Ihlenburg1991,Oberai2000,Ainsworth2004,Ainsworth2009,Melenk2011,Zhu2013,Du2015} suggest that an effective approach to control pollution is to increase the polynomial degree $p$ (sometimes called order) as a function of mesh size $h$ and wavenumber $k$. While significant efforts have been made to avoid polynomials in attempts to eliminate pollution errors, experimental evidence suggests that high-order finite element methods are competitive with these alternative approaches \cite{Lieu2016}. However, there are difficulties associated with the solution of the resulting linear systems when the polynomial degree increases \cite{Vos2010,Mitchell2015} which tend to limit the extent to which high-order methods are adopted in practice. To circumvent these difficulties, we describe an iterative domain decomposition method for which fast algorithms from spectral methods can be applied locally \cite{Shen1994,Shen2011}.
	
While low frequency problems can be effectively solved using Krylov subspace methods with preconditioners suitable for nearby static problems \cite{Yserentant1988} (such as domain decomposition methods \cite{Toselli2005} or multi-grid methods \cite{Hackbusch2003}), high frequency problems remain difficult to precondition effectively due to the complex-symmetric and/or indefinite nature of the arising discretized systems \cite{Thompson2006}. Efforts to formulate improved preconditioners for discretizations of the Helmholtz equation include extensions of domain decomposition methods, variants on multi-grid methods, shifted-Laplacian preconditioners, and combinations of these methods, e.g. multi-grid applied to a shifted-Laplacian preconditioner (see \cite{Erlangga2008} for an overview of such methods). More recently, a type of domain decomposition method called sweeping preconditioner has been proposed for second-order finite difference discretizations of the Helmholtz equation which can be applied with near-linear computational complexity \cite{Engquist2011,Poulson2013}.
	
The iterative method we propose for our high-order scheme is closely related to methods of finite element tearing and interconnecting (FETI) type \cite{Farhat1991}. In particular, we use ideas from dual-primal FETI (FETI-DP) \cite{Stefanica2002,Mandel2007,Klawonn2008} and their extension to the Helmholtz problem (FETI-DPH) \cite{Farhat2005}. Recent experimental evidence \cite{Gander2013} suggests that FETI-DPH can be effective for high frequency Helmholtz problems when compared to other domain decomposition methods if a suitably chosen coarse space of plane waves is used to augment the primal constraints.
	
In this work, we make two primary modifications to FETI-DPH. First, rather than augment the coarse space with carefully chosen plane waves, we formulate a spectral finite element method where continuity between element subdomains is imposed via a hierarchy of weak constraints. The method is globally conforming (unlike a mortar method \cite{Belgacem1999}), but we use the flexibility of the weak constraints to construct a non-conforming coarse space used in a corresponding FETI-DP domain decomposition method. We show experimentally that, by choosing the size of the coarse space based on a dispersion error criterion \cite{Ainsworth2004}, the number of iterations in our domain decomposition method depends weakly on $k$. Second, we use Robin boundary conditions \cite{Benamou1997,Farhat2000a} between subdomains to eliminate non-physical resonant frequencies which arise in the FETI-DPH method. We show that this increased robustness can come at a computational cost, increasing the number of iterations required for convergence.
	
The method is applicable to interior problems with Dirichlet or Robin boundary conditions, but we show that exterior Helmholtz problems can be treated by introducing perfectly matched layers (PMLs) \cite{Berenger1994}. With suitable parameter choices, PMLs do not adversely affect iteration counts for our iterative method. The weak continuity constraints between elements are sufficiently general so as to allow irregular mesh refinement and polynomial mismatch between adjacent elements making $h$ and $p$ refinement possible. We conclude the paper with examples which demonstrate that the iterative method is effective in the presence of both $h$ and $p$ refinements, and for both low and high frequency problems.

%

\section{Numerical Methods} \label{sec:numerical_methods}

We will describe our method as applied to the boundary value problem (BVP)
\begin{equation} \label{eq:prototype_pde}
	-\nabla\cdot\left(\vect{\alpha}\nabla\phi\right)+\beta\phi=f\qquad\textrm{in }\Omega,
\end{equation}
subject to boundary conditions 
\begin{align}
	\phi & =p\qquad\textrm{on }\Gamma_{D},\label{eq:prototype_dirichlet} \\
	\vect{n} \cdot \left(\vect{\alpha}\nabla\phi\right)+\gamma\phi & =q\qquad\textrm{on }\Gamma_{R},\label{eq:prototype_robin}
\end{align}
where: $\Omega$ is a specified bounded, $d$-dimensional domain; $\phi$ is a scalar function of $\vect{x}\in\Omega\subset\mathbb{R}^{d}$; $\vect{\alpha}\in\mathbb{C}^{d\times d}$ is a symmetric matrix whose entries depend upon spatial variables $\vect{x}$; $\beta$, $f$, $p$, $\gamma$, and $q$ are complex scalar functions of $\vect{x}$; and $\vect{n}\in\mathbb{R}^{d}$ is the outward pointing unit normal to $\Omega$. The functions $\vect{\alpha}$, $\beta$, $f$, $p$, $\gamma$, and $q$ can be discontinuous.
We assume that the two boundary components $\Gamma_{D}$ and $\Gamma_{R}$ are disjoint; that is, $\Gamma_{D}\cap\Gamma_{R}=\emptyset$. Furthermore, we also assume that the boundary of $\Omega$, denoted $\partial\Omega$, is given by the union of the two boundary components $\Gamma_{D}$ and $\Gamma_{R}$ (which may themselves be composed of disconnected components) so that the boundary of $\Omega$ is the disjoint union of Dirichlet and Robin boundary components. Throughout this paper, we illustrate the method using $d=1,2$, although extension to $d=3$ is natural.

\begin{remark}
	Note that this BVP is sufficient to describe time-harmonic acoustic scattering in two and three dimensions and electromagnetic scattering in two dimensions (we will describe examples of this type in Section \ref{sec:results}). For three-dimensional electromagnetic scattering, an analogous approach may be formulated for the weak form of the curl-curl equation \cite{Jin2014}. \remarkend
\end{remark}

In Section \ref{sec:preliminaries}, we describe the spectral element method in one dimension. We include a discussion of affine transformations of Legendre polynomials which helps extend the method to higher dimensions. In particular, we perform this extension in Section \ref{sec:extension} for two-dimensional problems on non-conforming meshes. In both sections, the discretization of the BVP results in a saddle point system of equations which must be solved. We explain how to solve this system via domain decomposition in Section \ref{sec:decomposition}. In this last section regarding numerical methods, we emphasize how to ensure that the method, when applied to the Helmholtz problem, converges in a number of iterations only weakly dependent on the wavenumber $k$.

\subsection{Preliminaries in One Dimension} \label{sec:preliminaries}

Our method in higher dimensions makes use of concepts developed in one dimension. It will be useful to describe a one-dimensional spectral method for solving \eqref{eq:prototype_pde} subject to \eqref{eq:prototype_dirichlet}-\eqref{eq:prototype_robin}. The example will be to find $\phi (x)$ such that
\begin{equation} \label{eq:prototype_pde_1D}
	-\frac{d}{dx} \left( \alpha (x) \frac{d\phi}{dx} \right) + \beta (x) \phi = f (x) \qquad x \in (-1,1),
\end{equation}
and
\begin{equation} \label{eq:prototype_bc_1D}
	\phi (-1) = p, \qquad \left[ \alpha \frac{d\phi}{dx} + \gamma \phi \right]_{x=1} = q,
\end{equation}
where $\alpha$, $\beta$, and $f$ are specified smooth functions and $p$, $\gamma$, and $q$ are specified constants. All of our computations will be done with respect to orthonormal Legendre polynomials $p_k$ satisfying the recurrence relation
\begin{equation} \label{eq:recurrence_relation}
	p_{k}(x) = \frac{\sqrt{(2k+1)(2k-1)}}{k} x p_{k-1}(x) - \frac{k-1}{k}\sqrt{\frac{2k+1}{2k-3}} p_{k-2}(x), \qquad k = 2,3,4,...
\end{equation}
with $p_0(x) = 1/\sqrt{2}$ and $p_1(x) = \sqrt{3/2}x$. Defining a vector containing the first $n+1$ such polynomials
\begin{equation}
	\vect{p}(x) =
	\begin{bmatrix}
    	p_0(x) \\
    	p_1(x) \\
        p_2(x) \\
        \vdots \\
        p_n(x)
	\end{bmatrix}
\end{equation}
allows us to encode certain important operations on Legendre polynomials (from now on, orthonormality will be implied). Since the polynomials are orthonormal,
\begin{equation}
	\int_{-1}^1 \vect{p}(x) \vect{p}(x)^T dx = \vect{I}
\end{equation}
where $\vect{I}$ is the identity matrix and integration is performed entry-wise. Similarly, by virtue of the derivative property 
\begin{equation} \label{eq:derivative_property}
	\sqrt{2k+1}p_k = \frac{1}{\sqrt{2k+3}}\frac{d}{dx}p_{k+1} - \frac{1}{\sqrt{2k-1}}\frac{d}{dx}p_{k-1}
\end{equation}
we have that differentiation can be encoded as
\begin{equation}
	\frac{d}{dx}\vect{p}(x) = \vect{D} \vect{p}(x)
\end{equation}
with entries
\begin{equation}
	d_{ij} = 
	\begin{cases}
		\sqrt{(2i - 1)(2j-1)} & i+j \textrm{ odd} \textrm{ and } i > j, \\
    	0 & \textrm{otherwise}.
	\end{cases}
\end{equation}

In the following, we choose integrated Legendre polynomials as basis functions in Section \ref{sec:basis_functions}. This choice results in sparse matrices when discretizing \eqref{eq:prototype_pde_1D} subject to boundary conditions \eqref{eq:prototype_bc_1D} via the method of weighted residuals when $\alpha$ and $\beta$ are constant. We then describe Legendre expansions in Section \ref{sec:legendre_expansions} to treat the forcing function $f$. These Legendre expansions, together with an expression for the integral of triple products of Legendre polynomials (as described in Section \ref{sec:triple_integrals}) allow us to extend sparsity results to cases where $\alpha$ and $\beta$ are not constant. Combining the ideas in these three sections leads to a spectral method, as described in Section \ref{sec:simple_spectral_method}, which we extend to a spectral finite element method in Section \ref{sec:spectral_FEM}. We conclude Section \ref{sec:preliminaries} with a discussion of affine transformations of Legendre polynomials in Section \ref{sec:affine_transformation} which will be useful in higher dimensions when imposing non-conforming continuity constraints.

\subsubsection{Integrated Legendre Polynomials} \label{sec:basis_functions}

In practice, we will not directly use Legendre polynomials as basis functions to represent $\phi$, but a particular linear combination of them. We define
\begin{equation} \label{eq:define_basis_functions}
	\vect{N}(x) = 
	\begin{bmatrix}
		p_0(x) \\
		\int \vect{p}(x) \, dx
	\end{bmatrix}
\end{equation}
where $\vect{p}(x)$ has only $n$ entries (rather than $n+1$ as before so that $\vect{N}(x)$ has $n+1$ entries). We set all constants of integration in \eqref{eq:define_basis_functions} to zero (this is acceptable because of the inclusion of $p_0$ which is a constant). Then 
\begin{equation}
	\vect{N}(x) = \vect{S} \vect{p}(x)
\end{equation}
where $\vect{S} = \diag{(\vect{g})} - \vect{S}_{DL} \diag{(\vect{g})} \vect{S}_{DL}$,
\begin{equation}
	g_i =
	\begin{cases}
		1 & i = 1, \\
		\displaystyle{\frac{1}{\sqrt{(2i-1)(2i-3)}}} & \textrm{otherwise,}
	\end{cases}
\end{equation}
and $(s_{DL})_{ij} = \delta_{i,j+1}$ where $\delta_{i,j}$ is the Kronecker delta. The definition of $\vect{S}$ follows from \eqref{eq:derivative_property}. $\vect{S}_{DL}$ is a shift matrix which shifts entries in a matrix down a row when multiplying from the left (adding zeros to the first row), and shifts entries left one column when multiplying from the right (adding zeros to the last column). This means that $\vect{S}$ is lower triangular with two non-zero diagonals (the main diagonal and the second sub-diagonal). Since $\vect{g}$ has no zero entries, $\vect{S}$ is invertible and its inverse can be applied to a vector with linear complexity via forward substitution. As a consequence, we can efficiently change basis between Legendre polynomials and integrals of Legendre polynomials. We choose to use integrated Legendre polynomials because of the following differentiation property:
\begin{align}
	\frac{d}{dx}\vect{N}(x) &= \frac{d}{dx}\vect{S}\vect{p}(x) \\
	&= \vect{S} \frac{d}{dx} \vect{p}(x) \\
	&= \vect{S} \vect{D} \vect{p}(x) \\
	&= \vect{S}_{DL} \vect{p}(x) \label{eq:differentiation_property}
\end{align}
in light of definition \eqref{eq:define_basis_functions}.

The polynomials \eqref{eq:define_basis_functions} are closely related to those described in the spectral method \cite{Shen1994} and $hp$ finite element \cite{Sherwin1999,Szabo1991} literature. In particular, we note that besides the two representations described thus far (integrals of Legendre polynomials or linear combinations of Legendre polynomials), there is also the equivalent description in terms of orthonormal Jacobi polynomials $p_k^{(\alpha,\beta)}(x)$:
\begin{equation} \label{eq:jacobi_form}
	N_{k}(x) = -\frac{(1 - x)(1 + x)}{\sqrt{k(k-1)}} p_{k-2}^{(1,1)}(x), \qquad k=2,3,4,...
\end{equation}
with $N_0(x) = 1/\sqrt{2}$ and $N_1(x) = x/\sqrt{2}$. For $k\ge 2$, these polynomials are eigenfunctions of 
\begin{equation}
	\frac{d^2}{dx^2}N_k + \frac{\lambda_k}{(1 - x)(1 + x)}N_k = 0
\end{equation}
with eigenvalues $\lambda_k = k(k-1)$ subject to homogeneous Dirichlet boundary conditions at $x = \pm 1$. Often, the first two polynomials $N_0$ and $N_1$ are replaced with interpolatory functions $l_0 = (1+x)/2$ and $l_1 = (1-x)/2$. This simplifies certain boundary expressions at the cost of complicating the sparsity of the discretized form of \eqref{eq:prototype_pde_1D}. In practice, we transform between the two representations via
\begin{equation}
	\begin{bmatrix}
    	N_0 \\
    	N_1
	\end{bmatrix}
	= \underbrace{\frac{1}{\sqrt{2}}
	\begin{bmatrix}
    	1 & \phantom{-} 1 \\
    	1 & -1
	\end{bmatrix}
	}_{\vect{Q}}
	\begin{bmatrix}
    	l_0 \\
    	l_1
	\end{bmatrix}.
\end{equation}
Note that $\vect{Q}$ is orthogonal and symmetric and thus involutory so that $\vect{Q}^{-1} = \vect{Q}$. Thus, if $\tilde{\vect{N}}(x)$ is the vector containing integrated Legendre polynomials with first two polynomials interpolatory, then 
\begin{equation} \label{eq:interpolatory_relation}
	\tilde{\vect{N}}(x) = 
	\underbrace{
	\begin{bmatrix}
    	\vect{Q} & \vect{0} \\
    	\vect{0} & \vect{I}
	\end{bmatrix}
	}_{\vect{B}}
	\vect{N}(x)
\end{equation}
and also $\vect{N}(x) = \vect{B} \tilde{\vect{N}}(x)$.

When dealing with Dirichlet or Robin boundary conditions, we will need to evaluate $\vect{N}(\pm 1)$. Equation \eqref{eq:jacobi_form} together with $N_0 = 1/\sqrt{2}$ and $N_1 = x/\sqrt{2}$ yield
\begin{align}
	\vect{N}(+1) &= \frac{1}{\sqrt{2}}(\vect{e}_1 + \vect{e}_2), \\
	\vect{N}(-1) &= \frac{1}{\sqrt{2}}(\vect{e}_1 - \vect{e}_2),
\end{align}
where $\vect{e}_i$ is the $i$th unit vector.

\subsubsection{Legendre Expansions} \label{sec:legendre_expansions}

To represent (potentially variable) coefficients $\alpha$, $\beta$, and forcing function $f$, we use Legendre polynomials as well. That is, we represent these functions using linear combinations of Legendre polynomials. For example,
\begin{equation}
	f(x)=\vect{f}^T \vect{p}(x) = \sum_{k=0}^{K_f} f_k p_k(x)
\end{equation}
such that 
\begin{equation} \label{eq:legendre_coefficients}
	\vect{f} = \int_{-1}^1 f(x) \vect{p}(x) \, dx
\end{equation}
($\alpha$ and $\beta$ are treated in a similar fashion). In practice, to compute such a Legendre expansion, we first interpolate $f$ at points $x_i = \cos{(i \pi /M)}$ with $i=0,1,...,M$ and $M=2^n$ for modest $n$ and use a fast inverse discrete cosine transform to compute the coefficients in a Chebyshev polynomial expansion representing the interpolating polynomial in $\mathcal{O}(M\log{M})$ operations. This process is repeated adaptively by increasing $n$ until the magnitude of coefficients at the tail end of the expansion has decayed to machine precision (or to a user specified tolerance). For a description of how to terminate such a process and to "chop" the expansion keeping only necessary coefficients to achieve a desired accuracy, see \cite{Aurentz2017}. Once the Chebyshev expansion is constructed, we convert the Chebyshev coefficients to Legendre coefficients using the fast algorithm described in \cite{Townsend2017}. If $K_f$ coefficients are needed to describe the Chebyshev expansion, then conversion to Legendre coefficients requires $\mathcal{O}(K_f(\log{K_f})^2)$ operations. When $K_f$ is known to be small, direct evaluation of \eqref{eq:legendre_coefficients} can be performed via Gauss or Clenshaw-Curtis quadrature and is more efficient than the fast transform. 

\subsubsection{Integrals of Triplets of Legendre Polynomials} \label{sec:triple_integrals}

Finally, it will be useful to characterize the integral of triple products of Legendre polynomials. In particular, we define the third-order tensor $\vect{\mathcal{T}}$ whose entries are given by
\begin{equation}
	\mathcal{T}_{ijk} = \int_{-1}^1 p_{i-1}(x) p_{j-1}(x) p_{k-1}(x) \, dx.
\end{equation}
Of particular interest are the matrices
\begin{equation}
	\vect{T}_k = \int_{-1}^1 p_k(x)\, \vect{p}(x)\vect{p}(x)^T \,dx
\end{equation}
related to frontal slices of $\vect{\mathcal{T}}$. The entries are known explicitly \cite{Adams1878} and are given by
\begin{equation}
	\int_{-1}^{1}p_{i}p_{j}p_{k}\,dx = 
	\begin{cases} 
		{\displaystyle C_{ijk}\frac{A\left(s-i\right)A\left(s-j\right)A\left(s-k\right)}{A\left(s\right)}} & i+j+k \textrm{ even},\, i+j\ge k, \textrm{ and } |i-j| \le k ,\\
		0 & \textrm{otherwise},
	\end{cases}
\end{equation}
where 
\begin{align}
	C_{ijk} & =\frac{2}{2s+1}\sqrt{\frac{2i+1}{2}}\sqrt{\frac{2j+1}{2}}\sqrt{\frac{2k+1}{2}},\\
	A\left(m\right) & =1\cdot\frac{3}{2}\cdot\frac{5}{3}\cdot\ldots\cdot\left(2-\frac{1}{m}\right), \qquad m=1,2,3,...,
\end{align}
$A(0) = 1$, and $s=(i+j+k)/2$.

Let us comment on the sparsity of $\vect{\mathcal{T}}$ in more depth. First, the condition $i+j+k$ even means that we will observe a checkerboard pattern of non-zeros which alternates with $k$ for each slice $\vect{T}_{k}$. Second, the condition $i+j\ge k$ means that for each slice, there is an anti-diagonal before and including which all anti-diagonal entries are zero (the $k$th anti-diagonal counting from the top left corner of each slice). Third, the condition $\left|i-j\right|\le k$ implies that each slice has a bandwidth $k$.

\subsubsection{A Simple Spectral Method} \label{sec:simple_spectral_method}

We now describe a spectral method for solving \eqref{eq:prototype_pde_1D} subject to \eqref{eq:prototype_bc_1D} which makes use of Legendre polynomials $\vect{p}(x)$ and integrated Legendre polynomials $\vect{N}(x)$. Multiplying \eqref{eq:prototype_pde_1D} by test function $\psi$ and integrating by parts yields the weak form
\begin{equation}
	\int_{-1}^1 \frac{d\psi}{dx} \alpha(x) \frac{d\phi}{dx} \, dx
	+ \int_{-1}^1 \psi \, \beta(x) \, \phi \, dx
	- \left[ \psi \alpha(x) \frac{d\phi}{dx} \right]_{-1}^1
	= \int_{-1}^1 \psi f \, dx.
\end{equation}
Substituting $\left[ \alpha(x) \frac{d\phi}{dx} \right]_{x=1}$ from \eqref{eq:prototype_bc_1D} and letting $\nu = \left[ \alpha(x) \frac{d\phi}{dx} \right]_{x=-1}$ yields
\begin{equation}
	\int_{-1}^1 \frac{d\psi}{dx} \alpha(x) \frac{d\phi}{dx} \, dx
	+ \int_{-1}^1 \psi \, \beta(x) \, \phi \, dx
	+ \psi(1) \, \gamma \, \phi(1) + \psi(-1) \, \nu
	= \int_{-1}^1 \psi f \, dx + \psi(1) \, q
\end{equation}
subject to $\phi(-1) = p$.

Let $\phi = \vect{N}(x)^T \vect{\phi}$ where $\vect{\phi}$ is an unknown vector of coefficients to be determined and assume that Legendre expansions for $\alpha$, $\beta$, and $f$ have been computed as described in Section \ref{sec:legendre_expansions} such that
\begin{equation}
	\alpha = \sum_{k=0}^{K_{\alpha}}\alpha_k p_k, \qquad \beta = \sum_{k=0}^{K_{\beta}}\beta_k p_k,
\end{equation}
and $f=\vect{p}(x)^T\vect{f}$. Repeating the weak form for each function in $\vect{N}(x)$ yields
\begin{equation}
	\vect{S}_{DL} \left[ \sum_{k=0}^{K_{\alpha}} \alpha_k \vect{T}_k \right] \vect{S}_{DL}^T \vect{\phi}
	+ \vect{S} \left[ \sum_{k=0}^{K_{\beta}} \beta_k \vect{T}_k \right] \vect{S}^T \vect{\phi}
	+  \frac{1}{\sqrt{2}}(\vect{e}_1 + \vect{e}_2) \, \gamma \, \frac{1}{\sqrt{2}}(\vect{e}_1 + \vect{e}_2)^T \vect{\phi}
	+ \frac{1}{\sqrt{2}}(\vect{e}_1 - \vect{e}_2) \, \nu 
	= \vect{S} \vect{f} + \frac{1}{\sqrt{2}}(\vect{e}_1 + \vect{e}_2) \, q
\end{equation}
subject to 
\begin{equation}
	\frac{1}{\sqrt{2}}(\vect{e}_1 - \vect{e}_2)^T \vect{\phi} = p.
\end{equation}
Letting
\begin{align}
	\vect{A} &= \vect{S}_{DL} \left[ \sum_{k=0}^{K_{\alpha}} \alpha_k \vect{T}_k \right] \vect{S}_{DL}^T
	+ \vect{S} \left[ \sum_{k=0}^{K_{\beta}} \beta_k \vect{T}_k \right] \vect{S}^T
	+ \frac{\gamma}{2}(\vect{e}_1 + \vect{e}_2) (\vect{e}_1 + \vect{e}_2)^T, \label{eq:local_A} \\ 
	\vect{C} &= \frac{1}{\sqrt{2}}(\vect{e}_1 - \vect{e}_2)^T, \\
	\vect{b} &= \vect{S} \vect{f} + \frac{1}{\sqrt{2}}(\vect{e}_1 + \vect{e}_2) \, q, \label{eq:local_b} \\
	d &= p,
\end{align}
yields the saddle point system
\begin{equation} \label{eq:saddle_point_system}
	\begin{bmatrix}
    	\vect{A} & \vect{C}^T \\
    	\vect{C} & 0 
	\end{bmatrix}
	\begin{bmatrix}
    	\vect{\phi} \\
    	\nu
	\end{bmatrix}
	=
	\begin{bmatrix}
    	\vect{b} \\
    	d
	\end{bmatrix}
\end{equation}
which can be solved for $\vect{\phi}$ and/or $\nu$ via numerous methods \cite{Benzi1999}. Notice how the sparsity of $\vect{A}$ depends crucially on the sum of frontal slices of tensor $\vect{\mathcal{T}}$. In fact, we have a banded matrix with bandwidth $K_{\vect{A}}=\max(K_{\alpha},\, K_{\beta}+2)$. Of course, if $K_{\vect{A}}$ is equal to or larger than the degree $L$ of the expansion for $\phi$ then the matrix $\vect{A}$ is full. When both $\alpha$ and $\beta$ are constant and $\gamma = 0$, $\vect{A}$ is pentadiagonal with first sub- and super-diagonal equal to zero. Solution of the linear system can be performed in $\mathcal{O}(L)$ operations (using a banded solver). Alternatively, a perfect shuffle permutation matrix $\vect{P}_{2,(L+1)/2}$ (assuming $L+1$ is even) separating odd and even degree polynomials in $\vect{N}(x)$ can transform the pentadiagonal matrix $\vect{A}$ into two tridiagonal matrices of half the size via $\vect{P}_{2,(L+1)/2}^T \vect{A} \vect{P}_{2,(L+1)/2}$ (each can then be solved using a banded solver).

\subsubsection{A Spectral Finite Element Method} \label{sec:spectral_FEM}

In the event that the bandwidth of $\vect{A}$ is too large, preventing computation of $\vect{\phi}$ with linear complexity, it is possible to adaptively subdivide the interval $(-1,1)$ into subdomains by computing local expansions of $\alpha$ and $\beta$ of smaller degree meeting some user-defined maximum bandwidth criterion locally, and then enforcing continuity of the global solution via additional constraint equations. Doing so results in a spectral finite element method with the same saddle point structure as in \eqref{eq:saddle_point_system} but with
\begin{equation} \label{eq:block_diag_matrices}
	\vect{A} =
	\begin{bmatrix}
    	\vect{A}_1 &            &        &            \\
    	           & \vect{A}_2 &        &            \\
        	       &            & \ddots &            \\
        	       &            &        & \vect{A}_N	
	\end{bmatrix}, \qquad
	\vect{\phi} = 
	\begin{bmatrix}
    	\vect{\phi}_1 \\
    	\vect{\phi}_2 \\
        \vdots \\
        \vect{\phi}_N
	\end{bmatrix}, \qquad
	\vect{b} = 
	\begin{bmatrix}
    	\vect{b}_1 \\
    	\vect{b}_2 \\
        \vdots \\
        \vect{b}_N
	\end{bmatrix},
\end{equation}
where each $\vect{A}_j$, $\vect{\phi}_j$, and $\vect{b}_j$ corresponds to discretization of a local element subdomain $(x_{j-1},x_j)$ (suppose, for simplicity, that $-1=x_0 < x_1 < \cdots < x_N=1$ with $j=1,2,...,N$). The structure of each local matrix $\vect{A}_j$ and vector $\vect{b}_j$ remains the same as in \eqref{eq:local_A} and \eqref{eq:local_b} but appropriate scale factors must multiply integral and derivative terms by virtue of transforming the interval $x\in(x_{j-1},x_j)$ to the canonical interval $u\in(-1,1)$ via
\begin{equation}
	x = \underbrace{\frac{x_{j}-x_{j-1}}{2}}_{J_{j}} u + \frac{x_{j}+x_{j-1}}{2}.
\end{equation}
In particular, the first term in \eqref{eq:local_A} must be multiplied by $\sign{(J_j)} J_j^{-1}$ while the second term in \eqref{eq:local_A} and the first term in \eqref{eq:local_b} must be multiplied by $\abs{J_j}$. For our example, Robin boundary terms should appear only in the local matrix and vector corresponding to the element adjacent to $x=1$.

Note that the constraint matrix $\vect{C}$ grows (by concatenating rows) to include equations enforcing inter-element continuity. For example, if two elements are defined on intervals $(x_{j-1},x_j)$ and $(x_{j},x_{j+1})$, then continuity at $x_j$ is imposed by constraint
\begin{equation}
	\begin{bmatrix}
    	0 & \cdots & 0 & \frac{1}{\sqrt{2}}(\vect{e}_1 + \vect{e}_2)^T & -\frac{1}{\sqrt{2}}(\vect{e}_1 - \vect{e}_2)^T & 0 & \cdots & 0
	\end{bmatrix}
	\vect{\phi} = 0
\end{equation}
with $\frac{1}{\sqrt{2}}(\vect{e}_1 + \vect{e}_2)^T$ appearing in the $j$th block column and $-\frac{1}{\sqrt{2}}(\vect{e}_1 - \vect{e}_2)^T$ in the $(j+1)$th block column. The dual variable $\nu$ and right hand side $d$ must grow in accordance with the number of additional constraints, becoming vectors $\vect{\nu}$ and $\vect{d}$.

This finite element construction is necessary for the spectral method to effectively handle situations where $\alpha$ and $\beta$ are discontinuous. Otherwise, their Legendre expansions exhibit Gibbs oscillations, and the sparsity of $\vect{A}$ is lost. In practice, we choose certain $x_j$ to coincide with points of discontinuity so that $\alpha$ and $\beta$ are continuous on element subdomains.

\subsubsection{Affine Transformations of Legendre Polynomials} \label{sec:affine_transformation}

In order to extend the methods of Section \ref{sec:spectral_FEM} to non-conforming meshes in higher dimensions, we require an additional one-dimensional construct. We will need the coefficients which relate Legendre polynomials $\vect{p}(x)$ to Legendre polynomials under an affine transformation of variable $\vect{p}(ax+b)$. In practice, we will assume $x\in(-1,1)$ and that $a$ and $b$ are chosen such that $ax+b$ defines a subinterval of $(-1,1)$. In particular, we seek the lower triangular matrix $\vect{L}$ such that
\begin{equation} \label{eq:def_affine_transformation}
	\vect{p}(ax+b) = \vect{L} \vect{p}(x).
\end{equation}
We will focus on the situation where $\vect{L}$ is square. Rather than compute $\vect{L}$ via
\begin{equation}
	\vect{L} = \int_{-1}^1 \vect{p}(ax+b) \vect{p}(x)^T \, dx,
\end{equation}
we can solve a Sylvester equation using recurrence to obtain the entries of $\vect{L}$. We will also show that a fast multiplication algorithm exists for products of $\vect{L}$ with vectors.

To do so, rewrite the recurrence relation \eqref{eq:recurrence_relation} by isolating the term $x p_{k-1}$. Then
\begin{equation} \label{eq:modified_recurrence}
	x \vect{p}(x) = \vect{J}_{n+1} \vect{p}(x) + \sqrt{\beta_{n+1}} p_{n+1}(x) \vect{e}_{n+1}
\end{equation}
with the truncated Jacobi matrix
\begin{equation}
	\vect{J}_{n+1} = 
	\begin{bmatrix}
    	0              & \sqrt{\beta_1} &     			 &        		  &                \\
    	\sqrt{\beta_1} & 0 			    & \sqrt{\beta_2} &        		  &                \\
        			   & \sqrt{\beta_2} & 0 		     & \ddots 		  &                \\
        			   &     			& \ddots		 & \ddots 		  & \sqrt{\beta_n} \\
        			   &   				&    			 & \sqrt{\beta_n} & 0 
	\end{bmatrix}
\end{equation}
and $\sqrt{\beta_k} = k/\sqrt{(2k+1)(2k-1)}$. Typically, \eqref{eq:modified_recurrence} is encountered when computing Gauss quadrature rules via eigenvalue routines \cite{Gautschi2004}. Instead, here we rewrite \eqref{eq:modified_recurrence} with argument $ax+b$ so that
\begin{equation}
	(ax+b) \vect{p}(ax+b) = \vect{J}_{n+1} \vect{p}(ax+b) + \sqrt{\beta_{n+1}} p_{n+1}(ax+b) \vect{e}_{n+1}
\end{equation}
then substitute \eqref{eq:def_affine_transformation} so that
\begin{equation}
	(ax+b) \vect{L} \vect{p}(x) = \vect{J}_{n+1} \vect{L} \vect{p}(x) + \sqrt{\beta_{n+1}} p_{n+1}(ax+b) \vect{e}_{n+1}.
\end{equation}
A second application of \eqref{eq:modified_recurrence} removes $x\vect{p}(x)$ giving
\begin{equation}
	a\vect{L} [\vect{J}_{n+1} \vect{p}(x) + \sqrt{\beta_{n+1}} p_{n+1}(x) \vect{e}_{n+1}] + b \vect{L} \vect{p}(x) = \vect{J}_{n+1} \vect{L} \vect{p}(x) + \sqrt{\beta_{n+1}} p_{n+1}(ax+b) \vect{e}_{n+1}.
\end{equation}
Finally, multiplying from the right by $\vect{p}(x)^T$ and integrating over $(-1,1)$ yields the matrix equation
\begin{equation} \label{eq:nearly_sylvester}
	a\vect{L} \vect{J}_{n+1} + b \vect{L} = \vect{J}_{n+1} \vect{L} + \vect{e}_{n+1} \underbrace{\left[ \sqrt{\beta_{n+1}} \int_{-1}^1 p_{n+1}(ax+b) \vect{p}(x)^T\, dx \right]}_{\displaystyle \vect{p}_{n+1}^T}.
\end{equation}
Letting $\tilde{\vect{A}} = \vect{J}_{n+1} - b\vect{I}$ and $\tilde{\vect{B}} = a\vect{J}_{n+1}$ shows that \eqref{eq:nearly_sylvester} corresponds to
\begin{equation} \label{eq:displacement_rank}
	\tilde{\vect{A}} \vect{L} - \vect{L} \tilde{\vect{B}} = - \vect{e}_{n+1} \vect{p}_{n+1}^T,
\end{equation}
which is a Sylvester equation for $\vect{L}$.

Since entry $l_{11} = 1$ (the affine transformation of a constant function returns the same constant function), $\tilde{\vect{A}}$ and $\tilde{\vect{B}}$ are tridiagonal, and $\vect{e}_{n+1} \vect{p}_{n+1}^T$ is almost entirely zero, we can solve for the rows of $\vect{L}$ sequentially without computing $\vect{p}_{n+1}$ (enlarge $\tilde{\vect{A}}$, $\tilde{\vect{B}}$, and $\vect{L}$ by one row and column and stop the recurrence relation one iteration prematurely). If $\vect{l}_k^T$ is the $k$th row of $\vect{L}$, then $\vect{l}_1^T = \vect{e}_1^T$ and
\begin{equation}
	\vect{l}_{k+1}^T = \frac{1}{\tilde{a}_{k,k+1}} \left( \vect{l}_{k}^T \tilde{\vect{B}} - \tilde{a}_{k,k-1} \vect{l}_{k-1}^T - \tilde{a}_{k,k} \vect{l}_{k}^T \right)
\end{equation}
for $k=1,2,...$ (zero indices are treated by ignoring zero-indexed terms). This algorithm requires $\mathcal{O}(n^2)$ operations to compute the entries of $\vect{L}$. This is comparable with the algorithm \cite{Salzer1973}.

When the size of $\vect{L}$ is large, we can avoid explicitly computing its entries and only compute its action on a vector. To do so, we notice that \eqref{eq:displacement_rank} indicates that $\vect{L}$ has displacement rank 1, suggesting that fast matrix vector products are possible. Indeed, notice that $\tilde{\vect{A}}$ and $\tilde{\vect{B}}$ are simultaneously diagonalizable. By computing the eigenvalue decomposition $\vect{J}_{n+1}\vect{V} = \vect{V} \vect{\Lambda}$, we directly obtain the eigenvectors $\vect{V}$ of both matrices. Their eigenvalues are $\vect{\Lambda}_A = \vect{\Lambda} - b\vect{I}$ and $\vect{\Lambda}_B = a\vect{\Lambda}$ respectively. Then letting $\vect{L} = \vect{V}\tilde{\vect{L}}\vect{V}^T$, $\tilde{\vect{e}} = \vect{V}^T \vect{e}_{n+1}$, $\tilde{\vect{p}} = \vect{V}^T \vect{p}_{n+1}$, and multiplying the Sylvester equation from the left by $\vect{V}^T$ and from the right by $\vect{V}$ yields
\begin{equation}
	\vect{\Lambda}_A \tilde{\vect{L}} - \tilde{\vect{L}} \vect{\Lambda}_B = - \tilde{\vect{e}} \tilde{\vect{p}}^T.
\end{equation}
This means that
\begin{equation}
	\tilde{l}_{ij} = -\frac{\tilde{e}_i\tilde{p}_j}{(\lambda_A)_i - (\lambda_B)_j}
\end{equation}
and that $\tilde{\vect{L}}$ is Cauchy-like so that its matrix-vector product can be computed using the fast multipole method (FMM) \cite{Carrier1988}. Similarly, the matrix-vector products with $\vect{V}$ and $\vect{V}^T$ can also be accelerated by FMM since $\vect{J}_{n+1}$ is tridiagonal \cite{Gu1995}. For this process to work, we also need to be able to compute $\vect{p}_{n+1}$ efficiently. Recall that this vector is a scaled vector of Legendre coefficients for the function $p_{n+1}(ax+b)$. We can compute this vector efficiently by using $\mathcal{O}(1)$ evaluation of Legendre polynomials \cite{Bogaert2012} to sample $p_{n+1}(ax+b)$ and use those samples as in Section \ref{sec:legendre_expansions} to compute the Legendre coefficients with $\mathcal{O}(n (\log n)^2)$ operations.

\subsection{Extension to Higher Dimensions} \label{sec:extension}

There are several complications that arise when extending the spectral finite element method in Section \ref{sec:spectral_FEM} to higher dimensions. We describe how to perform such an extension by starting with a spectral method on the canonical domain $\Omega = (-1,1)^2$, as described in Section \ref{sec:2d_spectral_method}. We then consider a related method on a domain $\Omega$ whose geometry is described by transfinite interpolation \cite{Gordon1971} with corresponding map $\vect{x}:(-1,1)^2 \rightarrow \Omega$ and $\vect{u} \in (-1,1)^2$, as described in Section \ref{sec:curvilinear_method}. Properties of this spectral method will suggest certain mesh design criteria, described in Section \ref{sec:mesh_generation}, that we adhere to when constructing the corresponding spectral finite element method. Finally, in Section \ref{sec:noncomforming_constraints}, we describe how to impose weak continuity constraints between adjacent elements in both conforming and non-conforming configurations.

\subsubsection{A Simple Spectral Method} \label{sec:2d_spectral_method}

We begin by solving \eqref{eq:prototype_pde} subject to \eqref{eq:prototype_dirichlet} and \eqref{eq:prototype_robin} on $\Omega = (-1,1)^2$. We will refer to $\vect{x} = [x_1 \; x_2]^T$ and boundary components
\begin{align}
	\Gamma_{1} &=\left\{ \vect{x}\in\mathbb{R}^{2}:\, x_{1}=-1,\, x_{2}\in\left(-1,1\right)\right\} , \\
	\Gamma_{2} &=\left\{ \vect{x}\in\mathbb{R}^{2}:\, x_{1}=+1,\, x_{2}\in\left(-1,1\right)\right\} , \\
	\Gamma_{3} &=\left\{ \vect{x}\in\mathbb{R}^{2}:\, x_{1}\in\left(-1,1\right),\, x_{2}=-1\right\} , \\
	\Gamma_{4} &=\left\{ \vect{x}\in\mathbb{R}^{2}:\, x_{1}\in\left(-1,1\right),\, x_{2}=+1\right\}.
\end{align}
Boundary component $\Gamma_{1}$ corresponds to the left edge, $\Gamma_{2}$ to the right edge, $\Gamma_{3}$ to the bottom edge, and $\Gamma_{4}$ to the top edge of $\Omega$. We assume that $\Gamma_{D}$ and $\Gamma_{R}$ satisfy $\partial\Omega=\Gamma_{D}\cup\Gamma_{R}$, $\Gamma_{D}\cap\Gamma_{R}=\emptyset$, and that they are each comprised of a union of a subset of $\Gamma_{1}$, $\Gamma_{2}$, $\Gamma_{3}$, and $\Gamma_{4}$. This last assumption eliminates the possibility of a single edge sharing more than one type of boundary condition. Rather than explicitly specify which edges belong to $\Gamma_D$ and $\Gamma_R$, we will describe how to address either type of boundary condition on any of $\Gamma_{1}$, $\Gamma_{2}$, $\Gamma_{3}$, or $\Gamma_{4}$.

We leverage as much as possible from the one-dimensional construction by exploiting the tensor product structure of $\Omega$. That is, we approximate the solution $\phi$ by a linear combination of products of integrated Legendre polynomials:
\begin{equation}
	\phi (\vect{x}) = \vect{N}(x_1)^T \vect{\Phi} \vect{N}(x_2)
\end{equation}
or, in vectorized form, 
\begin{align}
	\phi (\vect{x}) &= \Big( \vect{N}(x_2)^T \otimes \vect{N}(x_1)^T \Big) \vecm{(\vect{\Phi})} \\
	&= \Big( \vect{N}(x_2) \otimes \vect{N}(x_1) \Big)^T \vecm{(\vect{\Phi})}
\end{align}
where $\otimes$ denotes the Kronecker product between two matrices and $\vecm{(\cdot)}$ denotes the vectorization operator. To save on notation, we define $\vect{N}(\vect{x}) = \vect{N}(x_2) \otimes \vect{N}(x_1)$ and $\vect{\phi} = \vecm{(\vect{\Phi})}$ so that $\phi(\vect{x}) = \vect{N}(\vect{x})^T\vect{\phi}$. Partial differentiation of $\phi$ with respect to $x_1$ and $x_2$ respectively is given by
\begin{equation}
	\frac{\partial}{\partial x_1}\phi (\vect{x}) = \Big( \vect{N}(x_2) \otimes \vect{S}_{DL}\vect{p}(x_1) \Big)^T \vect{\phi}, \qquad 
	\frac{\partial}{\partial x_2}\phi (\vect{x}) = \Big( \vect{S}_{DL} \vect{p}(x_2) \otimes \vect{N}(x_1) \Big)^T \vect{\phi},
\end{equation}
by differentiation property \eqref{eq:differentiation_property}.

The boundary functions $p_i$, $\gamma_i$, and $q_i$ for each possible edge $\Gamma_i$ are represented by one-dimensional Legendre expansions, as in Section \ref{sec:legendre_expansions}. We also represent each function in $\vect{\alpha}$, $\beta$, $f$, by expansions in terms of products of Legendre polynomials. However, using $f$ as an example, we require that these expansions be of the form
\begin{equation}
	f(\vect{x}) = \sum_{k=1}^{K} \sigma_k u_k(x_1) v_k(x_2).
\end{equation}
By doing so, we can compute one-dimensional Legendre expansions for each function $u_k(x_1) = \vect{p}(x_1)^T \vect{u}_k$ and $v_k(x_2) = \vect{v}_k^T \vect{p}(x_2)$ as in Section \ref{sec:legendre_expansions}. Suppose that each vector of coefficients $\vect{u}_k$ is zero padded so that they all have the same dimensions (and likewise for $\vect{v}_k$), then
\begin{align}
	f(\vect{x}) &= \sum_{k=1}^{K} \sigma_k \left(\vect{p}(x_1)^T \vect{u}_k\right) \left(\vect{v}_k^T \vect{p}(x_2)\right) \label{eq:sum_of_rank_1_functions} \\
				&= \vect{p}(x_1)^T \left[ \sum_{k=1}^{K} \sigma_k  \vect{u}_k \vect{v}_k^T \right] \vect{p}(x_2) \\
				&= \vect{p}(x_1)^T \underbrace{\vect{U} \vect{\Sigma} \vect{V}^T}_{\displaystyle{\hat{\vect{F}}}} \vect{p}(x_2)
\end{align}
with $\vect{\Sigma}$ diagonal with diagonal entries $\sigma_k$.

In practice, if we know in advance that the dimension of vectors $\vect{u}_k$ and $\vect{v}_k$ will be small, then the entries of $\hat{\vect{F}}$ can be computed by Gauss or Clenshaw-Curtis quadrature by
\begin{equation}
	\hat{\vect{F}} = \vect{P} \diag{(\vect{w})} \vect{F} \diag{(\vect{w})} \vect{P}^T
\end{equation}
where $\vect{w}$ is a vector of quadrature weights $w_i$ of length $n+1$, the entries of $\vect{F}$ are samples of $f$ at quadrature points $x_i$ such that $f_{ij} = f(x_i,x_j)$, and $\vect{P} = [\vect{p}(x_0) \; \vect{p}(x_1) \; \cdots \; \vect{p}(x_n)]$. Then the singular value decomposition (SVD) of $\hat{\vect{F}}$ yields the factors $\vect{U}$, $\vect{\Sigma}$, and  $\vect{V}^T$.

For more complicated functions $f$, we use the potentially more efficient low rank algorithm described in \cite{Townsend2013} to compute an expansion of the form \eqref{eq:sum_of_rank_1_functions} with Legendre polynomials replaced by Chebyshev polynomials. If $K_{1}$ is the degree required to represent functions in $x_1$ and $K_2$ the degree for functions in $x_2$, then $\mathcal{O}\big(K^2(K_1 + K_2) + K^3\big)$ operations are required to produce the expansion. We then apply the algorithm of \cite{Townsend2017} (as in one dimension) to convert each one-dimensional Chebyshev expansion into its corresponding Legendre expansion which costs an additional $\mathcal{O}\Big(K\big(K_1(\log{K_1})^2 + K_2(\log{K_2})^2\big)\Big)$ operations. Compared with the $\mathcal{O}\big(K_1K_2(K_1 + K_2) + K_2^3\big)$ operations of the SVD, this more sophisticated approach can be significantly less expensive when $K$ is small compared to $K_1$ and $K_2$. Note that, when using the low rank algorithm, matrices $\vect{U}$ and $\vect{V}$ need not possess orthonormal columns, nor does $\vect{\Sigma}$ have diagonal entries satisfying $\sigma_1 \ge \sigma_2 \ge \cdots \ge \sigma_{K} \ge 0$ as one expects from the SVD.

Having constructed valid Legendre expansions, we find the weak form of \eqref{eq:prototype_pde} by multiplying by test function $\psi$ and integrating by parts to obtain
\begin{equation}
	\int_{\Omega} \nabla \psi \cdot (\vect{\alpha} \nabla \phi) \, d\Omega
	+ \int_{\Omega} \psi \, \beta \, \phi \, d\Omega
	- \oint_{\partial \Omega} \psi \,  \vect{n} \cdot (\vect{\alpha} \nabla \phi) \, d\Gamma
	= \int_{\Omega} \psi f \, d\Omega
\end{equation}
then use \eqref{eq:prototype_robin} so that
\begin{equation}
	\int_{\Omega} \nabla \psi \cdot (\vect{\alpha} \nabla \phi) \, d\Omega
	+ \int_{\Omega} \psi \, \beta \, \phi \, d\Omega
	+ \int_{\Gamma_R} \psi \, \gamma \, \phi \, d\Gamma
	- \int_{\Gamma_D} \psi \,  \vect{n} \cdot (\vect{\alpha} \nabla \phi) \, d\Gamma
	= \int_{\Omega} \psi f \, d\Omega
	+ \int_{\Gamma_R} \psi q \, d\Gamma
\end{equation}
subject to \eqref{eq:prototype_dirichlet}. Letting $\nu = - \vect{n} \cdot (\vect{\alpha} \nabla \phi)$ on $\Gamma_D$ and testing \eqref{eq:prototype_dirichlet} with $\psi_D$ yields
\begin{align}
	\left\{ \int_{\Omega} \nabla \psi \cdot (\vect{\alpha} \nabla \phi) \, d\Omega
	+ \int_{\Omega} \psi \, \beta \, \phi \, d\Omega
	+ \int_{\Gamma_R} \psi \, \gamma \, \phi \, d\Gamma \right\}
	+ \int_{\Gamma_D} \psi \nu \, d\Gamma
	&= \int_{\Omega} \psi f \, d\Omega
	+ \int_{\Gamma_R} \psi q \, d\Gamma \label{eq:operator_weak_form} \\
	\int_{\Gamma_D} \psi_D \, \phi \, d\Gamma \phantom{	+ \int_{\Gamma_R} \psi \, \gamma \, \phi \, d\Gamma \Bigg\} + \int_{\Gamma_D} \psi \nu \, d\Gamma} &= \int_{\Gamma_D} \psi_D p \, d\Gamma \label{eq:bc_weak_form}
\end{align}
which we have typeset to emphasize how the following discretization will lead to a symmetric saddle point system. We use test functions $\vect{N}(\vect{x})$ in \eqref{eq:operator_weak_form} and test functions related to $\vect{p}(x_1)$ or $\vect{p}(x_2)$ (depending on which edges $\Gamma_i$ belong to $\Gamma_D$) in \eqref{eq:bc_weak_form}. Since $\vect{\alpha}$ is symmetric, we compute two-dimensional separable Legendre expansions for $\alpha_{11}$, $\alpha_{12}=\alpha_{21}$, $\alpha_{22}$, $\beta$, and $f$, each with their corresponding set of coefficients $\vect{\Sigma}$, $\vect{U}$, and $\vect{V}$. Since
\begin{equation}
\nabla \psi \cdot (\vect{\alpha} \nabla \phi) = \sum_{i=1}^{2}\sum_{j=1}^{2}\frac{\partial\psi}{\partial x_{i}}\alpha_{ij}\left(\vect{x}\right)\frac{\partial\phi}{\partial x_{j}},
\end{equation}
the matrices
\begin{align}
	\vect{S}_{11} &= \int_{\Omega} \left[\frac{\partial}{\partial x_1}\vect{N}(\vect{x})\right] \alpha_{11} \left[\frac{\partial}{\partial x_1}\vect{N}(\vect{x})\right]^T d\Omega, \label{eq:s_11} \\
	\vect{S}_{12} &= \int_{\Omega} \left[\frac{\partial}{\partial x_1}\vect{N}(\vect{x})\right] \alpha_{12} \left[\frac{\partial}{\partial x_2}\vect{N}(\vect{x})\right]^T d\Omega, \label{eq:s_12} \\
	\vect{S}_{22} &=  \int_{\Omega} \left[\frac{\partial}{\partial x_2}\vect{N}(\vect{x})\right] \alpha_{22} \left[\frac{\partial}{\partial x_2}\vect{N}(\vect{x})\right]^T d\Omega, \label{eq:s_22}
\end{align}
discretize $\int_{\Omega} \nabla \psi \cdot (\vect{\alpha} \nabla \phi) \, d\Omega$ via the sum of products $(\vect{S}_{11} + \vect{S}_{12} + \vect{S}_{12}^T + \vect{S}_{22})\vect{\phi}$. Similarly,
\begin{equation}
	\vect{M} = \int_{\Omega} \vect{N}(\vect{x}) \, \beta \, \vect{N}(\vect{x})^T d\Omega \label{eq:m}
\end{equation}
discretizes $\int_{\Omega} \psi \, \beta \, \phi \, d\Omega$ via the product $\vect{M} \vect{\phi}$.
Since these four matrices share similar structure, their entries can all be computed explicitly using
\begin{multline}
	\int_{\Omega} (\vect{X}_1 \otimes \vect{X}_3) \big( \vect{p}(x_2) \otimes \vect{p}(x_1)\big) \, 
	[ \vect{p}(x_1)^T \vect{U} \vect{\Sigma} \vect{V}^T \vect{p}(x_2) ]
	 \, \big( \vect{p}(x_2) \otimes \vect{p}(x_1) \big)^T(\vect{X}_2 \otimes \vect{X}_4)^T \, d\Omega = \\
	\sum_{k=1}^{K}\sigma_{k}\left(\vect{X}_1\left[\sum_{i=0}^{K_{2}}v_{ik}\vect{T}_{i}\right]\vect{X}_2^{T}\otimes\vect{X}_3\left[\sum_{i=0}^{K_{1}}u_{ik}\vect{T}_{i}\right]\vect{X}_4^{T}\right). \label{eq:complicated_integral}
\end{multline}
The matrices $\vect{X}_i$ depend on which of the four integrals must be computed and Table \ref{tab:volume_integrals} summarizes this dependence. Notice that by choosing to use tensor products of one-dimensional basis functions for testing and expanding the solution, as well as using separable Legendre expansions to represent $\vect{\alpha}$ and $\beta$, matrix-vector products with these matrices can be performed using only matrices arising from one-dimensional problems. This is performed using the relationship
\begin{equation}
	\sigma_{k}\left(\vect{X}_1\left[\sum_{i=0}^{K_{2}}v_{ik}\vect{T}_{i}\right]\vect{X}_2^{T}\otimes\vect{X}_3\left[\sum_{i=0}^{K_{1}}u_{ik}\vect{T}_{i}\right]\vect{X}_4^{T}\right) \vect{\phi}
	= \sigma_k \vecm{\left( \vect{X}_3\left[\sum_{i=0}^{K_{1}}u_{ik}\vect{T}_{i}\right]\vect{X}_4^{T} \vect{\Phi} \vect{X}_2\left[\sum_{i=0}^{K_{2}}v_{ik}\vect{T}_{i}\right]\vect{X}_1^{T} \right)}.
\end{equation}
When $K_1$ and $K_2$ are small, and the polynomial degree in each dimension is $L$, then each product, either with $\vect{X}_i$ or $\vect{T}_i$ (both of which are sparse) requires $\mathcal{O}(L^2)$ operations which is linear in the total number of unknowns in $\vect{\phi}$.

\begin{table}[!t]
	\centering
	\caption[]{Connection between matrices \eqref{eq:s_11}-\eqref{eq:m} and method of evaluation \eqref{eq:complicated_integral}.} 
	\label{tab:volume_integrals}

	\begin{tabular}{ccccc} \toprule
	    {} & {$\vect{X}_1$} & {$\vect{X}_2$} & {$\vect{X}_3$} & {$\vect{X}_4$} \\ \midrule
	    $ \vect{S}_{11} $  & $\vect{S}$ & $\vect{S}$ & $\vect{S}_{DL}$ & $\vect{S}_{DL}$ \\
	    $ \vect{S}_{12} $  & $\vect{S}$ & $\vect{S}_{DL}$ & $\vect{S}_{DL}$ & $\vect{S}$ \\
	    $ \vect{S}_{22} $  & $\vect{S}_{DL}$ & $\vect{S}_{DL}$ & $\vect{S}$ & $\vect{S}$ \\
	    $ \vect{M} $  & $\vect{S}$ & $\vect{S}$ & $\vect{S}$ & $\vect{S}$ \\ \bottomrule
	\end{tabular}
\end{table}

If $\vect{\Sigma}$, $\vect{U}$, and $\vect{V}$ contain the coefficients in a separable Legendre expansion for $f$, then the contribution of term $\int_{\Omega} \psi f \, d\Omega$ to the discretization is
\begin{align}
	\tilde{\vect{f}} &= \int_{\Omega}\vect{N}\left(\vect{x}\right)f\, d\Omega \\
	&= \vecm{\left(\left(\vect{S}\vect{U}\right)\vect{\Sigma}\left(\vect{S}\vect{V}\right)^{T}\right)}.
\end{align}

Discretization of $\int_{\Gamma_R} \psi \, \gamma \, \phi \, d\Gamma$ depends on the definition of boundary $\Gamma_R$. Here we state the outcome as a function of edge $\Gamma_i$. To do so, we compute one-dimensional Legendre expansions 
\begin{align}
	\gamma_i(x_2) &= \sum_{k=0}^{K_{\gamma}^{(i)}} \gamma_k^{(i)} p_k(x_2), \qquad i=1,2, \\
	\gamma_i(x_1) &= \sum_{k=0}^{K_{\gamma}^{(i)}} \gamma_k^{(i)} p_k(x_1), \qquad i=3,4,
\end{align}
then compute the matrices
\begin{equation}
	\vect{R}_i = \int_{\Gamma_{i}}\vect{N}\left(\vect{x}\right) \gamma_i \vect{N}\left(\vect{x}\right)^{T}d\Gamma =
	\begin{cases}
		\displaystyle
		\vect{S}\left[\sum_{k=0}^{K_{\gamma}^{(i)}}\gamma_{k}^{(i)}\vect{T}_{k}\right]\vect{S}^{T}\otimes\frac{1}{2}\left(\vect{e}_{1}+(-1)^i\vect{e}_{2}\right)\left(\vect{e}_{1} + (-1)^i \vect{e}_{2}\right)^{T} & i=1,2,\\
		\displaystyle
		\frac{1}{2}\left(\vect{e}_{1}+ (-1)^i \vect{e}_{2}\right)\left(\vect{e}_{1}+ (-1)^i \vect{e}_{2}\right)^{T}\otimes\vect{S}\left[\sum_{k=0}^{K_{\gamma}^{(i)}}\gamma_{k}^{(i)}\vect{T}_{k}\right]\vect{S}^{T} & i = 3,4.
	\end{cases}
\end{equation}
We define the index set $\mathcal{I} = \{1,2,3,4\}$ and specify the index of those edges which belong to $\Gamma_D$ by $\mathcal{D}$. Then the edges belonging to $\Gamma_R$ have indices $\mathcal{R} = \mathcal{I} \setminus \mathcal{D}$ and $(\sum_{i \in \mathcal{R}}\vect{R}_i)\vect{\phi}$ represents the discretized Robin boundary term.

We proceed similarly for the discretization of $\int_{\Gamma_R} \psi q \, d\Gamma$. We compute one-dimensional Legendre expansions $q_i(x_2) = \vect{p}(x_2)^T\vect{q}_i$ for $i=1,2,$ and $q_i(x_1) = \vect{p}(x_1)^T\vect{q}_i$ for $i=3,4,$ then compute the vectors
\begin{equation}
	\vect{r}_i = \int_{\Gamma_{i}}\vect{N}\left(\vect{x}\right)q_i\, d\Gamma =
	\begin{cases}
		\displaystyle
		\vect{S}\vect{q}_{i}\otimes\frac{1}{\sqrt{2}}\left(\vect{e}_{1} + (-1)^i \vect{e}_{2}\right) & i=1,2, \\
		\displaystyle
		\frac{1}{\sqrt{2}}\left(\vect{e}_{1} + (-1)^i \vect{e}_{2}\right)\otimes\vect{S}\vect{q}_{i} & i=3,4.
	\end{cases}
\end{equation}
The sum $\sum_{i\in\mathcal{R}}\vect{r}_i$ represents the contribution to the forcing term from the Robin boundary.

Finally, we consider the Dirichlet boundary terms $\int_{\Gamma_D} \psi_D \, \phi \, d\Gamma$ and  $\int_{\Gamma_D} \psi_D p \, d\Gamma$. We treat each edge $\Gamma_i$ separately and begin by computing one-dimensional Legendre expansions $p_i(x_2) = \vect{p}(x_2)^T\vect{p}_i$ for $i=1,2,$ and $p_i(x_1) = \vect{p}(x_1)^T\vect{p}_i$ for $i=3,4$ on each edge. Weighting by $\vect{\psi}$ set to one of either $\sqrt{2} \vect{S}^{-T} \vect{p}(x_2)$ (for $i=1,2$) or $\sqrt{2} \vect{S}^{-T} \vect{p}(x_1)$ (for $i=3,4$) gives matrices
\begin{equation}
	\vect{C}_i = \int_{\Gamma_{i}} \vect{\psi} \vect{N}\left(\vect{x}\right)^T d\Gamma =
	\begin{cases}
		\displaystyle
		\vect{I} \otimes \left(\vect{e}_1 + (-1)^i \vect{e}_2\right)^T & i=1,2,\\
		\displaystyle
		\left(\vect{e}_1 + (-1)^i \vect{e}_2\right)^T \otimes \vect{I} & i=3,4,
	\end{cases}
\end{equation}
and vectors 
\begin{equation}
	\vect{d}_i = \int_{\Gamma_{i}} \vect{\psi} p_i \, d\Gamma = \sqrt{2} \vect{S}^{-T} \vect{p}_i \quad i=1,2,3,4,
\end{equation}
which constrain the solution as $\vect{C}_i \vect{\phi} = \vect{d}_i$. Note that the choice to use weight functions scaled by $\sqrt{2} \vect{S}^{-T}$ leads to constraint matrices $\vect{C}_i$ comprised of only $\pm 1$ entries (this is not necessary and simply weighting by Legendre polynomials would suffice). To obtain a symmetric saddle point system, the function $\nu$ in term $\int_{\Gamma_D} \psi \nu \, d\Gamma$ must be expanded using the same weight functions $\vect{\psi}$ as for the corresponding Dirichlet boundary condition. For example, on edge $\Gamma_i$, we let $\nu_i = \vect{\psi}^T\vect{\nu}_i$ which produces terms of the form $\vect{C}_i^T \vect{\nu}_i$.

Putting these observations together, the resulting saddle point system has components
\begin{align}
	\vect{A} &= (\vect{S}_{11} + \vect{S}_{12} +\vect{S}_{12}^T +\vect{S}_{22}) + \vect{M} + \sum_{i\in \mathcal{R}} \vect{R}_i, \\
	\vect{b} &= \tilde{\vect{f}} + \sum_{i\in \mathcal{R}} \vect{r}_i,
\end{align}
and $\vect{C}$  (the concatenation of all $\vect{C}_i$ with $i\in\mathcal{D}$ by stacking rows), $\vect{d}$, and $\vect{\nu}$ (the same concatenation for $\vect{d}_i$ and $\vect{\nu}_i$ respectively). This transforms \eqref{eq:operator_weak_form} and \eqref{eq:bc_weak_form} into the system of equations
\begin{equation} \label{eq:2d_saddle_point_system}
	\begin{bmatrix}
    	\vect{A} & \vect{C}^T \\
    	\vect{C} & \vect{0} 
	\end{bmatrix}
	\begin{bmatrix}
    	\vect{\phi} \\
    	\vect{\nu}
	\end{bmatrix}
	=
	\begin{bmatrix}
    	\vect{b} \\
    	\vect{d}
	\end{bmatrix}.
\end{equation}
When $\vect{\alpha}$ and $\beta$ are constant and $\phi$ is represented by degree $L$ polynomials in $x_1$ and $x_2$, using the nullspace method \cite{Benzi1999} to eliminate $\vect{\nu}$ allows for the application of a fast solver \cite{Shen1994} to compute $\vect{\phi}$ in $\mathcal{O}(L^3)$ operations. In certain variable coefficient cases, it may be possible to achieve similar computational complexity using a Krylov subspace method for generalized Sylvester equations \cite{Bouhamidi2008} with the fast solver as preconditioner.

\begin{remark} \label{rem:rank_def_remark}
	If more than one edge $\Gamma_i$ is constrained by Dirichlet conditions, the saddle point system \eqref{eq:2d_saddle_point_system} may be singular. This situation arises depending on the particular set of indices in $\mathcal{D}$ and is a consequence of over-constraining vertices of $\Omega$ which results in redundant constraint equations and a constraint matrix $\vect{C}$ which does not have full row rank.
	
	Fortunately, correcting this rank deficiency is simple. One approach is to begin with an empty constraint matrix $\vect{C}$ and vector $\vect{d}$ and update them by considering each edge $\Gamma_{i}$ sequentially while updating a list of vertices $\left(-1,-1\right)$, $\left(1,-1\right)$, $\left(-1,1\right)$, and $\left(1,1\right)$, and the edges connected to them. This vertex to edge incidence list is a list of every vertex. For each vertex, the list indicates which edges are incident upon the vertex (for example, vertex $(-1,-1)$ has edges $\Gamma_1$ and $\Gamma_3$ incident upon it). To impose Dirichlet constraints, we sequentially check edges to see if a Dirichlet boundary condition should be imposed. If yes, we go to the vertex to edge incidence list and mark that edge $\Gamma_{i}$ in all locations it appears. We then add the constraints $\vect{C}_{i}$ and $\vect{d}_{i}$ by appending their rows to $\vect{C}$ and $\vect{d}$. For every list associated to a vertex that becomes completely marked, we have a redundant equation. If only one equation is redundant, we remove the first row from the constraint matrix $\vect{C}_{i}$ and vector $\vect{d}_{i}$ that we would otherwise append to the currently assembled constraint matrix and vector. If two equations are redundant, we remove the first two rows from $\vect{C}_i$ and $\vect{d}_i$. This procedure ensures that the final constraint matrix $\vect{C}$ has full rank. \remarkend
\end{remark}

\subsubsection{A Spectral Method on Curvilinear Quadrilaterals} \label{sec:curvilinear_method}

In this section we extend the spectral method from Section \ref{sec:2d_spectral_method} to domains $\Omega$ with curvilinear edges. Consider domains with $\vect{x} \in \Omega$ described by the transfinite interpolation map
\begin{multline}
	\vect{x}\left(\vect{u}\right)
	=\frac{(1-u_{1})}{2}\tilde{\vect{X}}_{1}\vect{p}\left(u_{2}\right) + \frac{(1+u_{1})}{2}\tilde{\vect{X}}_{2}\vect{p}\left(u_{2}\right) + \frac{(1-u_{2})}{2}\tilde{\vect{X}}_{3}\vect{p}\left(u_{1}\right) + \frac{(1+u_{2})}{2}\tilde{\vect{X}}_{4}\vect{p}\left(u_{1}\right) \\
	\hspace{3.5cm} -\frac{\left(1-u_{1}\right)\left(1-u_{2}\right)}{4}\tilde{\vect{X}}_{1}\vect{p}\left(-1\right) - \frac{\left(1+u_{1}\right)\left(1-u_{2}\right)}{4}\tilde{\vect{X}}_{2}\vect{p}\left(-1\right) \\
	-\frac{\left(1-u_{1}\right)\left(1+u_{2}\right)}{4}\tilde{\vect{X}}_{1}\vect{p}\left(1\right) - \frac{\left(1+u_{1}\right)\left(1+u_{2}\right)}{4}\tilde{\vect{X}}_{2}\vect{p}\left(1\right)
\end{multline}
where $u_1,u_2 \in (-1,1)$ and $\tilde{\vect{X}}_i$ for $i=1,2,3,4,$ are 2-by-$(L_i+1)$ matrices whose multiplication with Legendre polynomials $\vect{p}(u_j)$ describe parametric curves representing four components of the boundary of $\Omega$.

In practice, the matrices $\tilde{\vect{X}}_i$ can be computed by the methods described in Section \ref{sec:legendre_expansions} if the edges of $\Omega$ are specified as curves parametrized by arc length. If each edge is specified as the zero level set of implicit function $\Phi(\vect{x})$, we can still compute these matrices by constructing line segment
\begin{equation}
	\vect{y}\left(t\right)=\frac{1-t}{2}\vect{x}_{\textrm{start}}+\frac{1+t}{2}\vect{x}_{\textrm{end}}
\end{equation}
joining each pair of adjacent vertices ($\vect{x}_{\textrm{start}}$ and $\vect{x}_{\textrm{end}}$) and projecting the points $\vect{y}_k = \vect{y}(\cos{(k\pi/L_i)})$ for $k=0,1,...,L_i$ onto the boundary using initial iterate $\vect{x}_{k}^{(0)}=\vect{y}_{k}$ and iteration
\begin{equation} \label{eq:projection_iteration}
	\vect{x}_{k}^{(i+1)} = \vect{x}_{k}^{(i)} - \frac{\Phi(\vect{x}_{k}^{(i)})}{\|\nabla\Phi(\vect{x}_{k}^{(i)})\|_{2}^{2}}\nabla\Phi(\vect{x}_{k}^{(i)}).
\end{equation}
This iteration is performed to find each point $\vect{x}_k$ on the boundary from which the Legendre coefficients are computed (in the same way as in Section \ref{sec:legendre_expansions}). When the curved boundary satisfies an implicit function theorem with respect to the line segment $\vect{y}(t)$, this approach works well. The iterative method is a first order method for finding the closest point $\vect{x}_k$ on the boundary to the point $\vect{y}_k$ \cite{Persson2005}. If $\Phi$ is a signed distance function, the iteration terminates after one iteration. If the gradient of the implicit function is unknown, the gradient can be approximated by finite differences. Both of these approximations do not significantly affect the quality of the interpolation of the curve because we iterate on the degree $L_i$ to ensure that a given accuracy of boundary representation is met.

Partial derivatives of the transfinite map yield
\begin{multline}
	\frac{\partial}{\partial u_{1}} \vect{x}(\vect{u}) =
	-\frac{1}{2}\tilde{\vect{X}}_{1}\vect{p}\left(u_{2}\right)
	+\frac{1}{2}\tilde{\vect{X}}_{2}\vect{p}\left(u_{2}\right)
	+\frac{1-u_{2}}{2}\tilde{\vect{X}}_{3}\vect{D}\vect{p}\left(u_{1}\right) 
	+\frac{1+u_{2}}{2}\tilde{\vect{X}}_{4}\vect{D}\vect{p}\left(u_{1}\right) \\
	+\frac{1-u_{2}}{4}\tilde{\vect{X}}_{1}\vect{p}\left(-1\right)
	-\frac{1-u_{2}}{4}\tilde{\vect{X}}_{2}\vect{p}\left(-1\right)
	+\frac{1+u_{2}}{4}\tilde{\vect{X}}_{1}\vect{p}\left(1\right)
	-\frac{1+u_{2}}{4}\tilde{\vect{X}}_{2}\vect{p}\left(1\right)
\end{multline}
and 
\begin{multline}
	\frac{\partial}{\partial u_{2}} \vect{x}(\vect{u}) = 
	\frac{1-u_{1}}{2}\tilde{\vect{X}}_{1}\vect{\tilde{D}}\vect{p}\left(u_{2}\right)
	+\frac{1+u_{1}}{2}\tilde{\vect{X}}_{2}\vect{\tilde{D}}\vect{p}\left(u_{2}\right)
	-\frac{1}{2}\tilde{\vect{X}}_{3}\vect{p}\left(u_{1}\right)
	+\frac{1}{2}\tilde{\vect{X}}_{4}\vect{p}\left(u_{1}\right) \\
	+\frac{1-u_{1}}{4}\tilde{\vect{X}}_{1}\vect{p}\left(-1\right)
	+\frac{1+u_{1}}{4}\tilde{\vect{X}}_{2}\vect{p}\left(-1\right)
	-\frac{1-u_{1}}{4}\tilde{\vect{X}}_{1}\vect{p}\left(1\right)
	-\frac{1+u_{1}}{4}\tilde{\vect{X}}_{2}\vect{p}\left(1\right)
\end{multline}
which we use to populate the Jacobian matrix 
\begin{equation}
	\vect{J} = 
	\begin{bmatrix}
    	\displaystyle \frac{\partial}{\partial u_{1}} \vect{x}(\vect{u}) & \displaystyle \frac{\partial}{\partial u_{2}} \vect{x}(\vect{u})
	\end{bmatrix}.
\end{equation}
If weight functions and basis functions are defined such that they transform to $\vect{N}(\vect{u})$ after change of variables, then the matrices from Section \ref{sec:2d_spectral_method} continue to hold, with the exception that the Legendre expansions used to represent parameters $\vect{\alpha}$, $\beta$, $f$, $\gamma$, $q$, and $p$ must be replaced by expansions for their corresponding effective parameters.

That is, taking integrals from Section \ref{sec:2d_spectral_method} over $\vect{x} \in \Omega$ and changing variables to the canonical domain $\vect{u} \in (-1,1)^2$ yields effective parameters
\begin{align}
	\vect{\alpha}_{\textrm{eff}} &= \vect{J}^{-1} \vect{\alpha}\left(\vect{x}(\vect{u})\right) \vect{J}^{-T}\, |\det{(\vect{J})}|, \\
	\beta_{\textrm{eff}} &= \beta\left(\vect{x}(\vect{u})\right) \, |\det{(\vect{J})}|, \\
	f_{\textrm{eff}} &= f\left(\vect{x}(\vect{u})\right) \, |\det{(\vect{J})}|.
\end{align}
Similarly, for boundary integrals
\begin{equation}
	\gamma_{\textrm{eff}}^{(i)} = 
	\begin{cases}
		\gamma\left(\tilde{\vect{X}}_{i}\vect{p}\left(u_{1}\right)\right)\|\tilde{\vect{X}}_{i}\vect{D}\vect{p}\left(u_{1}\right)\|_{2} & i=1,2, \\
		\gamma\left(\tilde{\vect{X}}_{i}\vect{p}\left(u_{2}\right)\right)\|\tilde{\vect{X}}_{i}\vect{D}\vect{p}\left(u_{2}\right)\|_{2} & i=3,4,
	\end{cases}
\end{equation}
and
\begin{equation}
	q_{\textrm{eff}}^{(i)} = 
	\begin{cases}
		q\left(\tilde{\vect{X}}_{i}\vect{p}\left(u_{1}\right)\right)\|\tilde{\vect{X}}_{i}\vect{D}\vect{p}\left(u_{1}\right)\|_{2} & i=1,2, \\
		q\left(\tilde{\vect{X}}_{i}\vect{p}\left(u_{2}\right)\right)\|\tilde{\vect{X}}_{i}\vect{D}\vect{p}\left(u_{2}\right)\|_{2} & i=3,4.
	\end{cases}
\end{equation}
No changes are needed for $p$ since the term $\|\tilde{\vect{X}}_{i}\vect{D}\vect{p}(u_{j})\|_{2}$ can be absorbed in the definition of the boundary weight functions. This means that
\begin{equation}
	p_{\textrm{eff}}^{(i)} = 
	\begin{cases}
		p\left(\tilde{\vect{X}}_{i}\vect{p}\left(u_{1}\right)\right) & i=1,2, \\
		p\left(\tilde{\vect{X}}_{i}\vect{p}\left(u_{2}\right)\right) & i=3,4.
	\end{cases}
\end{equation}

\begin{remark} \label{rem:remark_transformations}
	Notice that in the special case where parameters $\vect{\alpha}$, $\beta$, $f$, $\gamma$, $q$, and $p$  are constant with respect to $\vect{x}$, the map $\vect{x}(\vect{u})$ renders the effective parameters variable in $\vect{u}$. This can seriously degrade the sparsity of the matrix $\vect{A}$ in Section \ref{sec:2d_spectral_method}. In fact, even a standard bilinear transformation (a specific simple case of the transfinite map) can have such an effect. However, affine transformations do not change the sparsity of $\vect{A}$ when parameters are constant. \remarkend
\end{remark}

\subsubsection{A Desirable Finite Element Mesh} \label{sec:mesh_generation}

To account for more complicated domains than can be described by the techniques of Section \ref{sec:curvilinear_method}, we partition the domain $\Omega$ into several subdomains $\Omega_j$ each described by its own transfinite interpolation map. As per Remark \ref{rem:remark_transformations}, a desirable finite element mesh is comprised mostly of subdomains whose map is affine. In this section, we describe a mesh generation procedure which subdivides $\Omega$ into subdomains described by affine maps, wherever possible, ideally to avoid bilinear or more general transfinite maps wherever possible. The method is applicable to domains with internal boundaries which are necessary when dealing with discontinuous parameters $\vect{\alpha}$ or $\beta$ and belongs to the family of superposition methods for quadrilateral mesh generation \cite{Schneiders1996}.

Rather than parametrize the boundary and internal interfaces, we assume that domain $\Omega$ is comprised of a disjoint union of subdomains $\hat{\Omega}_i$ (not to be confused with element subdomains $\Omega_j$), each described by
\begin{equation}
	\hat{\Omega}_i = \{ \vect{x} \in \mathbb{R}^2 : \Phi_i(\vect{x}) < 0 \}.
\end{equation}
The implicit functions $\Phi_i$ need not be signed distance functions, but we use them wherever possible (for a calculus of signed distance functions that can be used to build more complicated domains from simple implicit functions, see \cite{Osher2003}). We choose a bounding square large enough to contain $\Omega$ and recursively subdivide the square into four congruent squares (in the same manner as a quadtree). We do this for a user specified maximum level of recursive subdivisions chosen such that the squares are small enough to represent the fine features of the boundary (boundaries with high curvature require more subdivisions). For now, we work with this uniform mesh of elements to better approximate the boundaries of $\hat{\Omega}_i$, although the quadtree is used at the end of this subsection to allow for non-uniform refinement away from boundaries. In the following, we use as example a bounding box $(-2,2)^2$ with four levels of uniform refinement performed to build a mesh which can accurately accommodate an internal boundary corresponding to the unit circle.

First, we classify these elements into groups by sampling each implicit function at the vertices of the mesh. If an element has all four of its vertices in one subdomain $\hat{\Omega}_i$, we classify this element as belonging to group $i$. Those elements that lie entirely outside $\Omega$ are discarded in the process. There will be small bands of unclassified elements which straddle the boundaries of the subdomains. To classify them, we perform a more careful check by computing an approximate area fraction of the element that lies inside each subdomain \cite{Owen2014}. If $|\Omega_j|$ is the area of $\Omega_j$, then the area fraction of $\Omega_j$ contained in $\hat\Omega_i$ is
\begin{align}
	A_{ij} &= \frac{1}{|\Omega_j|} \int_{\Omega_j} \mathbbm{1}_{\hat\Omega_i} d\Omega \\
	&= \frac{1}{|\Omega_j|} \int_{\Omega_j} [1 - H(\Phi_i(\vect{x}))] d\Omega
\end{align}
where $\mathbbm{1}_{\hat\Omega_i}$ is the unit indicator function on $\hat\Omega_i$ which we represent using the Heaviside step function $H$. In practice, we use a first order approximation to the Heaviside step function \cite{Osher2003} and numerically integrate to compute $A_{ij}$. Since $\sum_i A_{ij} + A_{0j} = 1$, we also compute the area fraction $A_{0j}$ corresponding to the exterior of $\Omega$. We then classify these boundary elements by largest area fraction. 

\begin{figure*}[!t]
    \centering
    \begin{subfigure}[t]{0.475\textwidth}
        \centering
        \includegraphics[width=\textwidth]{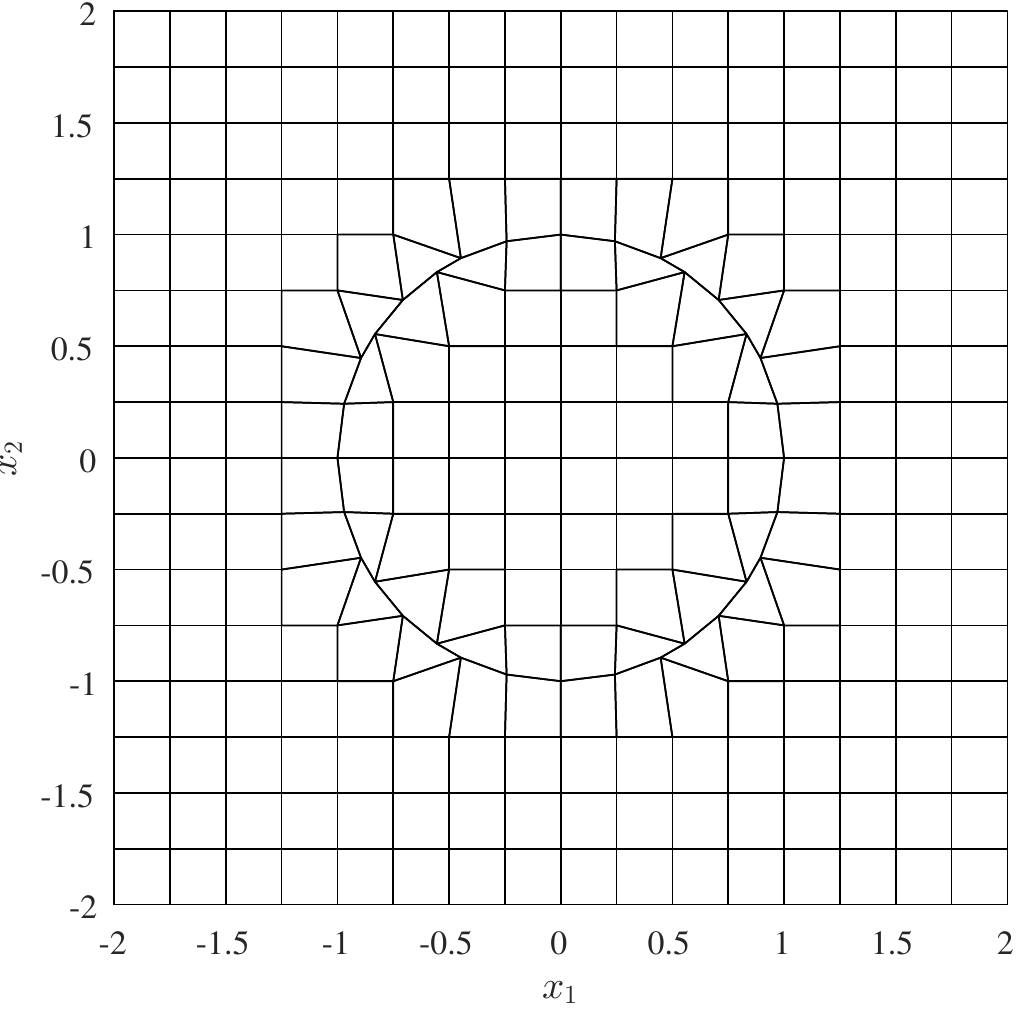}
        \caption{Projection of vertices from a uniform grid onto the internal boundary.\label{fig:mesh_gen_1}}
    \end{subfigure}
    \hfill
    \begin{subfigure}[t]{0.475\textwidth}
        \centering
        \includegraphics[width=\textwidth]{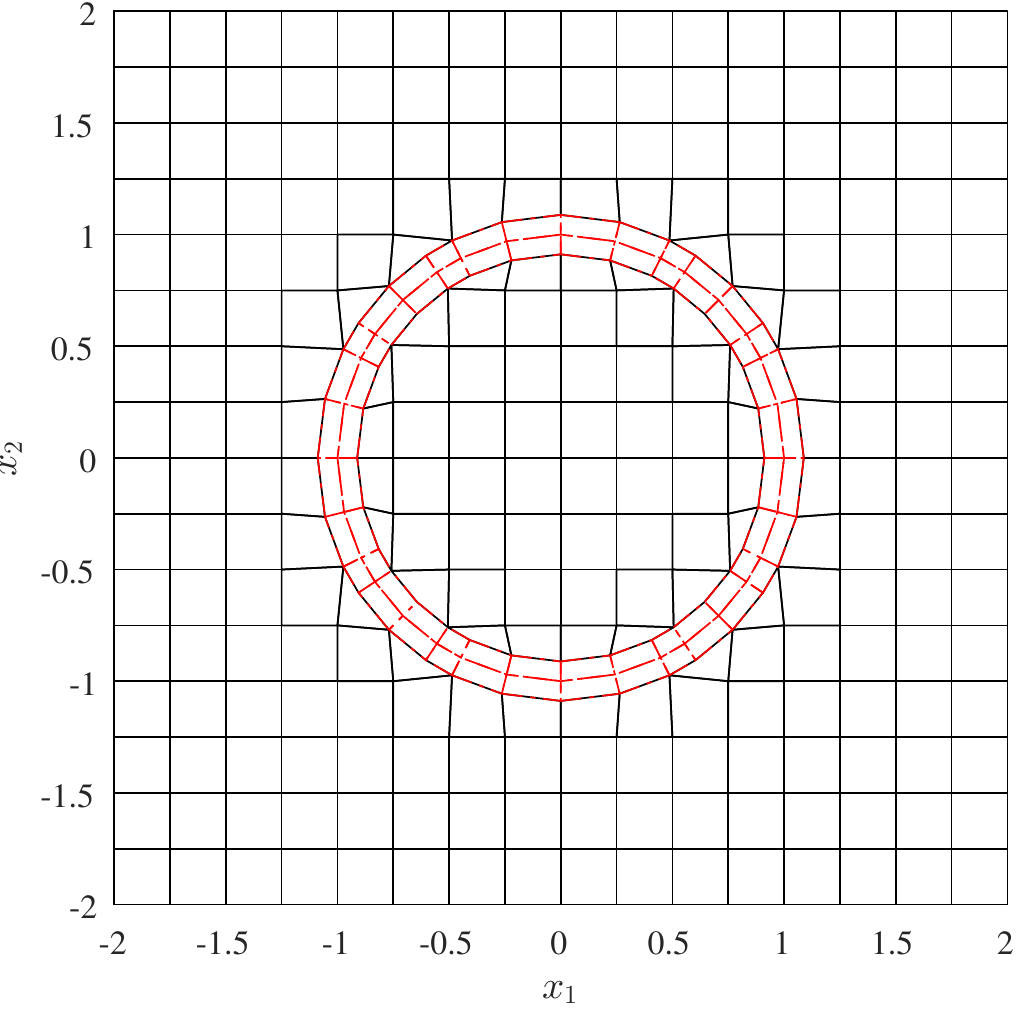}
        \caption{Double layer of pillow elements added to the mesh. New elements are shown with red dashed edges.\label{fig:mesh_gen_2}}
    \end{subfigure}
    \vskip\baselineskip
    \begin{subfigure}[t]{0.475\textwidth}
        \centering
        \includegraphics[width=\textwidth]{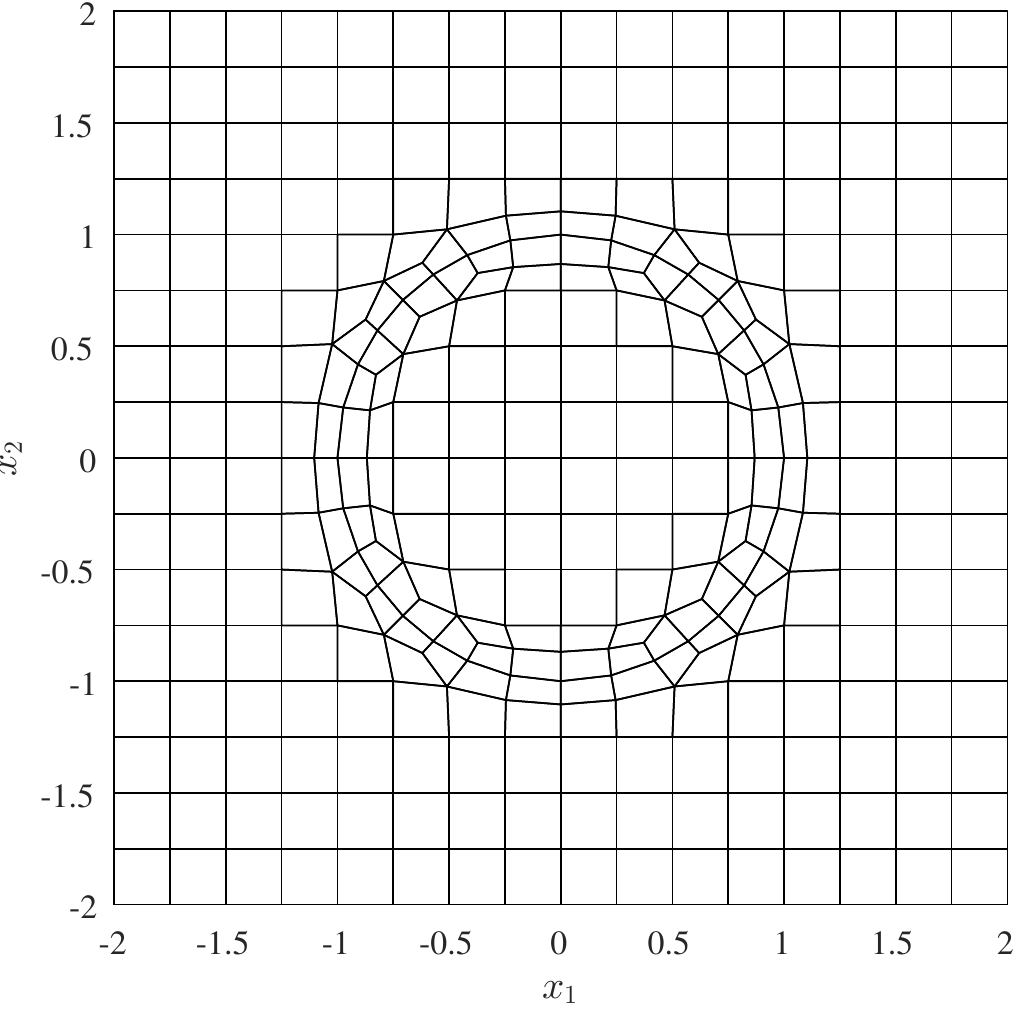}
        \caption{Conforming mesh after Laplacian smoothing and mesh optimization. \label{fig:mesh_gen_3}}
    \end{subfigure}
    \hfill
    \begin{subfigure}[t]{0.475\textwidth}
        \centering
        \includegraphics[width=\textwidth]{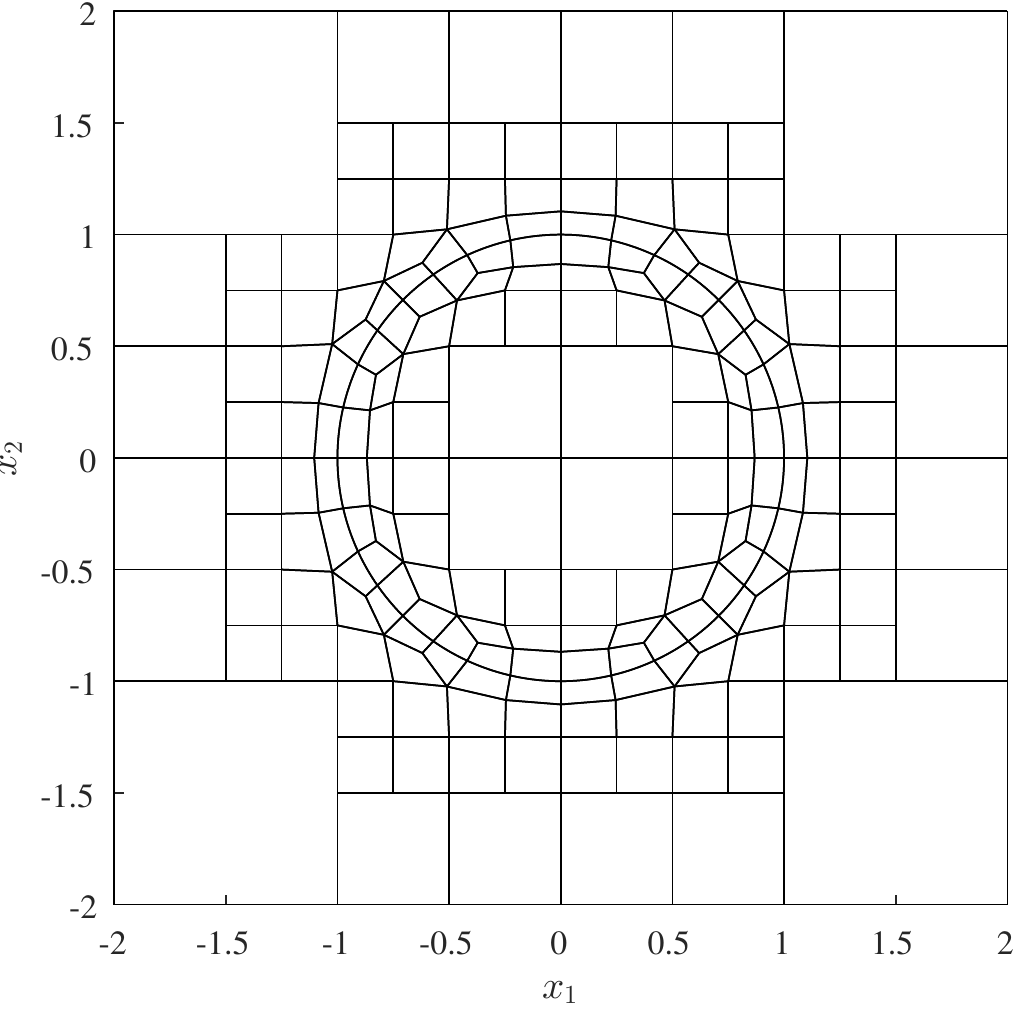}
        \caption{Final non-conforming mesh after quadtree coarsening and introduction of curvilinear edges.\label{fig:mesh_gen_4}}
    \end{subfigure}
    \caption{Mesh generation for an internal boundary given by the unit circle.}
\end{figure*}

The classification of elements is used to determine which vertices to project onto the boundaries of $\hat\Omega_i$. A vertex sharing four elements of the same classification is considered fixed while others are projected. When more than one boundary is present local to the vertex, the decision of which boundary to project to is made by comparing the distance of a point to its approximate projection onto each boundary of $\hat\Omega_i$ (the projection is computed using \eqref{eq:projection_iteration}). A user is also able to specify vertices (e.g. corner vertices of the boundary) that must be part of the final mesh. We satisfy such criteria at this stage by substituting the closest movable vertices with these fixed vertices. Figure \ref{fig:mesh_gen_1} illustrates the mesh after this projection step.

Next, pillow layers of elements are added along all interfaces \cite{Owen2014}. This is done to ensure that the quality of elements near interfaces is not compromised. Pillow layer elements are formed by duplicating the interface vertices on either side of the interface. If $h$ is the original spacing of the uniform grid, then we project the duplicated vertices away from interfaces along the gradients of $\Phi_i$ a distance $\sqrt{2} h / 4$. The duplicated vertices maintain the original connectivity of the interface vertices on either side of the interface and are then connected to the interface vertices themselves to form pillow layer elements that are conforming at the interface. Figure \ref{fig:mesh_gen_2} illustrates the mesh after this pillowing step.

Then a smoothing and mesh optimization step follows. We apply Laplacian smoothing \cite{Ohtake2001} to vertices $\vect{x}_m$ adjacent to and on boundaries by computing their new locations 
\begin{align}
	\vect{x}_{\textrm{new},m} = \sum_{n \in \mathcal{N}_m} \vect{x}_n
\end{align}
where $\mathcal{N}_m$ is the set of indices corresponding to vertices $\vect{x}_n$ sharing an edge connecting to vertex $\vect{x}_m$. Any boundary vertices prior to smoothing are projected back onto their respective boundaries (this means that they can move along the boundary). When boundaries are not convex (e.g. at re-entrant corners), it is possible that Laplacian smoothing produces invalid elements (whose Jacobian determinant corresponding to the transfinite map changes sign).

In such cases, we follow the smoothing step by a local optimization step which returns valid elements where invalid elements were produced \cite{Knupp2001}. Inverted elements are found by computing four signed areas $A_{ij}^{(t)}$ corresponding to triangles within element $j$. If any of the areas is negative, we mark the element as possibly inverted. For each possibly inverted element, we solve a local minimization problem for each vertex of the element which attempts to relocate the vertex so that the signed areas of all elements sharing the vertex become positive. For a given vertex, we solve
\begin{align}
	\vect{x}_{\textrm{untangled}} = \textrm{arg}\min_{\vect{x}} \sum_{\substack{ i=1,2,3,4 \\ j\in\mathcal{J}}} |A_{ij}^{(t)} - \beta A| - (A_{ij}^{(t)} - \beta A)
\end{align}
where $\mathcal{J}$ denotes the set of elements sharing the vertex, $\beta$ is a tolerance used to prevent untangling the element but producing one with zero area (set to the square root of the tolerance that the optimization problem is solved to), and $A$ is the average area of the elements in set $\mathcal{J}$.

Then a final local shape-based optimization is used to improve the quality of the elements that were invalid \cite{Knupp2010}. For each untangled vertex, we solve the local optimization problem
\begin{align}
	\vect{x}_{\textrm{optimized}} = \textrm{arg}\min_{\vect{x}} \sum_{\substack{ i=1,2,3,4 \\ j\in\mathcal{J}}} \|\vect{T}_{ij}\|_F \|\vect{T}_{ij}^{-1}\|_F
\end{align}
with initial iterate $\vect{x}_\textrm{untangled}$. Here $\vect{T}_{ij}$ corresponds to the matrix whose two columns represent two edges of each triangle with area $A_{ij}^{t}$. This tends to produce elements with roughly the same shape (in terms of angles and aspect ratio) and disregards size and orientation properties. We repeat the Laplacian smoothing and optimization steps no more than five times. Figure \ref{fig:mesh_gen_3} illustrates the mesh after three iterations of smoothing and optimization.

At this stage, we have a boundary-fitted mesh that is uniform away from boundaries which consists of quadrilaterals with straight edges. The mesh is conforming, but may be over refined away from fine features of boundaries. Since we started with a quadtree refinement (although only used elements at a uniform level of the tree), we can produce a non-conforming mesh by aggregating elements away from boundaries. We do so only for groups of four adjacent elements generated together by steps of the (quadtree-oriented) recursive subdivision algorithm that belong to a single subdomain $\hat\Omega_i$ and that have all remained unchanged in the mesh. Finally, we use the method in Section \ref{sec:curvilinear_method} to determine transfinite interpolation maps for each element. Only elements on boundaries require the projection step to resolve possibly curved boundaries. Figure \ref{fig:mesh_gen_4} illustrates this final graded and curvilinear mesh.

\subsubsection{Non-Conforming Continuity Constraints} \label{sec:noncomforming_constraints}

To finalize the description of the spectral finite element method, we show how to impose continuity constraints between adjacent elements arising from the mesh generation procedure in Section \ref{sec:mesh_generation}. For clarity, we focus on continuity between one fine element connected to a coarser element (one full edge $\Gamma_i$ of the fine element shares a portion of edge $\Gamma_j$ of the coarse element) then explain how this simple configuration can be applied to enforce continuity on a general mesh. We enforce continuity in the weak sense in the same way that Dirichlet boundary conditions were enforced in Section \ref{sec:2d_spectral_method}. That is, we require
\begin{equation} \label{eq:nonconforming_constraint}
	\int_{\Gamma_i} \vect{\psi} \left( \phi_c(\vect{x}) - \phi_f(\vect{x}) \right) \, d\Gamma = 0
\end{equation}
where $\phi_c$ is the local expansion for $\phi$ corresponding to the coarse element, $\phi_f$ is the local expansion for $\phi$ corresponding to the fine element, and $\vect{\psi}$ is the same set of weight functions as in Section \ref{sec:curvilinear_method}.

To account for situations where the coarse and fine element do not have the same degree of local expansion (say degree $L_c$ and $L_f$ respectively), we work with zero-padded coefficient vectors so as to treat both elements as though they have the same degree. That is, we use degree $L = \max{(L_c,L_f)}$ expansions in canonical coordinates $\phi_c(\vect{u}) = \vect{N}(\vect{u})^T \vect{\phi}_{c,\textrm{pad}}$ and $\phi_f(\vect{u}) = \vect{N}(\vect{u})^T \vect{\phi}_{f,\textrm{pad}}$ where $\vect{\phi}_{c,\textrm{pad}} = \vecm{(\vect{P}_c \vect{\Phi}_c \vect{P}_c^T)}$ and $\vect{\phi}_{f,\textrm{pad}} = \vecm{(\vect{P}_f \vect{\Phi}_f \vect{P}_f^T)}$. The matrices $\vect{P}_c$ and $\vect{P}_f$ extend the coefficient matrices $\vect{\Phi}_c$ and $\vect{\Phi}_f$ by zeros appropriately. That is,
\begin{equation}
	\vect{P}_c = 
	\begin{bmatrix}
    	\vect{I}_{L_c+1} \\
        \vect{0}
	\end{bmatrix}, \qquad
	\vect{P}_f = 
	\begin{bmatrix}
    	\vect{I}_{L_f+1} \\
        \vect{0}
	\end{bmatrix},
\end{equation}
where $\vect{I}_{n}$ denotes the identity matrix in $\mathbb{R}^{n \times n}$ and $\vect{P}_c\in \mathbb{R}^{(L+1) \times (L_c+1)}$ and $\vect{P}_f\in \mathbb{R}^{(L+1) \times (L_f+1)}$. Note that, by construction, one of the two padding matrices will be the identity matrix. Using such a padding induces modified constraint matrices
\begin{equation} \label{eq:fine_constraints}
	\vect{C}_i^f =
	\begin{cases}
		\displaystyle
		\vect{P}_f \otimes \left(\vect{e}_1 + (-1)^i \vect{e}_2\right)^T \vect{P}_f & i=1,2,\\
		\displaystyle
		\left(\vect{e}_1 + (-1)^i \vect{e}_2\right)^T \vect{P}_f \otimes \vect{P}_f & i=3,4,
	\end{cases}
\end{equation}
similar to those from Section \ref{sec:2d_spectral_method}. The term $\int_{\Gamma_i}\vect{\psi}\phi_f(\vect{x})\,d\Gamma$ in constraint \eqref{eq:nonconforming_constraint} gives rise to the discretized form $\vect{C}_i^f\vect{\phi}_f$ with $\vect{\phi}_f=\vecm{(\vect{\Phi}_f)}$.

The term $\int_{\Gamma_i}\vect{\psi}\phi_c(\vect{x})\,d\Gamma$ is complicated by the fact that integration is performed only over a portion of the edge of the coarse element. However, Section \ref{sec:affine_transformation} provides the requisite basis transformation to account for this difficulty. That is, we define the change of basis matrix 
\begin{equation}
	\vect{L}_c^f = \vect{S} \vect{L} \vect{S}^{-1}
\end{equation}
where $\vect{L}$ is the matrix satisfying $\vect{p}(ax+b) = \vect{L} \vect{p}(x)$ for the appropriate choice of parameters $a$ and $b$ (notice that $\vect{N}(ax+b) = \vect{L}_c^f \vect{N}(x)$). The parameters $a$ and $b$ depend on the portion of $\Gamma_j$ shared with $\Gamma_i$ (for example, if $\Gamma_i$ corresponds to half of $\Gamma_j$, then $a=1/2$ and $b=\pm1/2$). In addition, we define $\vect{C}_i^c$ as in \eqref{eq:fine_constraints} with $\vect{P}_c$ replacing $\vect{P}_f$ so that $\int_{\Gamma_i}\vect{\psi}\phi_c(\vect{x})\,d\Gamma$ gives rise to the discretized form $(\vect{L}_c^f)^T \vect{C}_j^c \vect{\phi}_c$ where $\vect{\phi}_c=\vecm{(\vect{\Phi}_c)}$. The index $j$ in $\vect{C}_j^c$ must correspond to the edge $\Gamma_j$ of the coarse element that shares a portion of edge $\Gamma_i$ of the fine element ($j$ may be different from $i$). The resulting constraint equations corresponding to \eqref{eq:nonconforming_constraint} are
\begin{equation} \label{eq:final_constraint}
	(\vect{L}_c^f)^T \vect{C}_j^c \vect{\phi}_c - \vect{C}_i^f\vect{\phi}_f = \vect{0}.
\end{equation}
They simplify considerably when the polynomial degrees $L_c$ and $L_f$ match (making $\vect{P}_c = \vect{P}_f = \vect{I}$) and the edges $\Gamma_i$ and $\Gamma_j$ match (making $\vect{L}_c^f = \vect{I}$).

When more than two elements are used to discretize a given domain $\Omega$ as in Section \ref{sec:mesh_generation}, we formulate a spectral finite element method as in Section \ref{sec:spectral_FEM} where we obtain block diagonal matrix $\vect{A}$ and block vectors $\vect{\phi}$ and $\vect{b}$ (see \eqref{eq:block_diag_matrices}) using the methods of Sections \ref{sec:2d_spectral_method} and \ref{sec:curvilinear_method} applied to each element independently. This leaves assembling the global constraint matrix $\vect{C}$ using \eqref{eq:final_constraint} along each edge in the mesh.

In the absence of additional constraints, constraint equations of the form \eqref{eq:final_constraint} possess full row rank (assuming infinite precision) but may become rank deficient when taken together in a global constraint matrix imposing continuity along all element edges. A systematic procedure can be followed to restore full row rank in the global case (similar to the method described in Remark \ref{rem:rank_def_remark}). First, we construct a set of vertex to edge incidence lists (one for each level of the quadtree) which enumerate distinct vertices that belong to elements of a given level of the tree and track which edges of elements on that level are incident to each vertex. Then, we construct the global constraint matrix in a local fashion, visiting each element edge (beginning with all elements belonging to the finest level of the tree, then moving to the next coarser level, etc.) and imposing continuity to an adjacent leaf element in the tree. The tree data structure allows us to determine the parameters $a$ and $b$ in the affine map along each edge so as to compute the appropriate change of basis matrix $\vect{L}_c^f$. Each time continuity is imposed along an edge, we verify if all edges on that given level of the tree have been visited for a given vertex in the vertex to edge incidence list. For each vertex whose list has been completely visited, we have one redundant equation in the global constraint matrix, and can remove the lowest order equation from the new set of local constraints to restore full rank to the global system.

There is also the possibility of losing global full rank when multiple fine elements share a common edge with a single coarse element. In particular, when an edge must be constrained to match a lower polynomial degree on an adjacent element, we zero certain basis functions through the padding matrices $\vect{P}_c$ or $\vect{P}_f$. If a second set of constraint equations requires a similar reduction in the degree along the edge, naively imposing the local constraints leads to a second set of redundant constraints zeroing the same basis functions. Instead, we track which basis functions have been zeroed on an edge and discard those constraints which would lead to redundancy. These are always the higher order equations in the local constraint and do not interfere with the removal of the lowest order constraints described above.

\begin{figure}[!t]
	\centering
	\includegraphics[scale=0.75]{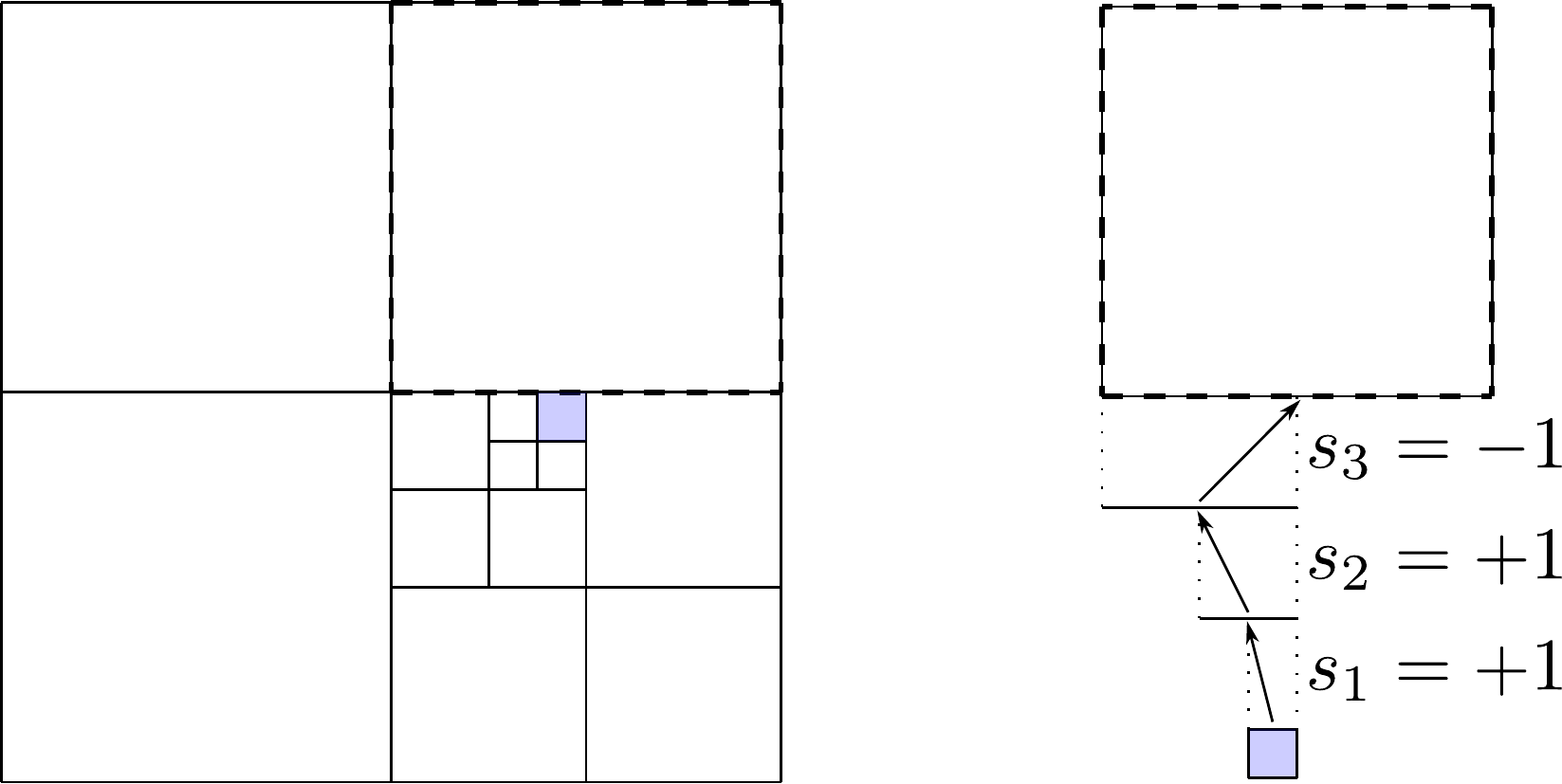}
	\caption{(left) Quadtree mesh where continuity between a fine element (shaded) and neighbouring coarse element above it (dashed) must be imposed. (right) Schematic illustrating the selection of signs $s_i$ for the associated edge configuration: $s_3$ is negative because the fine element shares a portion of the left bisection of the coarse edge, $s_2$ is positive because the fine element shares a portion of the right bisection of the previous bisection, and $s_1$ is positive because the fine element edge corresponds to the right bisection of the previous bisection. \label{fig:quadtree_sequence}}
\end{figure}

It remains to explain how parameters $a$ and $b$ are selected at each edge. By virtue of sequentially imposing constraints element by element starting at the finest level of the tree and moving to the next level of the tree after all finer element constraints have been imposed, we always impose constraints in the form \eqref{eq:final_constraint} (that is, always from the perspective of a fine element connecting to a coarser element). The quadtree lets us query the neighbour of an element on the same level of the tree (which may not be a leaf). We can then trace up the quadtree (by going to the parent of the neighbour, then its parent, etc.) until a leaf is found. If no leaf is found, this edge corresponds to a coarse edge relative to a previous finer level and has already been constrained. If a leaf is found, we must impose the constraint. We keep track of the sequence of nodes in the quadtree that were visited in finding the leaf to compute $a$ and $b$. Because each level of the quadtree is comprised of squares constructed by subdividing a square on the previous level into four squares of equal size, the affine transformation used to represent the coarse basis functions in terms of fine basis functions is of the form $ax+b = \{ [ ( x + s_1 ) /2 + s_2 ] /2 + \cdots \} /2$. The signs $s_i\in\{-1,+1\}$ in this composition of functions $(x + s_i)/2$ depend on the sequence of nodes used to find the neighbour leaf node starting from the fine neighbour. Figure \ref{fig:quadtree_sequence} illustrates one possible configuration of fine and coarse elements and the sequence of signs required to determine $a$ and $b$ for a given edge constraint. If there is a difference of $m$ levels between the fine element to the coarse leaf neighbour, then 
\begin{equation} \label{eq:affine_parameters}
	a = \frac{1}{2^m}, \qquad b = \frac{1}{2^m}\sum_{i=1}^m s_i 2^{i-1},
\end{equation}
where $s_1$ is the sign corresponding to the first step up the quadtree and $s_m$ corresponds to the $m$th step required to finally reach the neighbouring leaf.

With the global constraint matrix $\vect{C}$ assembled, we obtain a saddle point system of  form \eqref{eq:2d_saddle_point_system} where each block now depends upon multiple elements (rather than a single element). One benefit of such a construction is that the change of basis matrix $\vect{L}_c^f$ along each edge yields a quantitative way of assessing whether the saddle point system is invertible in finite precision. In particular, it quantifies whether the mismatch in adjacent polynomial degree and element size is too severe, leading to extreme ill-conditioning of the global saddle point system and suggests what type of refinement to avoid in order to compute accurate solutions.

This ill-conditioning arises because the matrix $\vect{L}_c^f$ tends to have decaying diagonal entries when the number of levels $m$ between coarse and fine element is sufficiently large and/or the polynomial degree is large enough. The diagonals decay because row $k+1$ of $\vect{L}$ corresponds to the coefficients in a Legendre expansion of the function $p_k(ax+b)$. When $x \in (-1,1)$ and $a$ and $b$ are chosen as in \eqref{eq:affine_parameters}, then $ax+b$ represents a small subinterval of $(-1,1)$ (particularly when the number of levels $m$ is large). As a consequence, $p_k(ax+b)$ varies slowly over $(-1,1)$ compared to $p_k(x)$ and only a small number of low degree Legendre polynomials make a significant contribution to the expansion (higher degree polynomials contribute, but substantially less) causing entries near the diagonal of $\vect{L}$ to decay in magnitude. Matrix $\vect{L}_c^f$ inherits this property from $\vect{L}$.

Since the constraints along edges involve the product of the transpose of $\vect{L}_c^f$ (which is upper triangular with entries in the final rows possibly decaying), corresponding rows of the global constraint matrix $\vect{C}$ may become near zero, causing an effective loss of full row rank in finite precision when certain combinations of polynomial degree and edge refinements are made. Introducing a 2:1 mesh refinement rule common to quadtree-based finite element meshes \cite{Frey2008} can alleviate this problem with low or moderate degree polynomials on neighbouring elements, but is ineffective at large polynomial degree (even at degree 32, the diagonal has decayed to approximately $10^{-10}$). Instead, we can monitor the magnitude of the diagonal and discard any constraints which have decayed below a user specified threshold. This results in a non-conforming finite element method. Alternatively, we can use the diagonal to indicate where in the mesh to prohibit further refinement of degree or element size.

\subsection{Solution of the Saddle Point System via Domain Decomposition} \label{sec:decomposition}

The discretization process described in Sections \ref{sec:2d_spectral_method}, \ref{sec:curvilinear_method}, \ref{sec:mesh_generation}, and \ref{sec:noncomforming_constraints} leads to a saddle point system
\begin{equation} \label{eq:second_saddle_point_system}
	\begin{bmatrix}
    	\vect{A} & \vect{C}^T \\
    	\vect{C} & \vect{0} 
	\end{bmatrix}
	\begin{bmatrix}
    	\vect{\phi} \\
    	\vect{\nu}
	\end{bmatrix}
	=
	\begin{bmatrix}
    	\vect{b} \\
    	\vect{d}
	\end{bmatrix}
\end{equation}
where $\vect{A}$ is block diagonal. In this section, we describe a domain decomposition algorithm which solves this system. While our derivation of the saddle point system employed the weak form and Galerkin's method, we note that the same saddle point system can be obtained discretizing an appropriate functional and finding the stationary point of the Lagrangian
\begin{equation}
	\mathcal{L}(\vect{\phi},\vect{\nu}) = \frac{1}{2} \vect{\phi}^T \vect{A} \vect{\phi} - \vect{\phi}^T \vect{b} + \vect{\nu}^T(\vect{C}\vect{\phi} - \vect{d})
\end{equation}
where $\vect{\phi}$ are primal variables and $\vect{\nu}$ corresponding dual variables (Lagrange multipliers). Consequently, we refer to unknowns $\vect{\phi}$ and $\vect{\nu}$ as primal and dual respectively in subsequent sections.

In Section \ref{sec:dual_primal_algorithm}, we describe a general dual-primal domain decomposition framework applicable to solving \eqref{eq:second_saddle_point_system} which splits the constraints into two sets, and explicitly enforces one set via a nullspace method. Then, in Section \ref{sec:basis_for_nullspace}, we describe how to choose a sparse basis for the nullspace required by the dual-primal method for the constraint matrices that arise in Section \ref{sec:extension}. Section \ref{sec:choosing_basis_for_nullspce} explains how this choice affects the dual-primal method, and describes how to exploit the resulting structure of the problem. Finally, Section \ref{sec:consideration_for_helmholtz} emphasizes how to adapt the domain decomposition algorithm to the Helmholtz problem to obtain convergence in a number of iterations only weakly dependent on the wavenumber $k$.

\subsubsection{A Dual-Primal Algorithm} \label{sec:dual_primal_algorithm}

Rather than solve only for the primal variables as in a standard finite element method, domain decomposition methods of FETI-DP type (DP stands for Dual-Primal) begin by directly enforcing a subset of the constraints in $\vect{C}$. Suppose that the constraint matrix is partitioned into two submatrices such that
\begin{equation} \label{eq:large_saddle}
	\begin{bmatrix}
    	\vect{A}   & \vect{C}_p^T & \vect{C}_d^T \\
    	\vect{C}_p & \vect{0}     & \vect{0}     \\
    	\vect{C}_d & \vect{0}     & \vect{0} 
	\end{bmatrix}
	\begin{bmatrix}
    	\vect{\phi} \\
    	\vect{\nu}_p \\
    	\vect{\nu}_d
	\end{bmatrix}
	=
	\begin{bmatrix}
    	\vect{b} \\
    	\vect{d}_p \\
    	\vect{d}_d
	\end{bmatrix}
\end{equation}
so that constraints in $\vect{C}_p$ are meant to be imposed first. Using a partial nullspace method, we let
\begin{equation} \label{eq:null_space_identity}
	\vect{\phi} = \vect{Z} \vect{\phi}_p + \vect{\phi}_z
\end{equation}
with $\vect{C}_p \vect{Z} = \vect{0}$ such that $\vect{Z}$ is a basis for the nullspace of $\vect{C}_p$. We call this a partial nullspace method because the standard nullspace method chooses $\vect{Z}$ as a basis for the nullspace of the full constraint matrix $\vect{C}$ \cite{Benzi1999}. Substituting \eqref{eq:null_space_identity} into \eqref{eq:large_saddle} and multiplying the first row by $\vect{Z}^T$ yields two systems
\begin{equation} \label{eq:assembled_saddle_point_system}
	\vect{C}_p \vect{\phi}_z = \vect{d}_p, \qquad
	\begin{bmatrix}
    	\hat{\vect{A}} & \hat{\vect{C}}{}^T \\
    	\hat{\vect{C}} & \vect{0} 
	\end{bmatrix}
	\begin{bmatrix}
    	\vect{\phi}_p \\
    	\vect{\nu}_d
	\end{bmatrix}
	=
	\begin{bmatrix}
    	\hat{\vect{b}} \\
    	\hat{\vect{d}}
	\end{bmatrix},
\end{equation}
where $\hat{\vect{A}} = \vect{Z}^T \vect{A} \vect{Z}$, $\hat{\vect{C}} = \vect{C}_d \vect{Z}$, $\hat{\vect{b}} = \vect{Z}^T(\vect{b} - \vect{A} \vect{\phi}_z)$, and $\hat{\vect{d}} = \vect{d}_d - \vect{C}_d \vect{\phi}_z$. The first system can be satisfied by $\vect{\phi}_z = \vect{C}_p^+ \vect{d}_p$ using the Moore-Penrose pseudoinverse $\vect{C}_p^+ = \vect{C}_p^T(\vect{C}_p\vect{C}_p^T)^{-1}$ since $\vect{C}_p$ has full row rank.

Rather than solve the second system directly, we eliminate the primal variables using the range space method \cite{Benzi1999} to obtain
\begin{equation} \label{eq:schur_complement_system}
	\hat{\vect{C}} \hat{\vect{A}}{}^{-1} \hat{\vect{C}}{}^T \vect{\nu}_d = \hat{\vect{C}} \hat{\vect{A}}{}^{-1} \hat{\vect{b}} - \hat{\vect{d}}
\end{equation}
and use a left preconditioned Krylov subspace method to solve for $\vect{\nu}_d$. Instead of the preconditioner
\begin{equation} \label{eq:lumped_preconditioner}
	\vect{P}^{-1} = (\hat{\vect{C}}{}^+)^T \hat{\vect{A}} \hat{\vect{C}}{}^+
\end{equation}
(see, for example, \cite{Elman1999}), we use a modified approach.
In particular, since 
\begin{equation}
	\vect{P}^{-1} \hat{\vect{C}} \hat{\vect{A}}{}^{-1} \hat{\vect{C}}{}^T
	= (\hat{\vect{C}}{}^+)^T \hat{\vect{A}} \underbrace{\hat{\vect{C}}{}^+ \hat{\vect{C}}}_{\textrm{projection}} \hat{\vect{A}}{}^{-1} \hat{\vect{C}}{}^T
\end{equation}
we replace $\hat{\vect{C}}{}^+$ in $\vect{P}^{-1}$ by
\begin{equation} \label{eq:dirichlet_preconditioner}
	\hat{\vect{C}}{}^{\star} = \vect{W} \hat{\vect{C}}{}^T(\hat{\vect{C}} \vect{W} \hat{\vect{C}}{}^T)^{-1}
\end{equation}
which preserves the projection property of $\hat{\vect{C}}{}^{\star} \hat{\vect{C}}$. When $\vect{W} = \vect{I}$, then $\hat{\vect{C}}{}^{\star} = \hat{\vect{C}}{}^+$ and the projection is an orthogonal projection onto the range of $\hat{\vect{C}}{}^T$. Other choices of $\vect{W}$ result in projections that are not orthogonal, but that produce more effective preconditioners (that is, they reduce the number of iterations required by the preconditioned Krylov subspace method to converge to a given tolerance). Once $\vect{\nu}_d$ has been computed, we find $\vect{\phi}_p$ by solving
\begin{equation}
	\hat{\vect{A}} \vect{\phi}_p = \hat{\vect{b}} - \hat{\vect{C}}{}^T \vect{\nu}_d
\end{equation}
and compute \eqref{eq:null_space_identity} to obtain the original primal unknown. If needed, the Lagrange multipliers $\vect{\nu}_p$ can be obtain using
\begin{equation}
	\vect{\nu}_p = (\vect{C}_p^+)^T(\vect{b} - \vect{C}_d^T \vect{\nu}_d - \vect{A} \vect{\phi}).
\end{equation}

\subsubsection{A Sparse Basis for the Nullspace} \label{sec:basis_for_nullspace}

In order to complete our description of the domain decomposition method, we must specify how to partition the constraints into matrices $\vect{C}_p$ and $\vect{C}_d$ and how to construct the basis $\vect{Z}$ for the nullspace of $\vect{C}_p$. One particularly important constraint on the basis $\vect{Z}$ is that it should be sparse, otherwise we risk taking the original sparse finite element problem and transforming it into a dense problem. Fortunately, such a basis exists and can be computed in a straightforward manner.

Choosing the subset of constraints $\vect{C}_p$ depends on how we wish to perform domain decomposition. In the following, we will consider each element to belong to its own domain in the decomposition algorithm. We restrict our attention to this case because we are describing a spectral element method aimed at computing highly accurate solutions with potentially large polynomial degree on each element. This is not typical of domain decomposition algorithms applied to finite element methods of low polynomial degree because the number of unknowns for a single element is small. Instead, several elements are grouped together into domains and the solutions of the resulting local problems are performed using direct methods in parallel. While grouping several elements into a single domain is possible using our method, it is not clear how to directly apply the fast solver \cite{Shen1994} for local problem solution in such cases.

\begin{remark} \label{rem:remark_graded_domains}
	An interesting problem that we have not addressed arises when performing domain decomposition using domains of a single element for problems with corner singularities whose mesh has been obtained iteratively through $hp$ refinement. In such situations, the mesh tends to be graded near singularities of the solution, requiring small elements of low degree near singularities and large elements of high degree away from them. As such, problems with load balancing can occur when using a single-element domain decomposition method in parallel. The key is modifying matrix $\vect{C}_p$ to group elements of low degree into domains separate from other domains with single elements of high degree. \remarkend
\end{remark}

In a typical FETI-DP method, the constraint matrix $\vect{C}_p$ corresponds to constraints enforcing continuity at so-called cross points of the decomposition (vertices where several domains meet) and, in three dimensions, averages along boundaries between domains \cite{Toselli2005}. By virtue of imposing continuity constraints in the weak sense of \eqref{eq:nonconforming_constraint}, we have a hierarchy of constraints along each edge whose lowest order constraints fit into the standard FETI-DP framework. To enforce continuity at vertices between elements, include the first constraint equation from \eqref{eq:final_constraint} along each edge in $\vect{C}_p$. Note that care must be taken in situations where certain constraints were discarded (as described in Section \ref{sec:noncomforming_constraints}) in assembling $\vect{C}$. In particular, having been discarded, no constraint from the associated edge should be added to $\vect{C}_p$.

However, in contrast with a standard FETI-DP method, the hierarchy of constraints along each edge means that we can include however many constraints we wish along a given edge in $\vect{C}_p$. In practice, we keep this number, called $l$, fixed across the whole mesh (with the understanding that if the first $l_{\textrm{discard}}$ constraints have been discarded along an edge, we only include the remaining $l-l_{\textrm{discard}}$ constraints corresponding to that edge). Note that if $l$ exceeds the polynomial degree for all elements (all constraints are to be included), then $\vect{C}_p = \vect{C}$ and the domain decomposition method reduces to the nullspace method (that is, we globally assemble the finite element problem and there is no domain decomposition to speak of). If $l=0$ (no constraints are to be included), then $\vect{C}_d = \vect{C}$ and the domain decomposition method reduces to the range space method (but then the domain decomposition method does not possess a coarse space and the number of iterations required for the Krylov subspace method to converge can be large).

Having defined $\vect{C}_p$, we make use of the orthogonal projection $\vect{I} - \vect{C}_p^+ \vect{C}_p$ onto the nullspace of $\vect{C}_p$ to define a sparse basis for the nullspace. Unfortunately, this matrix has more columns than the number of columns in such a basis, a consequence of the rank-nullity theorem. However, if we change basis, choosing an appropriate set of linearly independent columns is straightforward. We use the connection \eqref{eq:interpolatory_relation} between integrated Legendre polynomials and the basis with first two polynomials interpolatory to define the block diagonal matrix $\vect{B}_z = \diag{(\vect{B}_j)}$ with block entries $\vect{B}_j = \vect{B} \otimes \vect{B}$. The size of matrices $\vect{B}$ in $\vect{B}_j$ should be chosen to agree with the local polynomial degree of each element $j$ so that matrix products of $\vect{B}_z$ conform with $\vect{\phi}$. Then the constraint equations $\vect{C}_p \vect{\phi} = \vect{d}_p$ are equivalent to $\vect{C}_p \vect{B}_z^2 \vect{\phi} = \vect{d}_p$ since $\vect{B}$ is involutory. Defining $\tilde{\vect{\phi}} = \vect{B}_z \vect{\phi}$ leads to constraints 
\begin{equation}
	\vect{C}_p \vect{B}_z \tilde{\vect{\phi}} = \vect{d}_p.
\end{equation}
This alternative representation of the constraints is useful because the coefficients in $\tilde{\vect{\phi}}$ correspond to basis functions that: interpolate the vertices of each element (vertex functions); are non-zero on edges of each element but zero at vertices (edge functions); are zero on all edges of each element (interior functions). It is easier to determine which columns of the orthogonal projection 
\begin{equation}
	\vect{I} - (\vect{C}_p \vect{B}_z)^+ (\vect{C}_p \vect{B}_z) = \vect{B}_z (\vect{I} - \vect{C}_p^+ \vect{C}_p) \vect{B}_z
\end{equation}	
to discard because of the interpretation of unknowns $\tilde{\vect{\phi}}$. If we keep columns indexed by the set $\mathcal{J}$, then a sparse basis for the nullspace of $\vect{C}_p \vect{B}_z$ is
\begin{equation}
	\tilde{\vect{Z}} = [ \vect{B}_z (\vect{I} - \vect{C}_p^+ \vect{C}_p) \vect{B}_z ] \, \vect{I}(:,\mathcal{J})
\end{equation}
where $\vect{I}(:,\mathcal{J})$ corresponds to the columns of the identity matrix indexed by set $\mathcal{J}$. Given $\tilde{\vect{Z}}$, a sparse basis for the nullspace of $\vect{C}_p$ is 
\begin{align}
	\vect{Z} &= \vect{B}_z \tilde{\vect{Z}} \\
		     &= (\vect{I} - \vect{C}_p^+ \vect{C}_p) \vect{B}_z \, \vect{I}(:,\mathcal{J}).
\end{align}

Choosing the set $\mathcal{J}$ can be performed systematically because of the direct correspondence between unknowns $\tilde{\vect{\phi}}$ and columns of $\vect{B}_z (\vect{I} - \vect{C}_p^+ \vect{C}_p) \vect{B}_z$. Starting from the finest (i.e. lowest) level of the quadtree, we visit each element. For each element, we visit each edge (we keep track of each edge that has been visited). We then check if the edge belongs to a boundary where a boundary condition is meant to be imposed. If it does, and the edge corresponds to a Dirichlet boundary condition, we discard the first $l-1$ edge unknowns corresponding to that edge. If the edge corresponds to a Robin boundary condition, we keep the first $l-1$ edge unknowns corresponding to that edge. If the edge does not correspond to a boundary condition and it has not been visited yet, we find the current element's neighbour that shares the current edge. If the neighbour is on the same level of the tree, then we keep the first $l-1$ edge unknowns of the element corresponding to that edge and discard the first $l-1$ edge unknowns of the neighbour corresponding to the shared edge. If the neighbour is on a coarser (i.e. higher) level of the tree, then we discard the first $l-1$ edge unknowns of the element corresponding to that edge and keep all edge unknowns of the neighbour corresponding to the shared edge.

We must also consider the vertex unknowns. While visiting the edges of the mesh (in the order described in the previous paragraph), if the neighbour element is on a coarser (i.e. higher) level, we discard the vertex unknowns of the current element corresponding to that shared edge. If the neighbour is on the same level or the current edge has multiple smaller neighbours on a finer (i.e. lower) level, then one vertex unknown per vertex must be kept, but all other vertex unknowns corresponding to the same vertex must be discarded. We classify all of these kept vertex and edge unknowns as type 2. They are associated with explicitly enforcing the constraints in $\vect{C}_p$ along edges and at vertices. All interior unknowns are kept and all edge unknowns that were not discarded or already classified as type 2 are kept and classified as type 1. They are unknowns that are only coupled through the matrix $\vect{C}_d$ or that are coupled only with other unknowns on an element-by-element basis (e.g. interior unknowns for a given element). We make sure that the splitting into two types of unknowns is reflected in the index set $\mathcal{J}$ so that 
\begin{equation}
	\vect{Z} = 
	\begin{bmatrix}
    	\vect{Z}_1 & \vect{Z}_2
	\end{bmatrix}.
\end{equation}
Algorithm \ref{alg:pseudocode} summarizes how to compute the index set using pseudocode.

\begin{algorithm}
	\begin{algorithmic}
	\ForAll{levels of the quadtree (starting from finest level)}
		\ForAll{elements on current level}
			\ForAll{edges of current element}
				\State // \textit{discard redundant edge unknowns}
			 	\If{edge belongs to boundary}
			 		\If{edge belongs to Dirichlet boundary}
			 			\State discard first $l-1$ edge unknowns
			 		\Else
			 			\State keep first $l-1$ edge unknowns
			 		\EndIf
			 	\Else
					\If{edge has not been visited}
						\State find neighbour element and edge
						\If{neighbour is on same level}
							\State keep first $l-1$ edge unknowns
							\State discard first $l-1$ edge unknowns of neighbour
						\ElsIf{neighbour is on coarser level}
							\State discard first $l-1$ edge unknowns
							\State keep all edge unknowns of neighbour
						\EndIf
						\State mark current edge and neighbour edge as visited
					\EndIf
				\EndIf
				\State // \textit{discard redundant vertex unknowns}
				\If{edge belongs to Dirichlet boundary \textbf{or} edge is adjacent to coarse neighbour}
					\State discard vertices
				\ElsIf{vertices have not already been visited}
					\State keep vertex unknowns and mark these vertices as visited
				\EndIf
			\EndFor
		\EndFor
	\EndFor
	\State set all interior unknowns \textbf{and} edge unknowns not yet kept or discarded as type 1
	\State set all kept edge and vertex unknowns as type 2
	\State \textbf{return} unknowns of type 1 and 2
	\end{algorithmic}
	
	\caption{Pseudocode to determine the index set $\mathcal{J}$. \label{alg:pseudocode}}
\end{algorithm}

\begin{figure}[!t]
	\centering
	\includegraphics[scale=0.9]{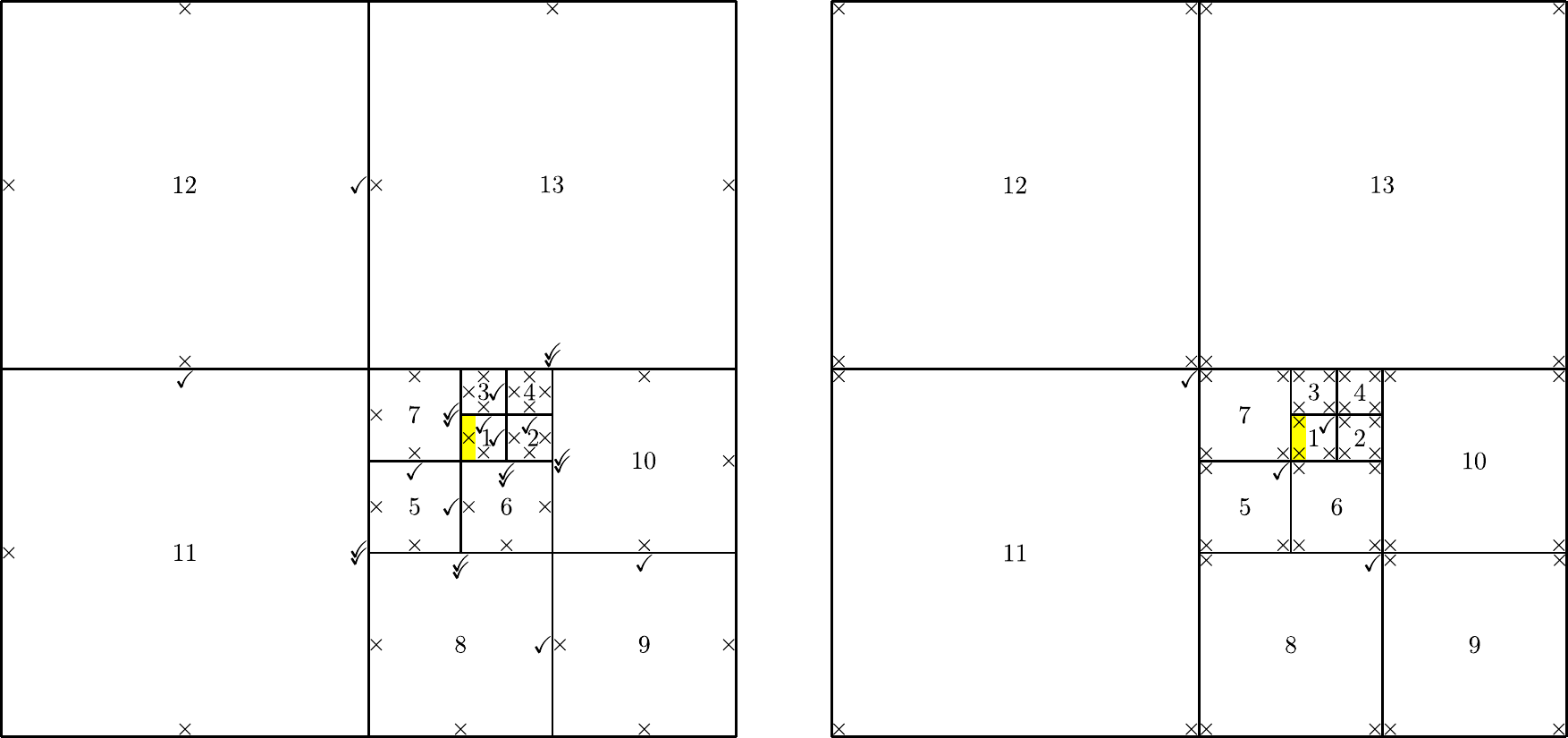}
	\caption{(left) Illustration of edge unknowns (columns of $\tilde{\vect{Z}}$) to keep when constructing $\vect{Z}$. The first $l-1$ edge unknowns are discarded on edges marked with an $\times$ and kept on edges marked with a $\checkmark$. A double $\checkmark$ indicates edges where all edge unknowns are kept. (right) Illustration of vertex unknowns to keep. An $\times$ indicates that the corresponding vertex unknown is discarded whereas a $\checkmark$ indicates that it is kept. The first edge and vertices visited by the algorithm are highlighted in yellow. \label{fig:edge_vertex_labeling}}
\end{figure}

As an example, Figure \ref{fig:edge_vertex_labeling} illustrates which edge and vertex unknowns to keep in set $\mathcal{J}$ for a given quadtree mesh. For this example, we have assumed that the boundaries of the square have Dirichlet boundary conditions imposed. The particular set of unknowns shown in Figure \ref{fig:edge_vertex_labeling} is selected by following a $z$-ordering of the quadtree from the finest level up to the coarsest level (indicated by the numbering). The edges for each element are visited following the sequence: left, right, bottom, top (the same order as defined in Section \ref{sec:2d_spectral_method}). For clarity, the edge and vertex diagrams are separated, but in practice, the edges and vertices can be tracked simultaneously.

\subsubsection{Consequences of this Choice of Basis for the Nullspace} \label{sec:choosing_basis_for_nullspce}

The partitioning of constraints into $\vect{C}_p$ and $\vect{C}_d$ and the choice of the associated basis for the nullspace $\vect{Z}$ in Section \ref{sec:basis_for_nullspace} affect the partially assembled saddle point system \eqref{eq:assembled_saddle_point_system} in Section \ref{sec:dual_primal_algorithm}. In particular, the splitting in $\vect{Z}$ induces the splitting
\begin{equation} \label{eq:block_matrices}
	\hat{\vect{A}} = 
	\begin{bmatrix}
    	\vect{A}_{11} & \vect{A}_{12} \\
    	\vect{A}_{12}^T & \vect{A}_{22}
	\end{bmatrix}, \qquad
	\hat{\vect{b}} = 
	\begin{bmatrix}
    	\vect{b}_{1} \\
    	\vect{b}_{2}
	\end{bmatrix}, \qquad
	\hat{\vect{C}} = 
	\begin{bmatrix}
    	\vect{C}_{1} & \vect{C}_{2}
	\end{bmatrix},
\end{equation}
with $\vect{A}_{ij} = \vect{Z}_i^T \vect{A} \vect{Z}_j$, $\vect{b}_i = \vect{Z}_i^T (\vect{b} - \vect{A} \vect{\phi}_z)$, and $\vect{C}_j = \vect{C}_d \vect{Z}_j$. By construction, $\vect{A}_{11}$ is block diagonal (one block for each element) with each block corresponding to edge and interior unknowns that have not been constrained by $\vect{C}_p$. We exploit this block diagonal structure using block matrix inversion
\begin{equation} \label{eq:block_matrix_inversion}
	\begin{bmatrix}
    	\vect{A}_{11} & \vect{A}_{12} \\
    	\vect{A}_{12}^T & \vect{A}_{22}
	\end{bmatrix}^{-1} = 
	\begin{bmatrix}
    	\vect{A}_{11}^{-1} + \vect{A}_{11}^{-1} \vect{A}_{12} \vect{K}^{-1} \vect{A}_{12}^T \vect{A}_{11}^{-1} & -\vect{A}_{11}^{-1} \vect{A}_{12} \vect{K}^{-1} \\
    	- \vect{K}^{-1} \vect{A}_{12}^T \vect{A}_{11}^{-1} & \vect{K}^{-1}
	\end{bmatrix}
\end{equation}
where $\vect{K} = \vect{A}_{22} - \vect{A}_{12}^T \vect{A}_{11}^{-1} \vect{A}_{12}$. In particular, substituting \eqref{eq:block_matrices} and \eqref{eq:block_matrix_inversion} into \eqref{eq:schur_complement_system} yields
\begin{multline} \label{eq:complicated_schur_system}
	\left\{ \vect{C}_1 [\vect{A}_{11}^{-1} + \vect{A}_{11}^{-1} \vect{A}_{12} \vect{K}^{-1} \vect{A}_{12}^T \vect{A}_{11}^{-1}] \vect{C}_1^T - \vect{C}_2 \vect{K}^{-1} \vect{A}_{12}^T \vect{A}_{11}^{-1} \vect{C}_1^T - \vect{C}_1 \vect{A}_{11}^{-1} \vect{A}_{12} \vect{K}^{-1} \vect{C}_2^T + \vect{C}_2 \vect{K}^{-1} \vect{C}_2^T \right\} \vect{\nu}_d = \\ 
	\vect{C}_1 [\vect{A}_{11}^{-1} + \vect{A}_{11}^{-1} \vect{A}_{12} \vect{K}^{-1} \vect{A}_{12}^T \vect{A}_{11}^{-1}] \vect{b}_1 - \vect{C}_2 \vect{K}^{-1} \vect{A}_{12}^T \vect{A}_{11}^{-1} \vect{b}_1 - \vect{C}_1 \vect{A}_{11}^{-1} \vect{A}_{12} \vect{K}^{-1} \vect{b}_2 + \vect{C}_2 \vect{K}^{-1} \vect{b}_2 - \hat{\vect{d}}
\end{multline}
where matrix-vector products with $\vect{A}_{11}^{-1}$ can be applied by solving local problems in parallel.

We choose the matrix $\vect{W}$ in the preconditioner
\begin{equation} \label{eq:general_preconditioner}
	\vect{P}^{-1} = (\hat{\vect{C}} \vect{W} \hat{\vect{C}}{}^T)^{-T} \hat{\vect{C}} \vect{W}^T \hat{\vect{A}} \vect{W} \hat{\vect{C}}{}^T(\hat{\vect{C}} \vect{W} \hat{\vect{C}}{}^T)^{-1}
\end{equation}
from Section \ref{sec:dual_primal_algorithm} as
\begin{equation} \label{eq:weight_matrix}
	\vect{W} = 
	\begin{bmatrix}
    	\vect{W}_{1} & -\vect{A}_{11}^{-1} \vect{A}_{12} \vect{W}_2 \\
    	\vect{0}     & \vect{W}_2
	\end{bmatrix}
\end{equation}
which is partitioned in accordance with $\hat{\vect{A}}$. Substituting \eqref{eq:block_matrices} and \eqref{eq:weight_matrix} gives components of $\vect{P}^{-1}$
\begin{equation} \label{eq:first_precond_comp}
	\hat{\vect{C}} \vect{W} \hat{\vect{C}}{}^T = \vect{C}_1 \vect{W}_1 \vect{C}_1^T + \vect{C}_2 \vect{W}_2 \vect{C}_2^T - \vect{C}_1 \vect{A}_{11}^{-1} \vect{A}_{12} \vect{W}_2 \vect{C}_2^T
\end{equation}
and
\begin{equation} \label{eq:second_precond_comp}
	\hat{\vect{C}} \vect{W}^T \hat{\vect{A}} \vect{W} \hat{\vect{C}}{}^T =
	\vect{C}_1 \vect{W}_1^T \vect{A}_{11} \vect{W}_1 \vect{C}_1^T + \vect{C}_2 \vect{W}_2^T \vect{K} \vect{W}_2 \vect{C}_2^T.
\end{equation}
When $\vect{W}_1 = \vect{I}$ and $\vect{W}_2 = \vect{0}$, this corresponds to the lumped preconditioner for FETI-DP. To obtain the Dirichlet preconditioner, we choose $\vect{W}_2=0$ and $\vect{W}_1$ to be block diagonal with blocks
\begin{equation}
	\vect{W}^{(j)} = 
	\begin{bmatrix}
    	\vect{I}                                            & \vect{0} \\
    	-(\vect{A}_{ii}^{(j)})^{-1} (\vect{A}_{ei}^{(j)})^T & \vect{0}
	\end{bmatrix}
\end{equation}
where
\begin{equation}
	\vect{A}^{(j)} = 
	\begin{bmatrix}
    	\vect{A}_{ee}^{(j)} & \vect{A}_{ei}^{(j)} \\
    	(\vect{A}_{ei}^{(j)})^T & \vect{A}_{ii}^{(j)}
	\end{bmatrix}
\end{equation}
are the diagonal blocks of $\vect{A}_{11}$ partitioned according to edge and interior unknowns. Finally, if there are jumps in the parameter $\vect{\alpha}$ between subdomains, we modify the weight matrix to include diagonal scaling
\begin{equation}
	\vect{W}^{(j)} = 
	\begin{bmatrix}
    	\vect{I}                                            & \vect{0} \\
    	-(\vect{A}_{ii}^{(j)})^{-1} (\vect{A}_{ei}^{(j)})^T & \vect{0}
	\end{bmatrix}
	\begin{bmatrix}
    	\diag{(\vect{A}_{ee}^{(j)})}^{-1} & \vect{0} \\
    	\vect{0}                    & \vect{I}
	\end{bmatrix}
\end{equation}
where $\diag{(\vect{A}_{ee}^{(j)})}$ is the diagonal matrix with diagonal entries corresponding to diagonal entries of $\vect{A}_{ee}^{(j)}$. If $\beta$ is non-zero or there are Robin boundary conditions which contribute to $\vect{A}^{(j)}$, such a scaling will not improve convergence of the iterative scheme. Instead, when forming the preconditioner, we take only the part of $\hat{\vect{A}}$ that arises from discretization of $-\nabla \cdot (\vect{\alpha} \nabla \phi)$. Using only this portion of the operator is a standard preconditioning technique when discretizing more complicated operators whose complicating terms involve lower order derivatives \cite{Yserentant1988}.

When the mesh is conforming, $\vect{C}_2 = 0$, and \eqref{eq:complicated_schur_system} simplifies to
\begin{equation}
	\vect{C}_1 [\vect{A}_{11}^{-1} + \vect{A}_{11}^{-1} \vect{A}_{12} \vect{K}^{-1} \vect{A}_{12}^T \vect{A}_{11}^{-1}] \vect{C}_1^T \vect{\nu}_d =
	\vect{C}_1 [\vect{A}_{11}^{-1} + \vect{A}_{11}^{-1} \vect{A}_{12} \vect{K}^{-1} \vect{A}_{12}^T \vect{A}_{11}^{-1}] \vect{b}_1 - \vect{C}_1 \vect{A}_{11}^{-1} \vect{A}_{12} \vect{K}^{-1} \vect{b}_2 - \hat{\vect{d}}.
\end{equation}
In addition, the preconditioner \eqref{eq:general_preconditioner} becomes
\begin{equation}
	\vect{P}^{-1} = (\vect{C}_1 \vect{W}_1^T \vect{C}_1^T)^{-1} \vect{C}_1 \vect{W}_1^T \vect{A}_{11} \vect{W}_1 \vect{C}_1^T (\vect{C}_1 \vect{W}_1 \vect{C}_1^T)^{-1}.
\end{equation}
With both the lumped and Dirichlet preconditioner, the matrix $\vect{C}_1 \vect{W}_1 \vect{C}_1^T$ in $\vect{P}^{-1}$ is diagonal, facilitating its inverse. If, in addition, $l=1$, then we have a typical FETI-DP algorithm for the spectral element method. When the mesh is non-conforming, $\vect{C}_2$ is non-zero and we choose $\vect{W}_2=\vect{I}$ (we can choose $\vect{W}_1$ to be either the lumped or Dirichlet preconditioner). This choice of $\vect{W}_2$ complicates the preconditioner. However, we have found that zeroing the terms $-\vect{C}_1 \vect{A}_{11}^{-1} \vect{A}_{12} \vect{W}_2 \vect{C}_2^T$ and $\vect{C}_2 \vect{W}_2^T \vect{K} \vect{W}_2 \vect{C}_2^T$ in \eqref{eq:first_precond_comp} and \eqref{eq:second_precond_comp} respectively does not adversely affect convergence measured by reduction of the preconditioned relative residual (although they may have an effect on the error in the solution as described in Remark \ref{rem:remark_edge_errors}). Removing the term $\vect{C}_2 \vect{W}_2 \vect{C}_2^T$ from \eqref{eq:first_precond_comp} does adversely affect convergence and we keep it at the cost of \eqref{eq:first_precond_comp} no longer being diagonal. It is possible to recover this diagonal property if the elements sharing non-conforming edges are grouped together in a single domain of the domain decomposition (but the local fast solver would not be used).

\begin{remark} \label{rem:remark_edge_errors}
	While ignoring $\vect{C}_2 \vect{W}_2^T \vect{K} \vect{W}_2 \vect{C}_2^T$ in \eqref{eq:second_precond_comp} does not adversely affect the reduction of the preconditioned relative residual, we have found that for high accuracy applications, one should include this term. For example, in Section \ref{sec:dielectric_cylinder} we solve a Helmholtz problem with a non-conforming mesh to an accuracy of 10 digits by including $\vect{C}_2 \vect{W}_2^T \vect{K} \vect{W}_2 \vect{C}_2^T$. Without this term in the preconditioner, the preconditioned relative residual decreases in the same way, but the solution is only accurate to 4 digits on non-conforming edges in the mesh (and far more accurate away from them). \remarkend
\end{remark}

\subsubsection{Considerations for the Helmholtz Problem} \label{sec:consideration_for_helmholtz}

The Poisson problem with $\vect{\alpha}$ positive definite, $\beta = 0$, and $\gamma = 0$, subject to Dirichlet boundary conditions yields symmetric positive definite system \eqref{eq:schur_complement_system} with associated symmetric positive definite preconditioner $\vect{P}^{-1} = \vect{L}_{\vect{P}} \vect{L}_{\vect{P}}^T$ (we never compute the factorization of $\vect{P}^{-1}$ explicitly). In such situations, we use the preconditioned conjugate gradients method (PCG) as Krylov subspace method. On a conforming mesh with all elements possessing degree $p$ and the number of assembled constraints $l=1$, the condition number
\begin{equation} \label{eq:condition_number_laplace}
	\kappa_2 \big( \vect{L}_{\vect{P}}^T \hat{\vect{C}} \hat{\vect{A}} \hat{\vect{C}}{}^T \vect{L}_{\vect{P}} \big) \le C \big(1 + \log{(p^2)} \big)^2
\end{equation}
with $C>0$ a constant independent of degree $p$, mesh size $h$, and parameter $\vect{\alpha}$ (similar to \cite{Klawonn2008}). In practice, this bound translates to a number of iterations of PCG that depends weakly on the polynomial degree, and consequently, the size of the discrete problem. However, this bound does not continue to hold for the Helmholtz problem when $\beta = - k^2$ and the wavenumber $k>0$ becomes large. It is possible to recover such behaviour by increasing the number of assembled constraints $l$ as a function of $h$ and $k$.

In particular, the key is to make sure that dispersion errors when $kh \gg 1$ are controlled for the coarse problem $\hat{\vect{A}} \vect{\phi}_p = \hat{\vect{b}}$ which arises from \eqref{eq:assembled_saddle_point_system} by ignoring the additional constraints $\hat{\vect{C}} \vect{\phi}_p = \hat{\vect{d}}$ and associated dual variables $\vect{\nu}_d$. On a uniform mesh with mesh size $h$, this is possible as long as we choose
\begin{equation} \label{eq:dispersion_criterion}
	l > \frac{1}{2} \big[ kh + (kh)^{1/3} - 1 \big].
\end{equation}
In practice, we set $l$ (which must be an integer) to the ceiling of the right hand side of \eqref{eq:dispersion_criterion}. We do not allow $l$ to be smaller than 1 so that the partial nullspace method actually provides a loosely coupled coarse problem ($l=0$ has no such coupling). The constraint \eqref{eq:dispersion_criterion} is adopted from Theorem 3.3 in \cite{Ainsworth2004}. There, the error in the discrete dispersion relation for conforming finite element discretization of the Helmholtz equation is analyzed on a uniform grid and is found to enter a regime of superexponential decay when replacing $l$ in \eqref{eq:dispersion_criterion} by $p$, the degree of approximation used on the mesh. In practice, we observe experimentally that a similar phenomenon holds when solving $\hat{\vect{A}} \vect{\phi}_p = \hat{\vect{b}}$ in non-conforming cases even when $p$ is larger than $l$ as long as \eqref{eq:dispersion_criterion} is satisfied.

\begin{figure*}[!t]
    \centering
    \begin{subfigure}[t]{0.475\textwidth}
        \centering
        \includegraphics[width=\textwidth]{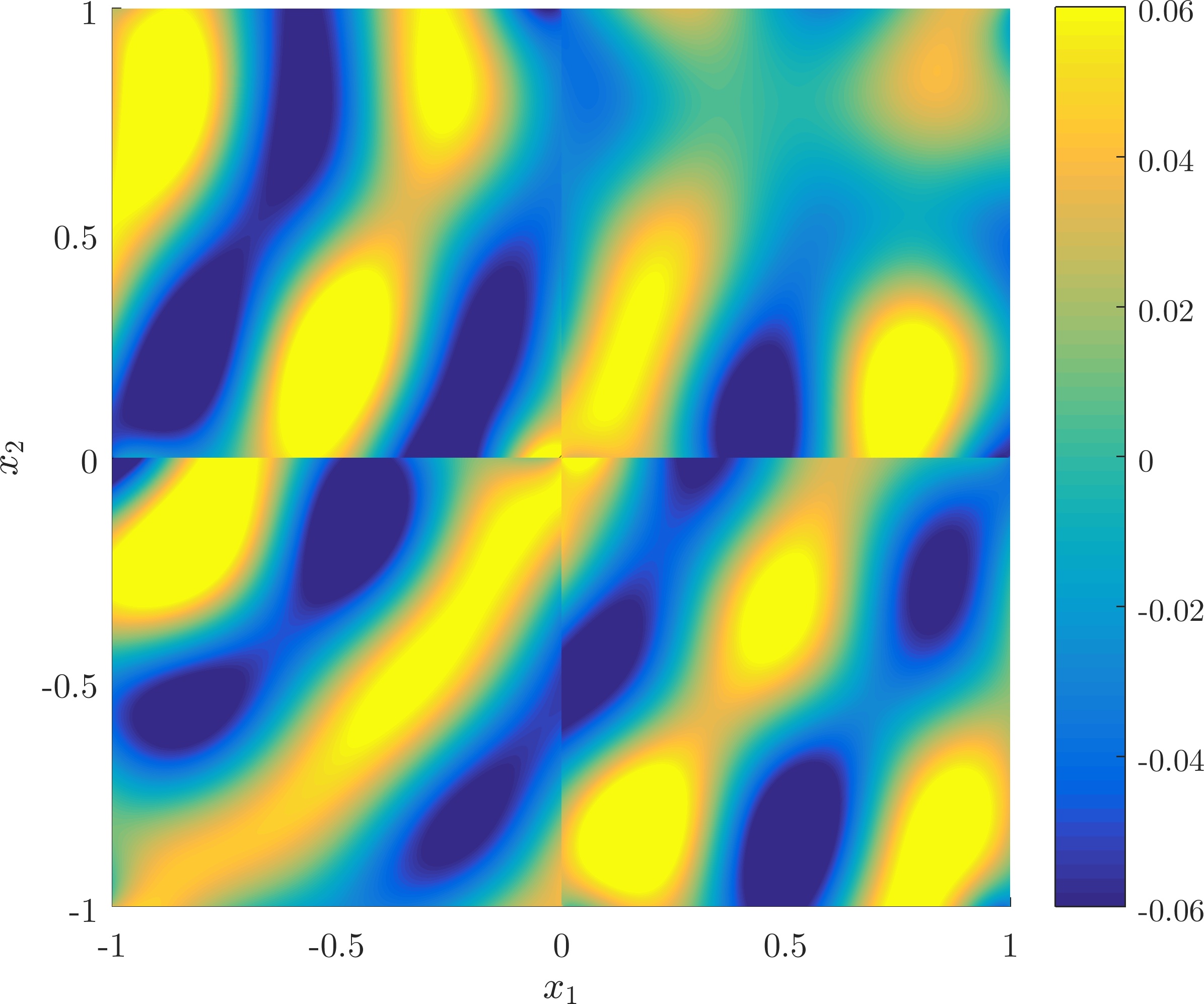}
        \caption{Real part of the solution to the coarse problem when $l=5$.}
    \end{subfigure}
    \hfill
    \begin{subfigure}[t]{0.475\textwidth}
        \centering
        \includegraphics[width=\textwidth]{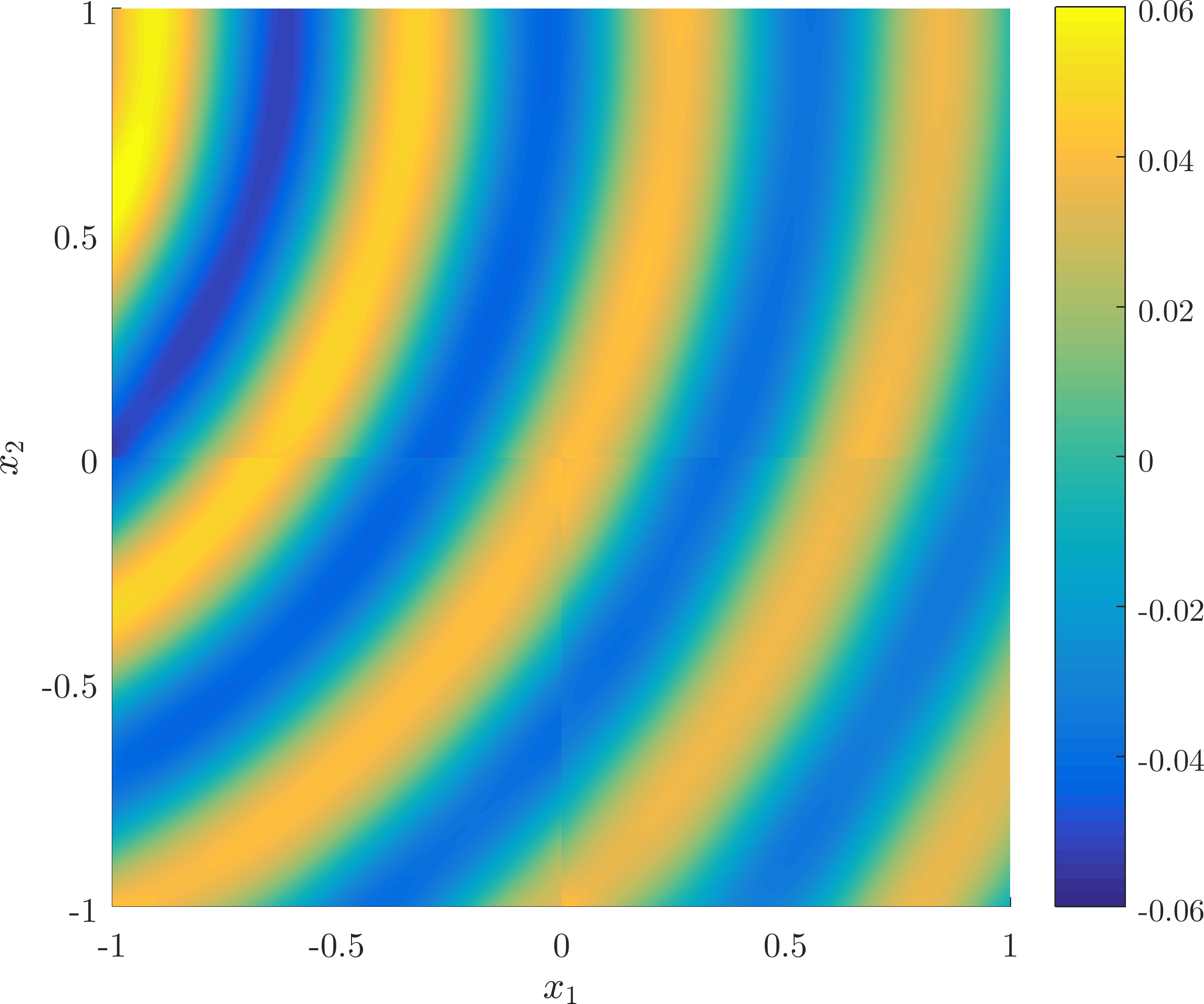}
        \caption{Real part of the solution to the coarse problem when $l=6$.}
    \end{subfigure}
    \vskip\baselineskip
    \begin{subfigure}[t]{0.475\textwidth}
        \centering
        \includegraphics[width=\textwidth]{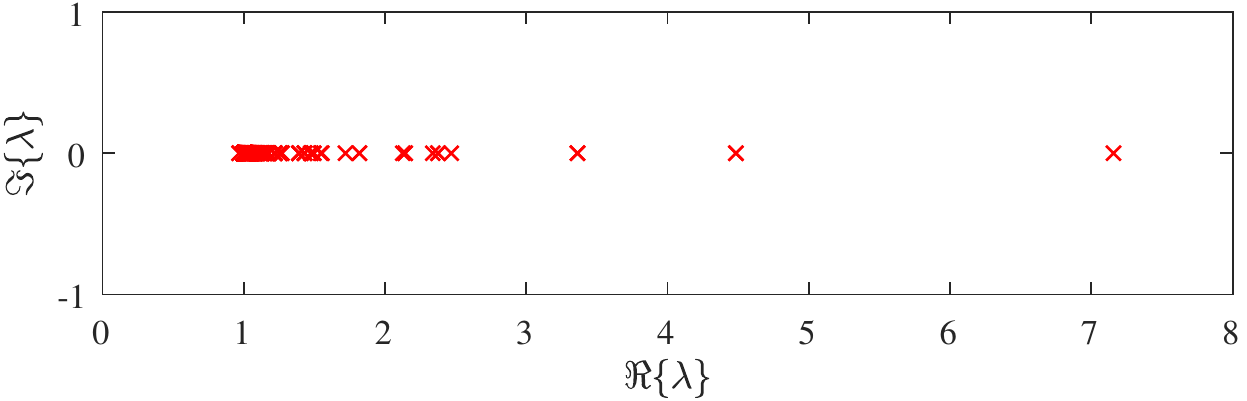}
        \caption{Eigenvalues of the preconditioned problem when $l=5$.}
    \end{subfigure}
    \hfill
    \begin{subfigure}[t]{0.475\textwidth}
        \centering
        \includegraphics[width=\textwidth]{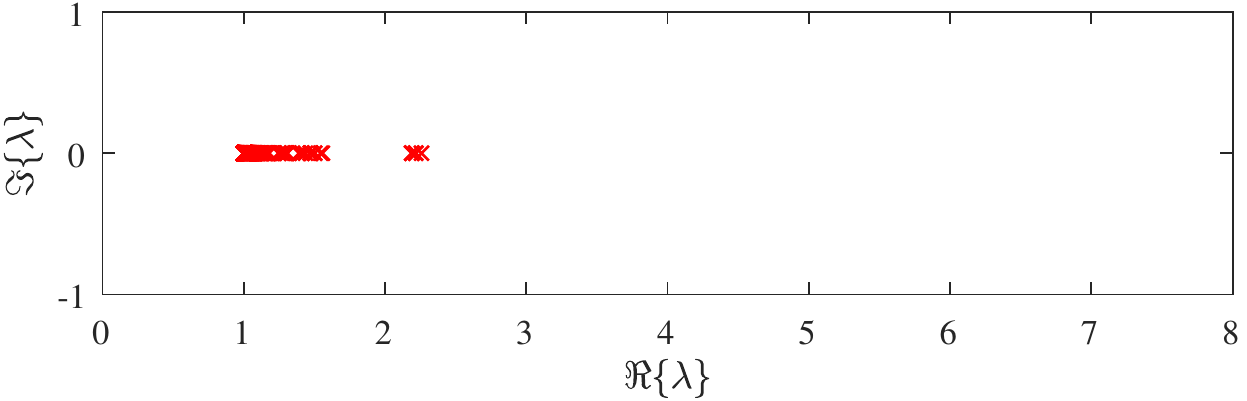}
        \caption{Eigenvalues of the preconditioned problem when $l=6$.}
    \end{subfigure}
    \caption{Comparison of the solution to the coarse problem $\hat{\vect{A}} \vect{\phi}_p = \hat{\vect{b}}$ for two values of $l$ and its effect on the eigenvalues of the associated preconditioned matrix $\vect{P}^{-1} \hat{\vect{C}} \hat{\vect{A}} \hat{\vect{C}}{}^T$. \label{fig:weak_continuity_l}}
\end{figure*}

To illustrate this phenomenon, we consider a hypothetical Helmholtz problem with $\vect{\alpha} = \vect{I}$, $\beta = -k^2$, $k=10.75$, and $f=0$ on domain $\Omega = (-1,1)^2$ with Dirichlet boundary conditions such that $p(\vect{x}) = -(\jmath /4) H_0^{(2)}(k \norm{\vect{x} - \vect{x}_c})$ on $\partial \Omega$ where $\jmath = \sqrt{-1}$ and $H_0^{(2)}$ is the Hankel function of the second kind of zeroth order. The solution $\phi(\vect{x})$ on $\Omega$ is $p(\vect{x})$ . We choose the center $\vect{x}_c = [-2,\, 1]^T$ in our test. We subdivide $\Omega$ into four square elements of equal size using degree 64 polynomials on each element to represent $\phi$. Figure \ref{fig:weak_continuity_l} illustrates the coarse problem solution for two values of $l=5,6$. Condition \eqref{eq:dispersion_criterion} is satisfied for $l \gtrsim 5.9785$ so that the coarse problem poorly approximates the true solution in the first case, but is suitable in the second. Figure \ref{fig:weak_continuity_l} also illustrates the eigenvalues of the associated preconditioned problem using these corresponding coarse spaces. When $1 \le l < 5$, these eigenvalues are less clustered (and can also be negative) and their corresponding coarse solutions even less effective. We have chosen to illustrate the case $l=5$ to show that the condition \eqref{eq:dispersion_criterion} is relatively sharp.

It is not possible to directly apply the theory of \cite{Ainsworth2004} to our setting because the analysis of the discrete dispersion relation there is only valid for an infinite grid of uniform square elements with continuity enforced between elements. Theorem 3.3 of \cite{Ainsworth2004} shows that in such a setting, the discrete dispersion error as a function of increasing $p$ enters a regime of decay for
\begin{equation}
	p > \frac{1}{2}[kh+(kh)^{1/3}-1]
\end{equation}
where $p$ is the polynomial degree of each element, $k$ is the wavenumber of the problem, and $h$ is the edge length of each element (the analysis holds for $kh\gg1$). For smaller values of $p$, the dispersion error oscillates and is $\mathcal{O}(1)$.

The main departure from the theory of \cite{Ainsworth2004} comes by only enforcing a hierarchy of edge moments to match between elements, rather than impose full continuity between elements (as in \cite{Ainsworth2004}). This is achieved by choosing $l<p$, and performing the partial assembly $\hat{\vect{A}}=\vect{Z}^{T}\vect{A}\vect{Z}$. One can think of the partially assembled problem as a continuous finite element method of degree $l$ augmented with additional basis functions up to degree $p$ for which continuity has not been enforced. This interpretation is valid because of the hierarchical nature of the basis functions. Under this interpretation, criterion \eqref{eq:dispersion_criterion} is required to hold for the partially assembled problem to leave the $\mathcal{O}(1)$ regime of dispersion error (the additional basis functions do not worsen the dispersion error). We choose $l$ to be the smallest integer that satisfies this relation to obtain a coarse problem that has entered the regime of decay. The coarse problem on its own need not be accurate since it has just exited the $\mathcal{O}(1)$ regime of dispersion error. However, it provides just enough coupling for the following step of the domain decomposition method (to solve the system $\hat{\vect{C}}\hat{\vect{A}}{}^{-1}\hat{\vect{C}}{}^{T}$ via a preconditioned Krylov subspace method) since the inversion of $\hat{\vect{A}}$ captures the wave-like behaviour of the problem (dispersion errors are controlled, although not to high accuracy). Full continuity and high accuracy of the solution are recovered by the Krylov subspace iterations.


\begin{remark}
	In our method, since there is a natural hierarchy of constraints along edges, there is a straightforward way of increasing the size of the coarse problem (that is, increasing $l$ and correspondingly the size of $\vect{K}$). In a sense, this is similar to the FETI-DPH method in that convergence can be achieved by increasing the size of the coarse problem. However, the methods differ in the way that the coarse problem is grown. In FETI-DPH, new constraints are added to $\vect{C}$ of the form $\vect{Q}_b^T \vect{C}$ which are then eliminated when applying the range space method (and consequently grow the coarse problem). The matrix $\vect{Q}_b$ consists of plane waves sampled along edges in the domain decomposition. In practice, because it is unclear \textit{a priori} how to choose the directions of these plane waves, many are chosen along each edge and rank-revealing QR factorizations are performed to select a set of orthogonal columns for $\vect{Q}_b$ such that it is not rank deficient. See \cite{Farhat2005} for details. \remarkend
\end{remark}

As with FETI-DPH, increasing the coarse space is not enough to achieve robust performance of the iterative method. Equation \eqref{eq:schur_complement_system} becomes indefinite for large enough $k$ so we also replace PCG with the preconditioned generalized minimum residual method (PGMRES). In addition, like FETI-DPH, the method described in this paper is also susceptible to spurious resonant frequencies associated with each domain in the domain decomposition. This is problematic because at such frequencies, the matrix $\vect{A}_{11}$ is singular, whereas the original saddle point problem is not. One possible solution is to choose the domains in the decomposition small enough so that all domains have resonances at frequencies larger than $k$ \cite{Farhat2005}.

Alternatively, instead of limiting the size of domains, it is possible to change the discretization in such a way that each subdomain problem becomes uniquely solvable without changing the primal solution \cite{Farhat2000a}. This involves adding Robin boundary terms to each local matrix with parameter $\gamma = \pm \jmath k$ and treating the corresponding parameter $q$ as a new dual variable replacing $\nu$. If the sign of $\gamma$ is chosen to be positive for an element and negative in its neighbour, then the new dual variable is related to the old one via $q = \nu + \jmath k \phi$. When there are more than two elements, it is necessary to choose signs so that every element has at least one neighbour with a different sign. If we define an undirected graph where each element in the mesh has an associated vertex in the graph and where edges in the graph represent the element adjacencies in the mesh, then assigning signs to each element is equivalent to solving the weak 2-colouring problem \cite{Naor1995}. We do so by constructing a spanning tree of the graph (choose an arbitrary initial element in the mesh and perform a breadth-first-search). Label elements in even levels of the spanning tree as positive and elements in odd levels as negative (or vice versa). When adding the local Robin boundary conditions for each element, we only add the condition on an edge adjacent to an element labelled with opposite sign. Following this process ensures that each element has a component of its boundary with Robin boundary condition with $\gamma = \pm \jmath k$ of a single chosen sign. Such problems are uniquely solvable \cite{Colton2013}.

The resulting saddle point system possesses the same structure as \eqref{eq:second_saddle_point_system}; only the $\vect{A}$ block is modified via Robin boundary terms (all other terms remain unchanged, although the interpretation of the dual variables has changed). However, such a modification makes the original real symmetric problem complex symmetric. In practice, we have found that applying the same domain decomposition approach to this modified saddle point system requires roughly twice as many iterations to converge than the unmodified approach. However, the modified method converges even at spurious resonant frequencies where the unmodified approach fails to converge at all. If one must use the modified approach for robustness to spurious resonant frequencies, it is possible to reduce the number of iterations by increasing $l$ at the cost of increasing the size of the coarse problem (this is also possible in the unmodified approach).

%

\section{Numerical Results and Discussion} \label{sec:results}

In this section we demonstrate convergence properties of the domain decomposition methods described in Section \ref{sec:numerical_methods} using certain Poisson and Helmholtz test problems then verify that the methods can be used to solve more challenging problems from electromagnetism. The Poisson test problems are used to confirm that the method described here converges in much the same way that a nodal FETI-DP method does \cite{Klawonn2008}, with the added benefit that the size of the coarse problem can be increased easily (through adding continuity constraints), decreasing the number of iterations required for convergence if needed. Similar tests are used for Helmholtz problems to show that increasing the size of the coarse problem becomes necessary to retain small numbers of iterations as the wavenumber is increased. Unless otherwise stated, we use the zero vector for initial iterate in all applications of PCG and PGMRES.

\subsection{Convergence Tests for the Poisson Equation} \label{sec:laplace_examples}

We begin by demonstrating the number of iterations and maximum eigenvalues of the preconditioned system when applied to solve \eqref{eq:prototype_pde} with $\vect{\alpha} = \rho (\vect{x}) \vect{I}$ and $\beta = 0$ on $\Omega = (-1,1)^2$ subject to zero Dirichlet boundary conditions on $\partial \Omega$. We use a test problem presented in Table 2 of \cite{Klawonn2008} where the element size and degree are varied to compare the convergence of our method with that of the FETI-DP method. This test shows that the number of iterations depends weakly on discontinuities in parameter $\rho$, as well as on the element degree, and the number of elements in the mesh. This favourable performance matches the performance presented in \cite{Klawonn2008} when $l=1$, and improves upon it when $l>1$ (at the expense of increasing the size of the coarse problem).

In the tests, the domain $\Omega$ is partitioned into a regular grid of $N = 2^{2n}$ elements (with $n=1,2,3,4$) with each element possessing a local degree $p$ expansion (with $p=2,3,4,8,16,32$). If the elements in the grid are numbered as $k=i+(j-1)2^n$ for $i,j=1,2,...,2^n$, then the parameter $\rho$ is chosen to be piecewise constant over each element with corresponding value $\rho_{ij} = 10^{(i-j)/4}$. For all problems, we set the right hand side $\vect{b}$ randomly (each entry sampled from the uniform distribution on the open interval $(0,1)$). We use the Dirchlet preconditioner with diagonal scaling with parameter $l=1,2,3,4$ and report the number of iterations required for the relative residual to be reduced by ten orders of magnitude. We also compute the maximum eigenvalue $\lambda_{\textrm{max}}$ of the preconditioned matrix $\vect{P}^{-1} \hat{\vect{C}} \hat{\vect{A}} \hat{\vect{C}}{}^T$. In all tests, the smallest eigenvalue is 1 to within at least two digits so we do not list it. Table \ref{tab:laplace_iterations} shows the number of iterations and maximum eigenvalue for each combination of degree $p$, number of elements $N$, and parameter $l$. Table \ref{tab:laplace_iterations} also provides the number of iterations and maximum eigenvalue for the FETI-DP method described in \cite{Klawonn2008} for direct comparison. Dashes in the table indicate problems where the partial nullspace step of the method globally assembles the finite element matrix, leaving no dual problem to solve (and consequently no iterations for PCG). In subsequent examples, we use large degree $p$ with small values of $l$ satisfying $p>l$ so that the problem is never globally assembled.

\begin{table}[!t]
	\centering
	\caption[]{Number of iterations for PCG to reach convergence and maximum eigenvalue of the preconditioned system as functions of degree $p$, number of elements $N$, and parameter $l$ for the Poisson problem.}
	\label{tab:laplace_iterations}
	\begin{tabular}{S[table-format = 2.0]S[table-format = 3.0]S[table-format = 2.0]S[table-format = 2.2]S[table-format = 2.0]S[table-format = 2.2]S[table-format = 2.0]S[table-format = 1.2]S[table-format = 2.0]S[table-format = 1.2]S[table-format = 2.0]S[table-format = 1.2]}
		\toprule
		{} & {} & \multicolumn{2}{c}{FETI-DP \cite{Klawonn2008}} & \multicolumn{2}{c}{$l=1$} & \multicolumn{2}{c}{$l=2$} & \multicolumn{2}{c}{$l=3$} & \multicolumn{2}{c}{$l=4$} \\
		\cmidrule(r){3-4} \cmidrule(r){5-6} \cmidrule(r){7-8} \cmidrule(r){9-10} \cmidrule(r){11-12}
		{$p$} & {$N$} & {Iterations} & {$\lambda_{\textrm{max}}$} & {Iterations} & {$\lambda_{\textrm{max}}$} & {Iterations} & {$\lambda_{\textrm{max}}$} & {Iterations} & {$\lambda_{\textrm{max}}$} & {Iterations} & {$\lambda_{\textrm{max}}$} \\ 
		\midrule
		 2 &   4 & 2 & 1.05 &  4 & 1.82 & {-} & {-} & {-} & {-} & {-} & {-} \\
		 2 &  16 & 6 & 1.46 & 11 & 1.93 & {-} & {-} & {-} & {-} & {-} & {-} \\
		 2 &  64 & 8 & 1.61 & 13 & 2.00 & {-} & {-} & {-} & {-} & {-} & {-} \\
		 2 & 256 & 8 & 1.62 & 13 & 2.02 & {-} & {-} & {-} & {-} & {-} & {-} \\
		\addlinespace
		 3 &   4 &  3 & 1.21 &  6 & 2.02 & 5 & 1.14 & {-} & {-} & {-} & {-} \\
		 3 &  16 &  8 & 2.10 & 13 & 2.45 & 7 & 1.16 & {-} & {-} & {-} & {-} \\
		 3 &  64 & 11 & 2.31 & 15 & 2.56 & 7 & 1.17 & {-} & {-} & {-} & {-} \\
		 3 & 256 & 12 & 2.36 & 15 & 2.59 & 7 & 1.17 & {-} & {-} & {-} & {-} \\
		\addlinespace
		 4 &   4 &  3 & 1.37 &  8 & 2.68 & 7 & 1.18 & 5 & 1.04 & {-} & {-} \\
		 4 &  16 & 10 & 2.65 & 15 & 3.12 & 8 & 1.26 & 6 & 1.07 & {-} & {-} \\
		 4 &  64 & 13 & 2.94 & 18 & 3.21 & 8 & 1.28 & 6 & 1.08 & {-} & {-} \\
		 4 & 256 & 14 & 3.00 & 18 & 3.23 & 8 & 1.29 & 6 & 1.08 & {-} & {-} \\
		\addlinespace
		 8 &   4 &  4 & 1.89 &  9 & 4.02 &  8 & 1.54 & 6 & 1.23 & 6 & 1.10 \\
		 8 &  16 & 12 & 4.37 & 18 & 4.90 & 11 & 1.70 & 8 & 1.30 & 6 & 1.12 \\
		 8 &  64 & 18 & 4.86 & 23 & 5.10 & 11 & 1.74 & 8 & 1.32 & 7 & 1.13 \\
		 8 & 256 & 19 & 4.97 & 23 & 5.15 & 11 & 1.75 & 8 & 1.33 & 7 & 1.13 \\
		\addlinespace
		16 &   4 &  5 & 2.57 & 11 & 5.84 & 10 & 2.08 &  8 & 1.61 & 7 & 1.38 \\
		16 &  16 & 15 & 6.63 & 20 & 7.25 & 13 & 2.30 & 10 & 1.71 & 8 & 1.44 \\
		16 &  64 & 21 & 7.38 & 27 & 7.59 & 13 & 2.37 & 10 & 1.74 & 8 & 1.45 \\
		16 & 256 & 25 & 7.53 & 28 & 7.67 & 13 & 2.39 & 11 & 1.75 & 8 & 1.46 \\
		\addlinespace
		32 &   4 &  6 &  3.42 & 13 &  8.14 & 11 & 2.79 &  9 & 2.17 &  8 & 1.86 \\
		32 &  16 & 17 &  9.44 & 22 & 10.22 & 15 & 3.08 & 12 & 2.30 & 10 & 1.94 \\
		32 &  64 & 25 & 10.52 & 33 & 10.72 & 15 & 3.16 & 13 & 2.34 & 11 & 1.96 \\
		32 & 256 & 31 & 10.74 & 33 & 10.84 & 16 & 3.18 & 13 & 2.35 & 11 & 1.96 \\
		\bottomrule
	\end{tabular} 
\end{table}

We note that for a given $l$ and $p$, the number of iterations converges to some fixed value as the number of elements $N$ increases. The same behaviour is observed in the maximum eigenvalue which controls how clustered the eigenvalues of the preconditioned system are to 1. We also note that for fixed $p$ and $N$, the number of iterations can be reduced by increasing $l$ (although the cost of each iteration increases because $l$ increases the size of the matrix $\vect{K}$ which must be factored). If $N$ and $l$ are fixed, increasing $p$ leads to a mild increase in the number of iterations.

\begin{figure}[!t]
	\centering
	\includegraphics[scale=1]{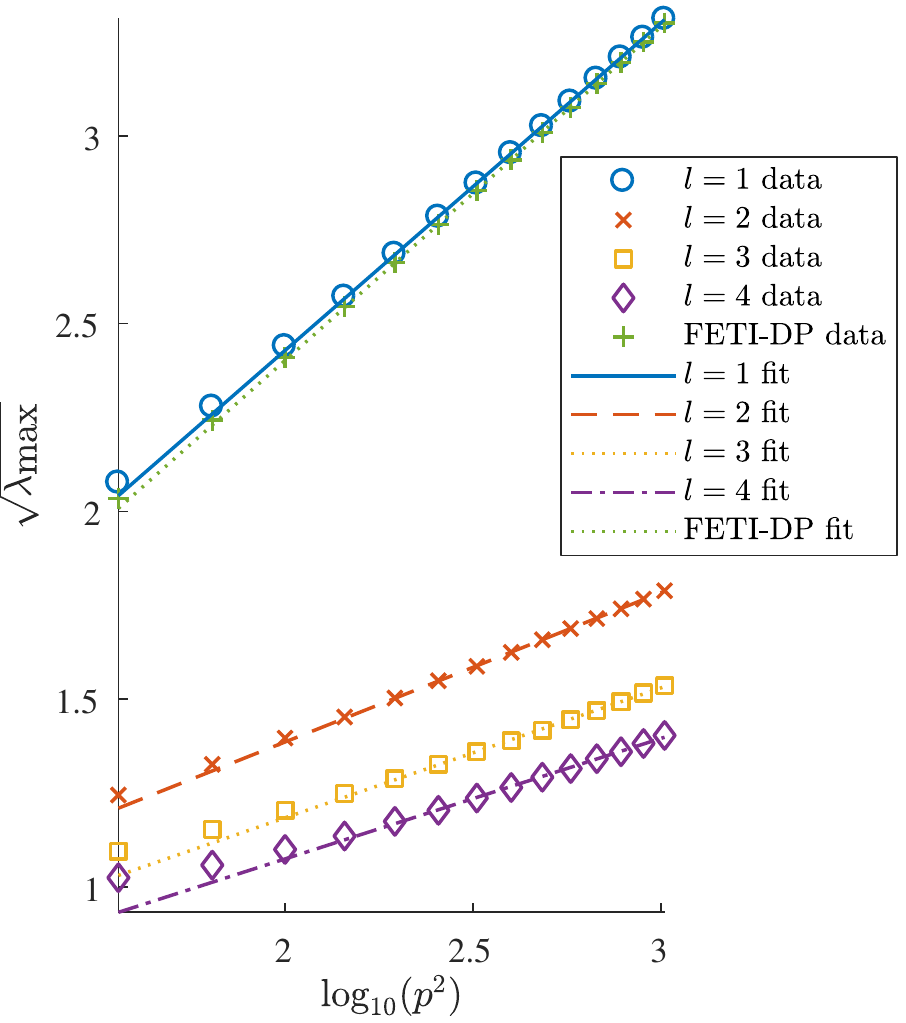}	
	\caption{Dependence of the maximum eigenvalue of the preconditioned system $\lambda_{\textrm{max}}$ on polynomial degree $p$ for the Poisson problem as a function of parameter $l$. The FETI-DP plots correspond to the work reported in \cite{Klawonn2008}. \label{fig:condition_number_versus_degree}}
\end{figure}

To explain this increase, we consider another test from Table 6 and Figure 4 in \cite{Klawonn2008}. This test demonstrates that \eqref{eq:condition_number_laplace} is satisfied. In particular, we now set $\vect{\alpha} = \vect{I}$ and fix $n=6$. We then compute $\lambda_{\textrm{max}}$ for $p$ ranging from 6 to 32 in increments of 2. We repeat this process for $l=1,2,3,4$. Figure \ref{fig:condition_number_versus_degree} illustrates how $\lambda_{\textrm{max}}$ increases as a function of degree $p$ for each possible $l$. Figure \ref{fig:condition_number_versus_degree} also includes a comparison to the FETI-DP method of \cite{Klawonn2008} which demonstrates that the $l=1$ case of our method behaves similarly to the FETI-DP method, and that $l>1$ improves upon FETI-DP. The figure also illustrates least squares fits to the data such that
\begin{equation}
	\sqrt{\lambda_{\textrm{max}}} \approx A + B \log_{10} (p^2)
\end{equation}
where we have used only data with $p \in \{14, 16, ..., 32 \}$ to obtain the fit (we do this because the data for $l=3,4$ does not appear to have entered the asymptotic regime for smaller values of $p$).

From these least squares computations, we can change base of the logarithm to obtain fits
\begin{equation}
	\lambda_{\textrm{max}} \approx C \big( 1 + \log_{D} (p^2) \big)^2
\end{equation}
with constants $C\approx0.473,0.354,0.244,0.191$ and bases $D\approx6.15,31.6,26.8,23.1$ for $l=1,2,3,4,$ respectively. The fit for the FETI-DP method yields $C\approx0.626$ and $D\approx5.08$. Since the smallest eigenvalue of the preconditioned system is 1 and $\lambda_{\textrm{max}}$ is real and positive, we have
\begin{equation}
	\kappa_2\big( \vect{L}_{\vect{P}}^T \hat{\vect{C}} \hat{\vect{A}} \hat{\vect{C}}{}^T \vect{L}_{\vect{P}} \big) = \lambda_{\textrm{max}}
\end{equation}
which agrees with \eqref{eq:condition_number_laplace}. Note that increasing $l$ from 1 to 2 appears to change the slope in Figure \ref{fig:condition_number_versus_degree} but that further increases in $l$ tend primarily to decrease the offset of the curve. This is reflected in constants $C$ and bases $D$ as well as in the number of iterations in Table \ref{tab:laplace_iterations} (the number of iterations are roughly halved when changing $l$ from 1 to 2 but only decrease mildly when going from 2 to 3 or 4). This means that for relatively small degree $p$, there is little value in increasing $l$ beyond 2 when solving Poisson problems. This result is not true when solving Helmholtz problems.

\subsection{Convergence Tests for the Helmholtz Equation} \label{sec:convergence_test_helmholtz}

We now perform similar tests for the Helmholtz equation and address the added complexities involved. We solve \eqref{eq:prototype_pde} with $\vect{\alpha} = \vect{I}$, $\beta = -k^2$, and $f=0$ on $\Omega = (-1,1)^2$ subject to the same Dirichlet boundary condition on $\partial \Omega$ described in Section \ref{sec:consideration_for_helmholtz}. We use the same discretizations as described in Section \ref{sec:laplace_examples} and vary element degree $p$, number of elements $N$, and parameter $l$ in the same ways. We repeat this exercise for several values of $k$ from 1 to 101 in increments of 10. We do not spatially vary $\vect{\alpha}$ so that the effect of $k$ on the number of iterations is clear. Table \ref{tab:helmholtz_iterations} shows the number of iterations required to reduce the relative residual by ten orders of magnitude for the cases $k=11,21,31$. We use the unmodified domain decomposition method without adding Robin boundary terms. Problems where the number of iterations exceeds 100 are marked with a dash. Problems where condition \eqref{eq:dispersion_criterion} is satisfied are marked in bold. In Table \ref{tab:helmholtz_iterations}, we do not show the eigenvalue with maximum absolute value as it is not necessarily an indicator of the condition number of the preconditioned system (now that the problem can be indefinite).

\begin{table}[!t]
	\centering
	\caption[]{Number of iterations for PGMRES to reach convergence as a function of degree $p$, number of elements $N$, parameter $l$, and wavenumber $k$ for the Helmholtz problem.}
	\label{tab:helmholtz_iterations}
	\sisetup{detect-weight=true,detect-inline-weight=math}
	\begin{tabular}{S[table-format = 2.0]S[table-format = 4.0]S[table-format = 2.0]S[table-format = 2.0]S[table-format = 2.0]S[table-format = 2.0]S[table-format = 2.0]S[table-format = 2.0]S[table-format = 2.0]S[table-format = 2.0]S[table-format = 2.0]S[table-format = 2.0]S[table-format = 2.0]S[table-format = 2.0]}
		\toprule
		{} & {} & \multicolumn{4}{c}{$k = 11$} & \multicolumn{4}{c}{$k = 21$} & \multicolumn{4}{c}{$k = 31$} \\
		\cmidrule(r){3-6} \cmidrule(r){7-10} \cmidrule(r){11-14}
		{$p$} & {$N$} & {$l=1$} & {$l=2$} & {$l=3$} & {$l=4$} & {$l=1$} & {$l=2$} & {$l=3$} & {$l=4$} & {$l=1$} & {$l=2$} & {$l=3$} & {$l=4$} \\ 
		\midrule
		8&4   & 61&42&29&22 & 69&58&47&35   & 72&58&50&40\\
		 &16  & 79&42&20& \bfseries 9  & {-}&{-}&69&42 & {-}&{-}&{-}&95\\
		 &64  & 88& \bfseries 27& \bfseries 10& \bfseries 7  & {-}&{-}& \bfseries 30& \bfseries 10 & {-}&{-}&98&34\\
		 &256 & \bfseries  {-}& \bfseries 15& \bfseries 9& \bfseries 7  & {-}& \bfseries 51& \bfseries 11& \bfseries 7   & {-}&{-}& \bfseries 26& \bfseries 9\\
		 &1024& \bfseries  {-}& \bfseries 12& \bfseries 8& \bfseries 6  & \bfseries  {-}& \bfseries 21& \bfseries 9& \bfseries 7    & {-}& \bfseries 54& \bfseries 11& \bfseries 7\\
		\addlinespace
		16&4   & 65&45&32&24 & {-}&93&76&61  & {-}&{-}&{-}&{-}\\
		  &16  & 85&43&21& \bfseries 12 & {-}&{-}&73&46 & {-}&{-}&{-}&{-}\\
		  &64  & {-}& \bfseries 32& \bfseries 13& \bfseries 9 & {-}&{-}& \bfseries 33& \bfseries 12 & {-}&{-}&{-}&36\\
		  &256 & \bfseries  {-}& \bfseries 19& \bfseries 11& \bfseries 9 & {-}& \bfseries 62& \bfseries 13& \bfseries 9   & {-}&{-}& \bfseries 30& \bfseries 11\\
		  &1024& \bfseries  {-}& \bfseries 15& \bfseries 11& \bfseries 8 & \bfseries  {-}& \bfseries 26& \bfseries 11& \bfseries 9   & {-}& \bfseries 67& \bfseries 13& \bfseries 9\\
		\addlinespace
		32&4   & 67&46&34&25  & {-}&94&78&63  & {-}&{-}&{-}&{-}\\
		  &16  & 91&46&24& \bfseries 14  & {-}&{-}&78&49 & {-}&{-}&{-}&{-}\\
		  &64  & {-}& \bfseries 34& \bfseries 15& \bfseries 11 & {-}&{-}& \bfseries 37& \bfseries 15 & {-}&{-}&{-}&38\\
		  &256 & \bfseries  {-}& \bfseries 22& \bfseries 13& \bfseries 11 & {-}& \bfseries 72& \bfseries 15& \bfseries 11  & {-}&{-}& \bfseries 33& \bfseries 13\\
		  &1024& \bfseries  {-}& \bfseries 17& \bfseries 13& \bfseries 11 & \bfseries  {-}& \bfseries 31& \bfseries 13& \bfseries 11  & {-}& \bfseries 79& \bfseries 15& \bfseries 11\\
		\addlinespace
		64&4   &69&47&35&26  & {-}&95&80&64  & {-}&{-}&{-}&{-}\\
		  &16  &{-}&53&27& \bfseries 15 & {-}&{-}&83&52 & {-}&{-}&{-}&{-}\\
		  &64  &{-}& \bfseries 38& \bfseries 16& \bfseries 14 & {-}&{-}& \bfseries 42& \bfseries 17 & {-}&{-}&{-}&43\\
		  &256 & \bfseries {-}& \bfseries 25& \bfseries 15& \bfseries 13 & {-}& \bfseries 75& \bfseries 18& \bfseries 14  & {-}&{-}& \bfseries 37& \bfseries 16\\
		  &1024& \bfseries {-}& \bfseries 20& \bfseries 15& \bfseries 12 & \bfseries  {-}& \bfseries 36& \bfseries 15& \bfseries 13  & {-}& \bfseries 89& \bfseries 18& \bfseries 13\\
		\bottomrule
	\end{tabular}
\end{table}

We note that often when $N$ is small (the first and sometimes second row for each new value of $p$ in Table \ref{tab:helmholtz_iterations}) the method converges but requires a relatively large number of iterations. This is not the regime we are interested in because these tend to be low accuracy solutions to the BVP. For high accuracy, condition \eqref{eq:dispersion_criterion} does a good job predicting which problems tend to converge with a small number of iterations, as desired. However, there are a small number of cases where this criterion is met, but the method fails to convergence in under 100 iterations. In our data set (which includes the data shown in Table \ref{tab:helmholtz_iterations} as well as data for $k$ up to 101) there are 26 such failures out of a total of 252 cases where $l$ meets the criterion. It is interesting to note that, in every failure case, increasing $l$ by 1 leads to convergence within 100 iterations. We attribute these failures to the fact that criterion \eqref{eq:dispersion_criterion} is based on a dispersion analysis on an infinite grid and holds only when $kh \gg 1$ \cite{Ainsworth2004}. For a more refined criterion, the same paper contains the exact dispersion relation on an infinite grid. We find that the numbers of iterations in Table \ref{tab:helmholtz_iterations} are small wherever the dispersion errors are small (taking care to replace $p$ with $l$ as in \eqref{eq:dispersion_criterion}). These observations suggest that for the preconditioner to be effective, dispersion errors must be controlled for the coarse problem. Similar behaviour is observed using the criteria from \cite{Melenk2011} where dispersion error is controlled on finite unstructured grids (although the criteria now depend on two implicit constants).

\begin{remark}
	It is important to consider the dispersion errors of the coarse problem rather than the global problem when determining the efficacy of the preconditioner. For example, if we replace $l$ by $p$ in criterion \eqref{eq:dispersion_criterion} (giving a global dispersion criterion) then there are 445 tests which fail to converge out of a total 724 cases satisfying this global dispersion criterion. Increasing $l$ by 1 in failure cases of this type does not lead to convergence. \remarkend
\end{remark}

\begin{figure}[!t]
	\centering
	\includegraphics[scale=1]{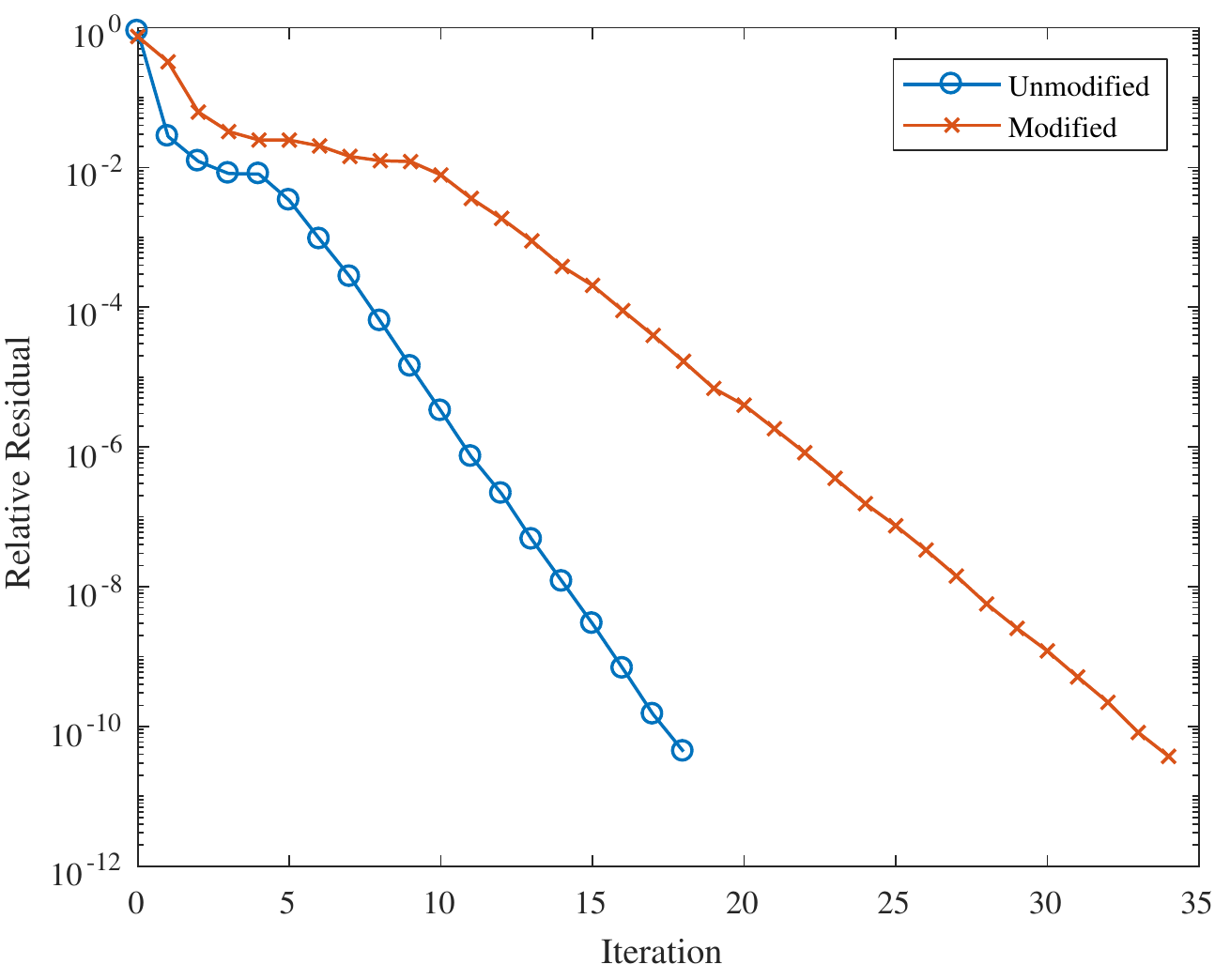}	
	\caption{Preconditioned relative residual $\norm{\vect{P}^{-1}\vect{r}_k}_2 / \norm{\vect{P}^{-1}\vect{r}_0}_2$ versus iteration number for the unmodified and modified Helmholtz problem. Vector $\vect{r}_k$ is the residual vector corresponding to \eqref{eq:schur_complement_system} at the $k$th iteration. \label{fig:residual_convergence}}
\end{figure}

The data in Table \ref{tab:helmholtz_iterations} was collected using grids and wavenumbers that avoid spurious resonant frequencies. In the following, we construct an example to illustrate that spurious resonant frequencies can adversely affect convergence. First, we solve the same BVP but with $N=64$, $p=32$, $l=3$, and $k=16.55$ with both the unmodified domain decomposition method and the Robin boundary modified method. This wavenumber is close to, but not exactly, an element resonant frequency $k \approx 16.56157163134991$ (which was computed numerically). Parameter $l$ was chosen to satisfy \eqref{eq:dispersion_criterion}. Figure \ref{fig:residual_convergence} illustrates the relative residual as a function of number of iterations for both methods. We terminate both methods when the relative residual has decreased by ten orders of magnitude. The behaviour in Figure \ref{fig:residual_convergence} is typical in the sense that the modified method requires more iterations to converge than the unmodified method. In both cases, the error in the computed solution measured in the $H^1$ norm is on the order of $10^{-16}$.

Next, we repeat the same experiment but with $k$ set to the spurious resonant frequency. The unmodified method fails catastrophically after one iteration and returns a solution with error on the order of $10^5$ (additional iterations can reduce the norm of the preconditioned residual but do not succeed in reducing this error). The modified method converges largely in the same way as in Figure \ref{fig:residual_convergence} with a solution whose error is on the order of $10^{-16}$.

\subsection{Electromagnetic Problems}

In the following examples, we solve Maxwell's source-free equations
\begin{align}
	\nabla \times \vect{E} &= -\frac{\partial}{\partial t} \vect{B}, \\ 
	\nabla \times \vect{H} &= \phantom{-} \frac{\partial}{\partial t} \vect{D}, \\
	\nabla \cdot \vect{D} &= 0, \\
	\nabla \cdot \vect{B} &= 0,
\end{align}
subject to constitutive relations $\vect{D} = \vect{\epsilon} \vect{E}$ and $\vect{B} = \vect{\mu} \vect{H}$ where $\vect{E}(\vect{x},t)$ is the electric field intensity, $\vect{H}(\vect{x},t)$ is the magnetic field intensity, $\vect{D}(\vect{x},t)$ is the electric flux density, $\vect{B}(\vect{x},t)$ is the magnetic flux density, and $\vect{\epsilon} = \epsilon_0 \vect{\epsilon}_r$ and $\vect{\mu} = \mu_0 \vect{\mu}_r$ are the permittivity and permeability tensors respectively, comprised of the permittivity of free space $\epsilon_0$, permeability of free space $\mu_0$, and relative permittivity $\vect{\epsilon}_r$ and permeability $\vect{\mu}_r$. These governing equations are subject to homogeneous boundary condition 
\begin{equation} \label{eq:pec_boundary_condition}
	\vect{n} \times \vect{E} = 0
\end{equation}
on perfect electric conductors (whose boundary has normal $\vect{n}$ directed into the conductor) and continuity conditions
\begin{align} \label{eq:tangential_continuity}
	\vect{n} \times (\vect{E}_2 - \vect{E}_1) &= 0, \\
	\vect{n} \times (\vect{H}_2 - \vect{H}_1) &= 0,
\end{align}
at discontinuities in tensors $\vect{\epsilon}$ and $\vect{\mu}$ where $\vect{E}_1$ and $\vect{H}_1$ are fields to one side of the discontinuity and $\vect{E}_2$ and $\vect{H}_2$ fields to the other side. The discontinuity is given along a boundary with normal $\vect{n}$ directed into the second region.

We assume time-harmonic solutions to Maxwell's equations (for example, $\vect{E}(\vect{x},t) = \Re \{ \hat{\vect{E}}(\vect{x}) e^{\jmath \omega t} \}$) so that the time-harmonic solutions are governed by curl-curl equations
\begin{align}
	\nabla \times (\vect{\mu}^{-1} \nabla \times \hat{\vect{E}}) &= \omega^2 \vect{\epsilon} \hat{\vect{E}}, \label{eq:curl_curl_E} \\
	\nabla \times (\vect{\epsilon}^{-1} \nabla \times \hat{\vect{H}}) &= \omega^2 \vect{\mu} \hat{\vect{H}}. \label{eq:curl_curl_H}
\end{align}
Since we strictly work with the time-harmonic quantities rather than their time-varying counterparts, we suppress the hat notation in this paper. Although the quantities in \eqref{eq:curl_curl_E}-\eqref{eq:curl_curl_H} possess units, we will work with a set of scaled quantities that are unitless using the change of variables
\begin{equation}
	\tilde{\vect{x}} = \frac{1}{L}\vect{x},\qquad
	\tilde{t} = \frac{1}{L\sqrt{\epsilon_0 \mu_0}}t,\qquad
	\tilde{\vect{H}} = \frac{1}{H}\vect{H},\qquad
	\tilde{\vect{E}} = \frac{1}{H}\sqrt{\frac{\epsilon_0}{\mu_0}}\vect{E},
\end{equation}
and corresponding angular frequency $\tilde{\omega} = L\sqrt{\epsilon_0 \mu_0}\omega$ where $L$ and $H$ are chosen constants (typically associated with the wavelength of the phenomena and the peak amplitude of $\vect{H}$, respectively, for a given problem). This change of variables results in permittivity $\tilde{\vect{\epsilon}} = \vect{\epsilon}_r$ and $\tilde{\vect{\mu}} = \vect{\mu}_r$. We assume that all examples in the paper have been scaled in this way and suppress the tilde notation.

All of our examples seek solutions to scattering problems on unbounded domains. To discretize such problems on a finite domain, we use PMLs aligned with the coordinate axes. We define 
\begin{equation}
	s_i = 1 - \jmath \frac{\sigma_i (\vect{x})}{\omega}
\end{equation}
with $\sigma_i(\vect{x})$ piecewise positive constant functions non-zero only where the PMLs are required and $i=1,2,3$. In addition, we let $\vect{S}_{\textrm{PML}} = \diag{(\vect{s})}$ and $\vect{\Lambda}_{\textrm{PML}} = \det{(\vect{S}_{\textrm{PML}})}\vect{S}_{\textrm{PML}}^{-2}$ so that, using the complex coordinate stretching approach to implementing PMLs \cite{Jin2014}, we  replace permittivity and permeability tensors with $\vect{\epsilon}_{\textrm{PML}} = \vect{\Lambda}_{\textrm{PML}}\vect{\epsilon}$ and $\vect{\mu}_{\textrm{PML}} = \vect{\Lambda}_{\textrm{PML}}\vect{\mu}$. Then $\vect{E}$ and $\vect{H}$ solve the original equations \eqref{eq:curl_curl_E}-\eqref{eq:curl_curl_H} in regions where $\sigma_i(\vect{x})$ are all zero, but decay exponentially to zero in the PMLs. We then terminate the domain of interest by using fictitious homogeneous boundary conditions outside the PMLs.

In addition, all of our examples will be translation invariant in the $x_3$ direction ($\vect{E}$, $\vect{H}$, $\vect{\epsilon}$, $\vect{\mu}$ are independent of $x_3$) and will have permittivity and permeability structured as
\begin{equation} \label{eq:anisotropic_dielectric}
	\vect{\epsilon} = 
	\begin{bmatrix}
    	\epsilon_{11} & \epsilon_{12} & 0             \\
    	\epsilon_{12} & \epsilon_{22} & 0             \\
        	        0 &             0 & \epsilon_{33}
	\end{bmatrix}, \qquad
	\vect{\mu} = 
	\begin{bmatrix}
    	\mu_{11} & \mu_{12} & 0        \\
    	\mu_{12} & \mu_{22} & 0        \\
        	   0 &        0 & \mu_{33}
	\end{bmatrix}.
\end{equation}
As a consequence, solutions to either of \eqref{eq:curl_curl_E}-\eqref{eq:curl_curl_H} can be decomposed into independent transverse magnetic and transverse electric modes (denoted by $\textrm{TM}_{x_3}$ mode and $\textrm{TE}_{x_3}$ mode, respectively). The $\textrm{TM}_{x_3}$ mode assumes a solution to \eqref{eq:curl_curl_E} with the third component of $\vect{H}$ and first and second components of $\vect{E}$ set to zero. Then a BVP of the form \eqref{eq:prototype_pde} for the third component of $\vect{E}$ can be solved (from which the first and second components of $\vect{H}$ can be obtained without solution of additional BVPs). A similar procedure can be performed to compute the $\textrm{TE}_{x_3}$ mode.

In our first three examples, we consider the $\textrm{TM}_{x_3}$ mode with both $\vect{\mu}$ and $\vect{\epsilon}$ diagonal and spatially varying. The third equation in \eqref{eq:curl_curl_E} becomes
\begin{equation} \label{eq:TM_z_equation}
	-\frac{\partial}{\partial x_1} \left( \frac{1}{\mu_{22}} \frac{\partial}{\partial x_1} E_{x_3} \right)
	-\frac{\partial}{\partial x_2} \left( \frac{1}{\mu_{11}} \frac{\partial}{\partial x_2} E_{x_3} \right) - \omega^2 \epsilon_{33} E_{x_3} = 0.
\end{equation}
We solve this equation by separating the total field $E_{x_3}$ into a known incident field $E_i$ and unknown scattered field $E_s$ such that $E_{x_3} = E_i + E_s$. Then \eqref{eq:TM_z_equation} becomes \eqref{eq:prototype_pde} with
\begin{equation}
	\vect{\alpha} =
	\begin{bmatrix}
    	\mu_{22}^{-1} & 0             \\
    	0             & \mu_{11}^{-1}
	\end{bmatrix}, \qquad
	\beta = -\omega^2 \epsilon_{33}, \qquad
	\phi = E_s, \qquad
	f = \nabla \cdot (\vect{\alpha} \nabla E_i) - \beta E_i,
\end{equation}
and we enforce zero Dirichlet boundary conditions on $\phi$ on an enclosing boundary beyond the PML. The techniques described in Section \ref{sec:numerical_methods} are used to compute the scattered field.

\subsubsection{Dielectric Cylinder} \label{sec:dielectric_cylinder}

We begin with a classical example of scattering from an infinitely long dielectric cylinder \cite{Jin2010} which we use to demonstrate the accuracy of the method. The cylinder is coaxial with the $x_3$ axis, has radius $a=1$, and is enclosed inside domain $\Omega = (-4,4)^2$. We choose $\epsilon_{33} = 4$ inside the cylinder and $\epsilon_{33} = 1$ outside. Both $\mu_{11}=\mu_{22}=1$ throughout the domain. We select an incident plane wave $E_i = e^{-\jmath \omega \vect{k}^T\vect{x}}$ with unit vector $\vect{k}$ so that
\begin{equation}
	f = \omega^2 (\epsilon_{33} - \vect{k}^T \vect{\alpha} \vect{k}) E_i
\end{equation}
and is non-zero only inside the cylinder. We set the direction $\vect{k} = \vect{e}_1$ corresponding to propagation of the incident wave in the positive $x_1$ direction. We choose $\omega = 4\cdot 2\pi$ so that the wavelength outside the cylinder is $\lambda = 1/4$ and the domain is $32\lambda \times 32\lambda$ in size.

To absorb the scattered wave, we include a PML of thickness $w=1$ adjacent to each edge of $\Omega$ with $\sigma_1$ non-zero in the two bands $x_1 \in (-4,-3)$ and $x_1 \in (3,4)$ (similarly, $\sigma_2$ is non-zero when $x_2 \in (-4,-3)$ and $x_2 \in (3,4)$). Each decay rate parameter takes the value 15 in their respective bands. Recall that the PML modifies the parameters $\epsilon_{33}$, $\mu_{11}$, and $\mu_{22}$ wherever $\sigma_i$ is non-zero.

\begin{figure*}[!tp]
    \centering
    \begin{subfigure}[t]{0.475\textwidth}
        \centering
        \includegraphics[width=0.775\textwidth]{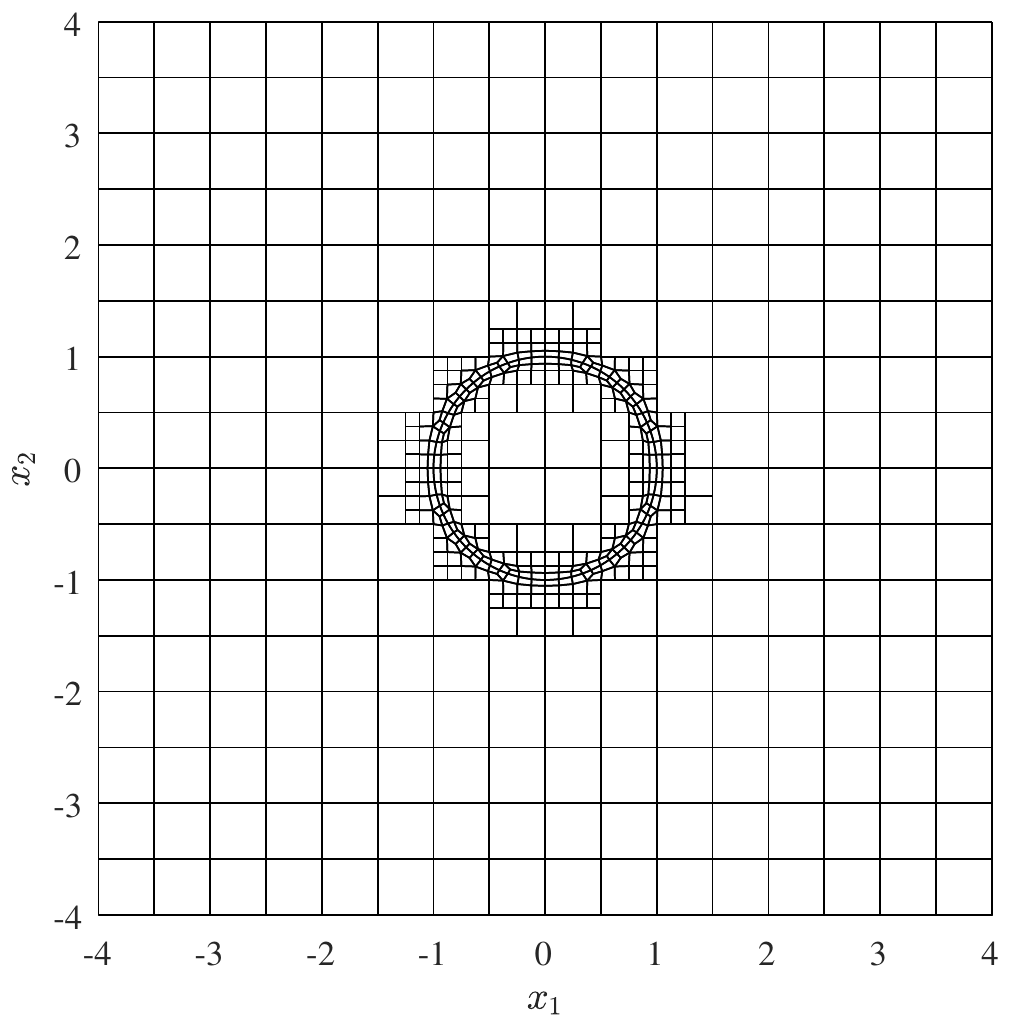}
        \caption{Mesh used to compute the scattered field. \label{fig:dielectric_cylinder_mesh}}
    \end{subfigure}
    \hfill
    \begin{subfigure}[t]{0.475\textwidth}
        \centering
        \includegraphics[width=\textwidth]{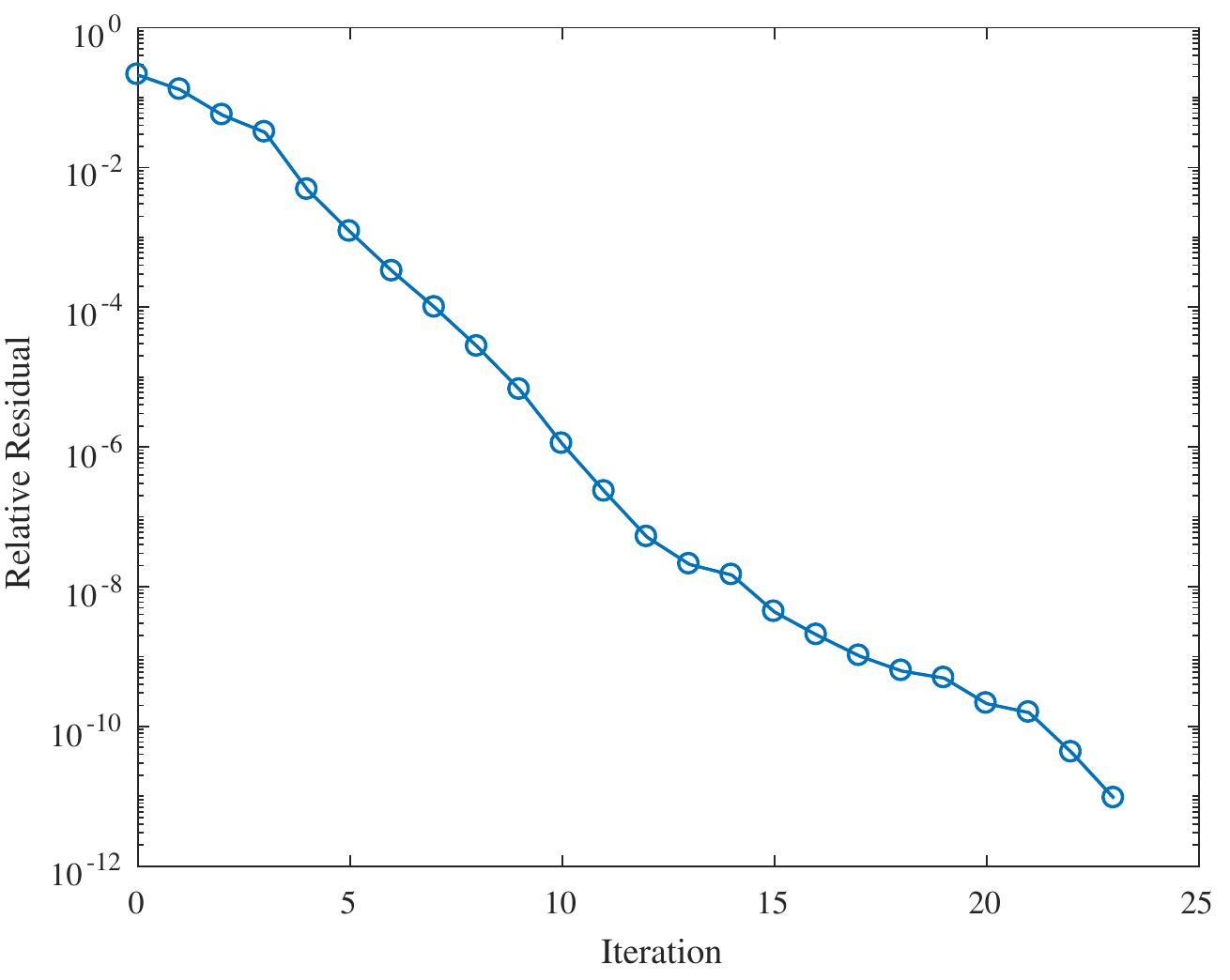}
        \caption{Preconditioned relative residual versus iteration number. \label{fig:dielectric_cylinder_residual}}
    \end{subfigure}
    
    \par\smallskip 
    
    \begin{subfigure}[t]{0.475\textwidth}
        \centering
        \includegraphics[width=\textwidth]{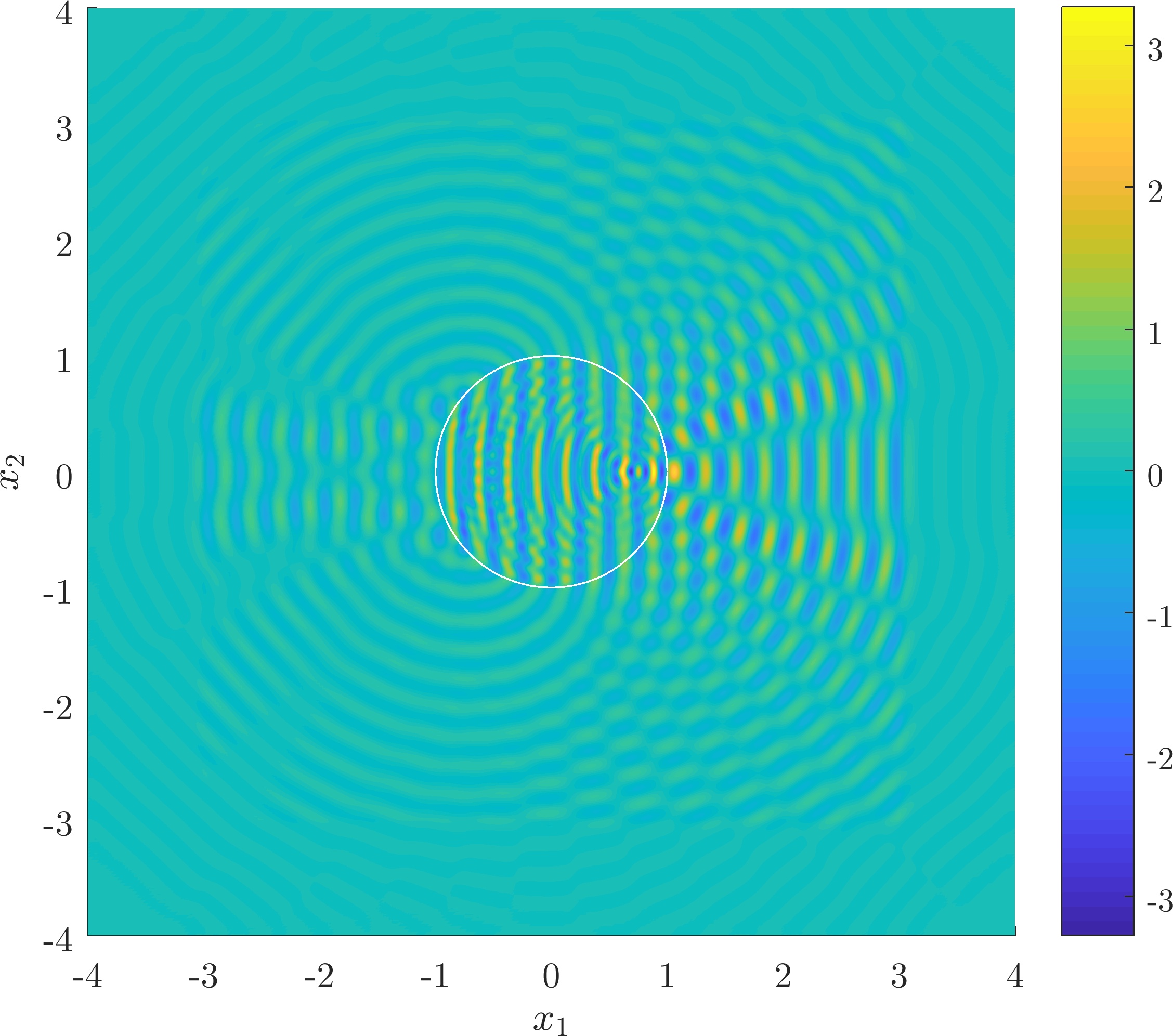}
        \caption{Real part of the scattered field. \label{fig:dielectric_cylinder_solution}}
    \end{subfigure}
    \hfill
    \begin{subfigure}[t]{0.475\textwidth}
        \centering
        \includegraphics[width=\textwidth]{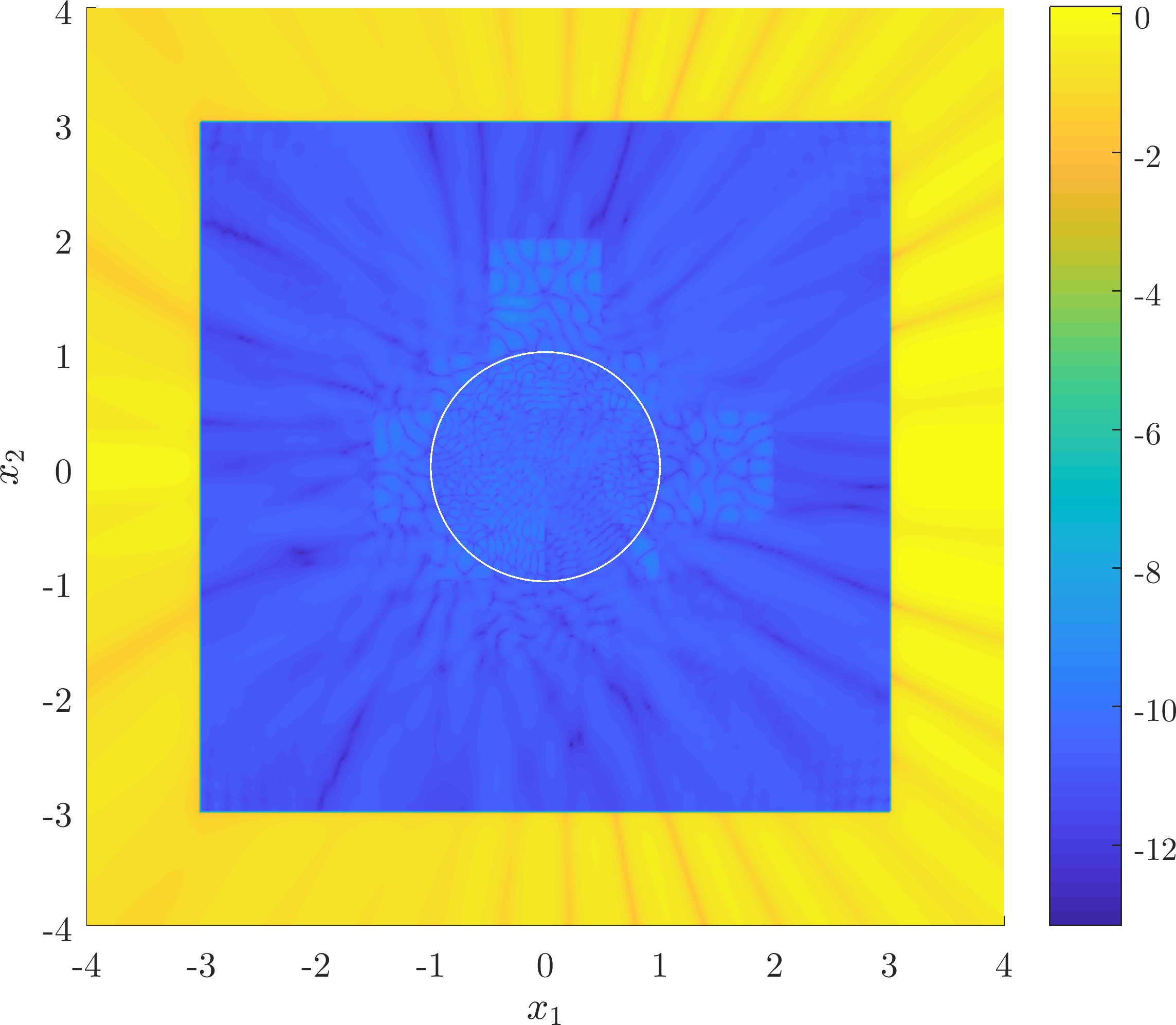}
        \caption{Logarithm of the absolute value of the error in the scattered field ($\log_{10}\abs{E_s - E_{s,\textrm{exact}}}$).\label{fig:dielectric_cylinder_error}}
    \end{subfigure}
    
    \par\medskip 
    
    \begin{subfigure}[t]{0.475\textwidth}
        \centering
        \includegraphics[width=\textwidth]{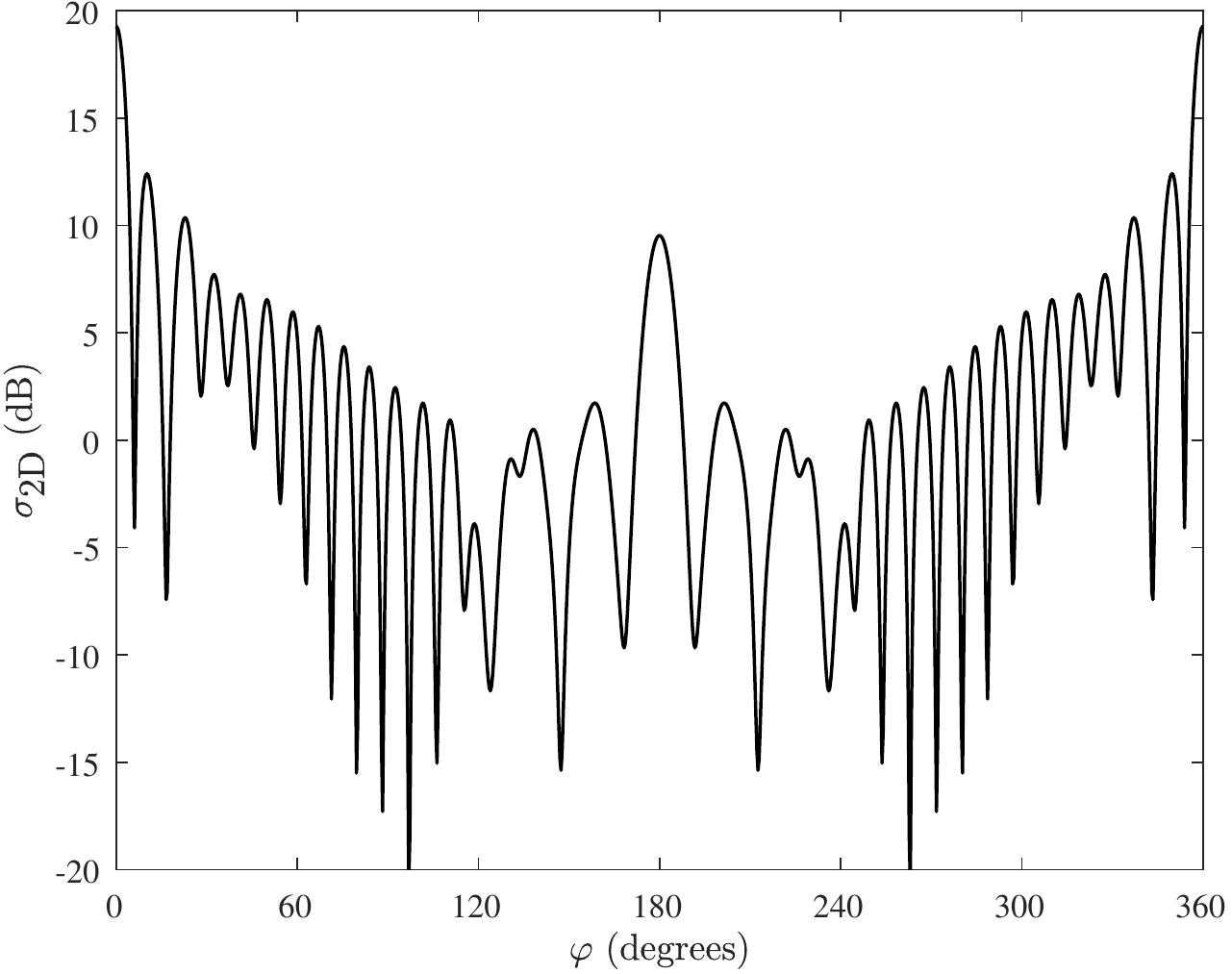}
        \caption{Scattering width (RCS). \label{fig:dielectric_cylinder_RCS}}
    \end{subfigure}
    \hfill
    \begin{subfigure}[t]{0.475\textwidth}
        \centering
        \includegraphics[width=\textwidth]{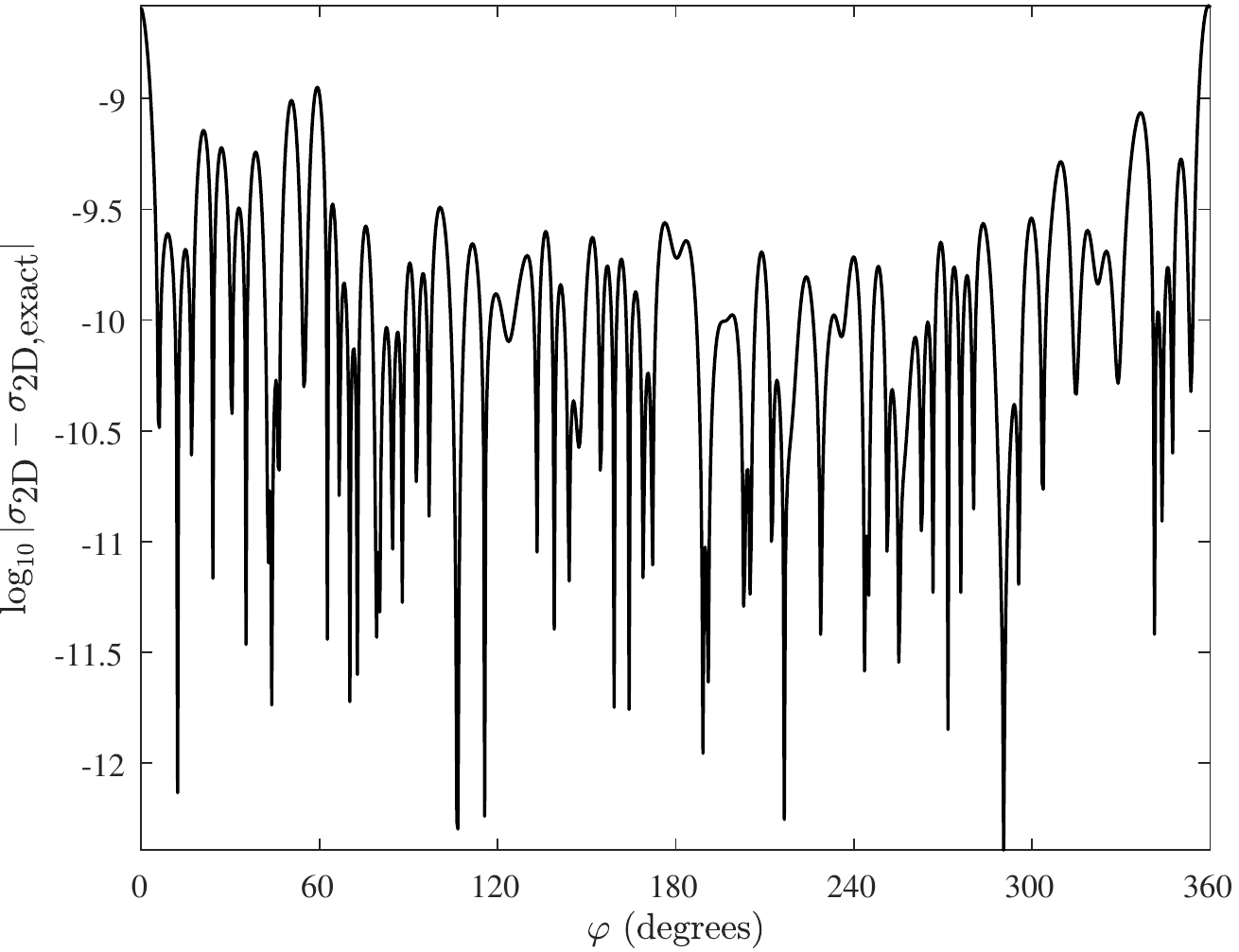}
        \caption{Logarithm of the absolute value of the error in the RCS. \label{fig:dielectric_cylinder_RCS_error}}
    \end{subfigure}

    \caption{Results for the dielectric cylinder problem described in Section \ref{sec:dielectric_cylinder}.}
\end{figure*}

We select a mesh (illustrated in Figure \ref{fig:dielectric_cylinder_mesh}) whose edges match boundaries in discontinuities in the parameters $\sigma_i$ and $\epsilon_{33}$ (that is, at boundaries of the PML or at the boundary of the cylinder) using the mesh generation technique of Section \ref{sec:mesh_generation}. In particular, elements in the quadtree near the cylinder boundary belong to the sixth level of refinement, whereas elements away from the boundary are kept with four levels of refinement. All boundary projections and Legendre expansions are computed to a tolerance of $10^{-12}$. Each element uses a degree 32 polynomial expansion to represent the scattered field. We solve the associated discrete problem using the domain decomposition method without Robin boundary modification (and do not use scaling) and terminate iteration when the relative residual has decreased by $10^{-10}$. We use the criterion \eqref{eq:dispersion_criterion} to determine the smallest acceptable value of $l$ then add one as a safety precaution (see the discussion in Section \ref{sec:convergence_test_helmholtz}). The mesh size $h$ in \eqref{eq:dispersion_criterion} is chosen as the longest edge in the mesh (this may be pessimistic in practice since many edges in the mesh may be smaller than $h$).

Figure \ref{fig:dielectric_cylinder_residual} demonstrates the relative residual reduction as a function of iteration number for this problem. There is a total of 640,332 unknowns in $\vect{\phi}$ and the coarse problem matrix $\vect{K}$ is square with 9,909 rows. Factorization of this matrix represents the main bottleneck of the domain decomposition method for high frequency problems because the size of the coarse problem grows as the frequency is increased (because $l$ is one factor that determines the size of $\vect{K}$ and $l$ must grow as the frequency grows to continue to satisfy \eqref{eq:dispersion_criterion}).

Figure \ref{fig:dielectric_cylinder_solution} illustrates the real part of the computed scattered field $E_s$ and Figure \ref{fig:dielectric_cylinder_error} shows the point-wise error $\abs{E_s - E_{s,\textrm{exact}}}$. The exact solution (see, for example, \cite{Jin2010}) is computed using the sum
\begin{equation} \label{eq:exact_solution}
	E_{s,\textrm{exact}}(r,\varphi) = 
	\begin{cases}
		\displaystyle
		2 \sideset{}{'}\sum_{i=0}^{\infty} a_i H_i^{(2)}(\omega r) \cos{(i \varphi)} & r \ge a \\
		\displaystyle
		-E_i + 2 \sideset{}{'}\sum_{i=0}^{\infty} c_i J_i(\omega_r r) \cos{(i \varphi)} & r < a
	\end{cases}
\end{equation}
where
\begin{align}
	a_i &= -\jmath^{-i} \frac{J_i'(\omega a) J_i(\omega_r a) - \sqrt{\epsilon_r} J_i(\omega a) J_i'(\omega_r a)}{H_i^{(2)\prime}(\omega a) J_i(\omega_r a) - \sqrt{\epsilon_r} H_i^{(2)}(\omega a) J_i'(\omega_r a)} \\
	c_i &= \frac{\jmath^{-(i+1)}}{\pi \omega a} \frac{2}{H_i^{(2)\prime}(\omega a) J_i(\omega_r a) - \sqrt{\epsilon_r} H_i^{(2)}(\omega a) J_i'(\omega_r a)}
\end{align}
and $\omega_r = \sqrt{\epsilon_r} \omega$, $\epsilon_r = 4$, $r = \norm{\vect{x}}_2$, $\varphi = \textrm{atan2}(x_2,x_1)$, and $J_i$ denotes the Bessel function of the first kind of order $i$. The primes on summation symbols in \eqref{eq:exact_solution} require halving the $i=0$ term and primes on Bessel and Hankel functions require taking the derivative with respect to their arguments. We truncated the sum after 75 terms which was enough for the series to converge in a disk large enough to include domain $\Omega$. In both the solution and error figures, we have shown the field in the PML region to emphasize that the scattered field decays to zero (in later examples, we do not show the PML region since the solution is not physically relevant there). Note that the error is computed to approximately 10 digits of accuracy throughout the physical domain (the error in the PML is large, but this is expected). To achieve such high accuracy, it is necessary to accurately capture the curvilinear boundary and use a PML with parameters $w$ and $\sigma_i$ chosen appropriately (as we have done here).

In addition to the scattered field, Figure \ref{fig:dielectric_cylinder_RCS} illustrates the computed radar cross section (RCS) of the cylinder and Figure \ref{fig:dielectric_cylinder_RCS_error} its corresponding error at 1000 evenly spaced observation angles. The RCS (sometimes called scattering width or echo width in two dimensions \cite{Jin2014}) is given by
\begin{align}
	\sigma_{\textrm{2D}}(\varphi,\varphi_i) &= \lim_{\norm{\vect{x}}_2 \rightarrow \infty} 2\pi \norm{\vect{x}}_2 \frac{\abs{E_s}^2}{\abs{E_i}^2} \\
	&= \frac{\omega}{4} \abs{E_{\textrm{far}}}^2 \label{eq:rcs_def}
\end{align}
with far field integral
\begin{equation} \label{eq:far_field}
	E_{\textrm{far}}(\varphi) = \oint_{\Gamma} \big[ \hat{\vect{x}}\cdot\vect{n}'E_s(\vect{x}') - (\jmath \omega)^{-1} \vect{n}^{\prime} \cdot \nabla' E_s(\vect{x}') \big] e^{\jmath \omega \hat{\vect{x}} \cdot \vect{x}'} \, d\Gamma'
\end{equation}
and $\hat{\vect{x}} = [\cos{(\varphi),\, \sin{(\varphi)}}]^T$. We integrate over the boundary of the cylinder $\Gamma$. Since the dielectric cylinder is invariant under rotations about the $x_3$ axis, we do not need to consider multiple incident field directions $\varphi_i$ and the solution in Figure \ref{fig:dielectric_cylinder_solution} is enough to characterize the RCS. 

In practice, we evaluate the far field integral by performing local integrals over elements with curvilinear boundaries belonging to the cylinder. Each local integral is computed using Clenshaw-Curtis quadrature. The exact RCS is computed by substituting the exact solution \eqref{eq:exact_solution} into \eqref{eq:far_field}. The integrand is periodic and continuous so we can use the Trapezoidal rule to compute the full boundary integral accurately. Figure \ref{fig:dielectric_cylinder_RCS_error} shows that the RCS determined from the computed scattered field $E_s$ is accurate to 9 digits.

\subsubsection{Eaton Lens} \label{sec:eaton_lens}

For our next example, we consider a lens problem based on one in \cite{Vico2016}. We use this example to demonstrate the method on a problem with more complicated spatial variation of permittivity and also to show how the method behaves when increasing the frequency of the problem relative to the size of the domain. Our lens example is a two-dimensional Eaton lens that is meant to bend electromagnetic beam fields by 90 degrees. Again, we work with the $\textrm{TM}_{x_3}$ mode. The lens has spatially varying permittivity 
\begin{equation}
	\epsilon_{33} = 
	\begin{cases}
		\displaystyle	
		\big[ n(\norm{\vect{x}}_2) \big]^2 & \norm{\vect{x}}_2 \le a \\
		1 & \textrm{otherwise}
	\end{cases}
\end{equation}
where the radially symmetric index of refraction $n$ satisfies
\begin{equation}
	n^2 = \frac{a}{n\norm{\vect{x}}}_2 - \sqrt{\left(\frac{a}{n\norm{\vect{x}}_2}\right)^2 - 1}.
\end{equation}
We choose radius $a=0.45$ and compute $n$ for a given $\norm{\vect{x}}_2$ when needed using Newton's method with initial iterate 1. Instead of an incident plane wave, we use an incident Gaussian beam $E_i = H_0^{(2)}(\omega \norm{\vect{x} - \vect{x}_c}_2)e^{-\omega / 2}$ with complex center $\vect{x}_c = [-0.76-0.5\jmath, \; 0.275]^T$ at frequency $\omega = 48 \cdot 2\pi$ which corresponds to a wavelength $\lambda = 1/48$ outside the lens. We solve for the scattered field in domain $\Omega = (-0.75,0.75)^2$ which is $72\lambda \times 72 \lambda$ in size.

The PML has thickness $w=0.1875$ and decay rates $\sigma_i = 37$. Since the index of refraction is singular at the origin, we truncate it at a radius of 1/30 and leave it constant inside that radius (choosing the constant so that $\epsilon_{33}$ is continuous). We resolve both this artificial radius and the radius $a$ using a mesh that is one refinement finer than the mesh in Section \ref{sec:dielectric_cylinder} to avoid having to compute large Legendre expansions for $\beta$ (which is no longer smooth, only continuous). We compute those coefficients to a reduced error tolerance of $10^{-6}$ and apply the domain decomposition method with a reduced relative residual reduction tolerance of $10^{-6}$ since we have already compromised on the accuracy of such a solution by truncating the index of refraction. Degree 24 polynomials are used for all elements. Otherwise, all other parameters are the same as in Section \ref{sec:dielectric_cylinder}.

\begin{figure*}[!t]
    \centering
    \begin{subfigure}[t]{0.475\textwidth}
        \centering
        \includegraphics[width=\textwidth]{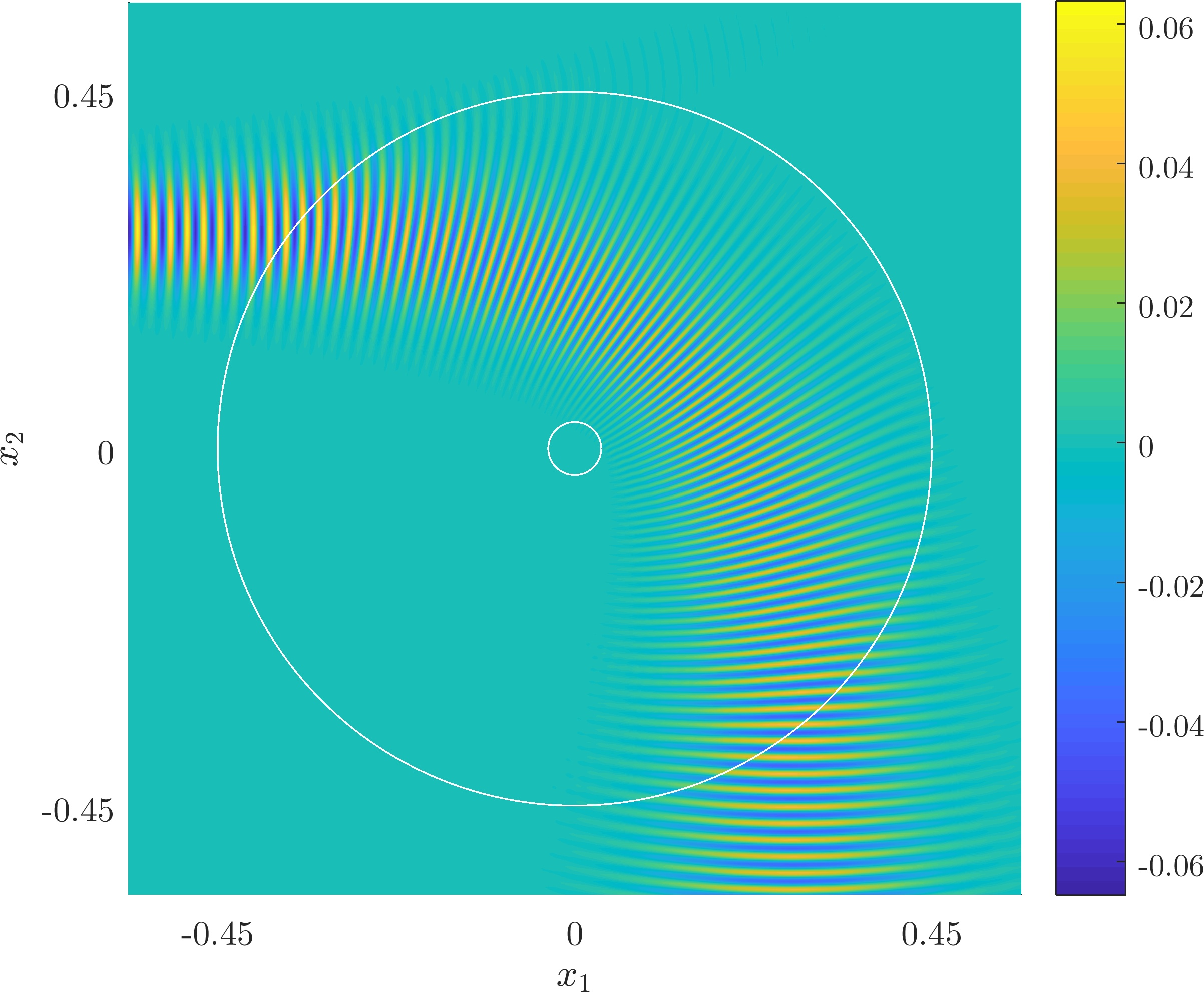}
        \caption{Real part of the total field. \label{fig:eaton_lens_solution}}
    \end{subfigure}
    \hfill
    \begin{subfigure}[t]{0.475\textwidth}
        \centering
        \includegraphics[width=\textwidth]{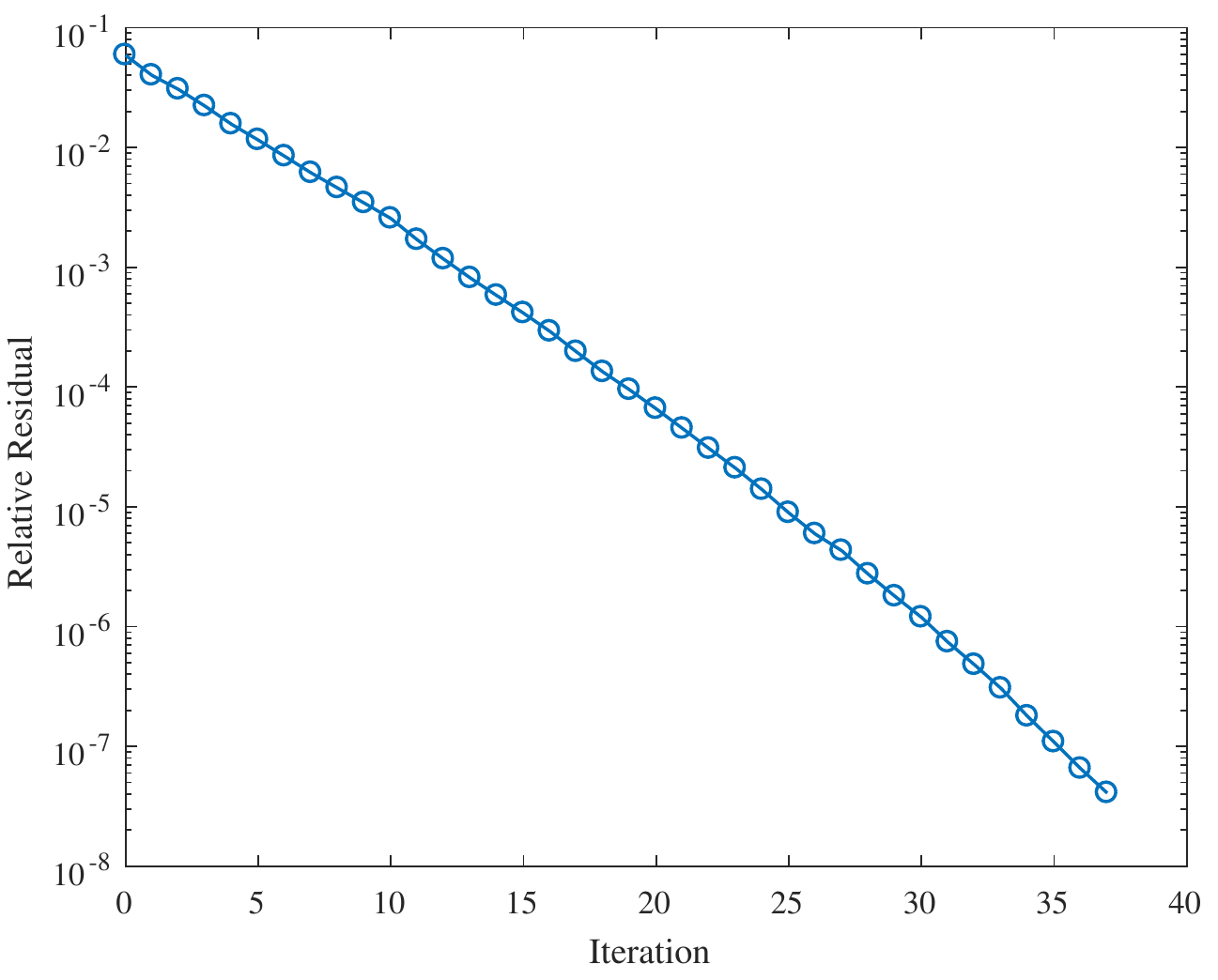}
        \caption{Preconditioned relative residual versus iteration number. \label{fig:eaton_lens_residual}}
    \end{subfigure}
    \caption{Eaton lens problem described in Section \ref{sec:eaton_lens}.}
\end{figure*}

Figure \ref{fig:eaton_lens_solution} illustrates the real part of the total field $E_i + E_s$ for the Eaton lens. Clearly, the Gaussian beam is bent by 90 degrees around the origin. Figure \ref{fig:eaton_lens_residual} shows the preconditioned relative residual as a function of the number of iterations. A modest number of iterations is required to compute the solution. There are 1,717,500 unknowns in $\vect{\phi}$ with 58,629 rows in matrix $\vect{K}$. This growth in the size of $\vect{K}$ relative to the dielectric cylinder problem in Section \ref{sec:dielectric_cylinder} is due to two factors: first, the increase in the problem size in terms of wavelength $\lambda$; second, the increase in the number of elements (by refining the grid we introduce more edges).

\subsubsection{Photonic Crystal Waveguide} \label{sec:photonic_waveguide}

Next, we consider a photonic crystal waveguide with geometry as described in \cite{Gillman2015}. In that paper, the dielectric rods that constitute the photonic crystal are spatially varying Gaussian cylinders. We choose to replace those cylinders with more conventional dielectric cylinders of constant permittivity (see, for example, \cite{Sharkawy2003}). Our method is capable of handling both problems, but the circular rods are more challenging (we discretize using a finer mesh to capture the circular boundary of each cylinder). We use this example to show how the method behaves when fine geometric features must be captured by the mesh.

\begin{figure*}[!t]
    \centering
    \begin{subfigure}[t]{\textwidth}
        \centering
        \includegraphics[width=\textwidth]{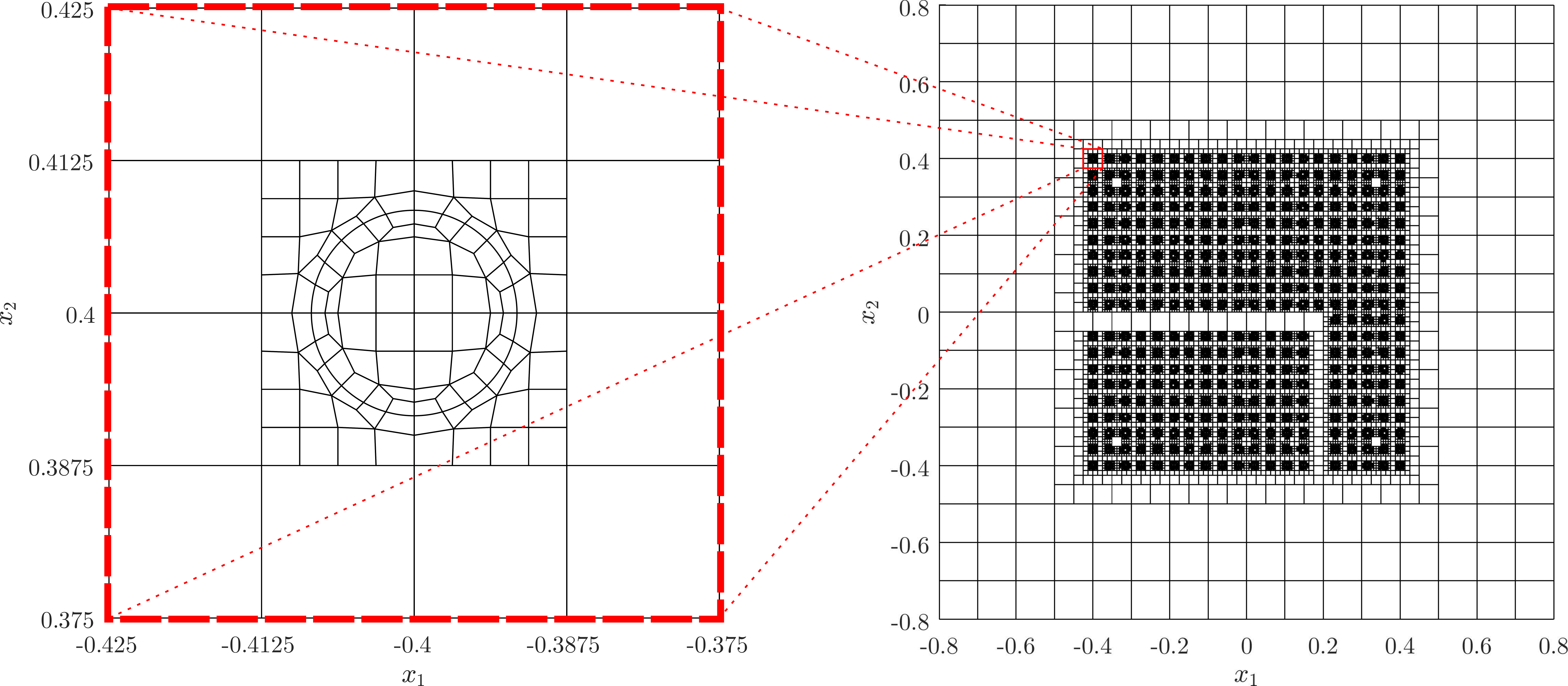}
        \caption{Mesh used to compute the scattered field. A detailed view near one dielectric rod appears on the left with the full mesh on the right. \label{fig:photonic_waveguide_mesh}}
    \end{subfigure}
    
    \vskip \baselineskip 
    
    \begin{subfigure}[t]{0.475\textwidth}
        \centering
        \includegraphics[width=0.95\textwidth]{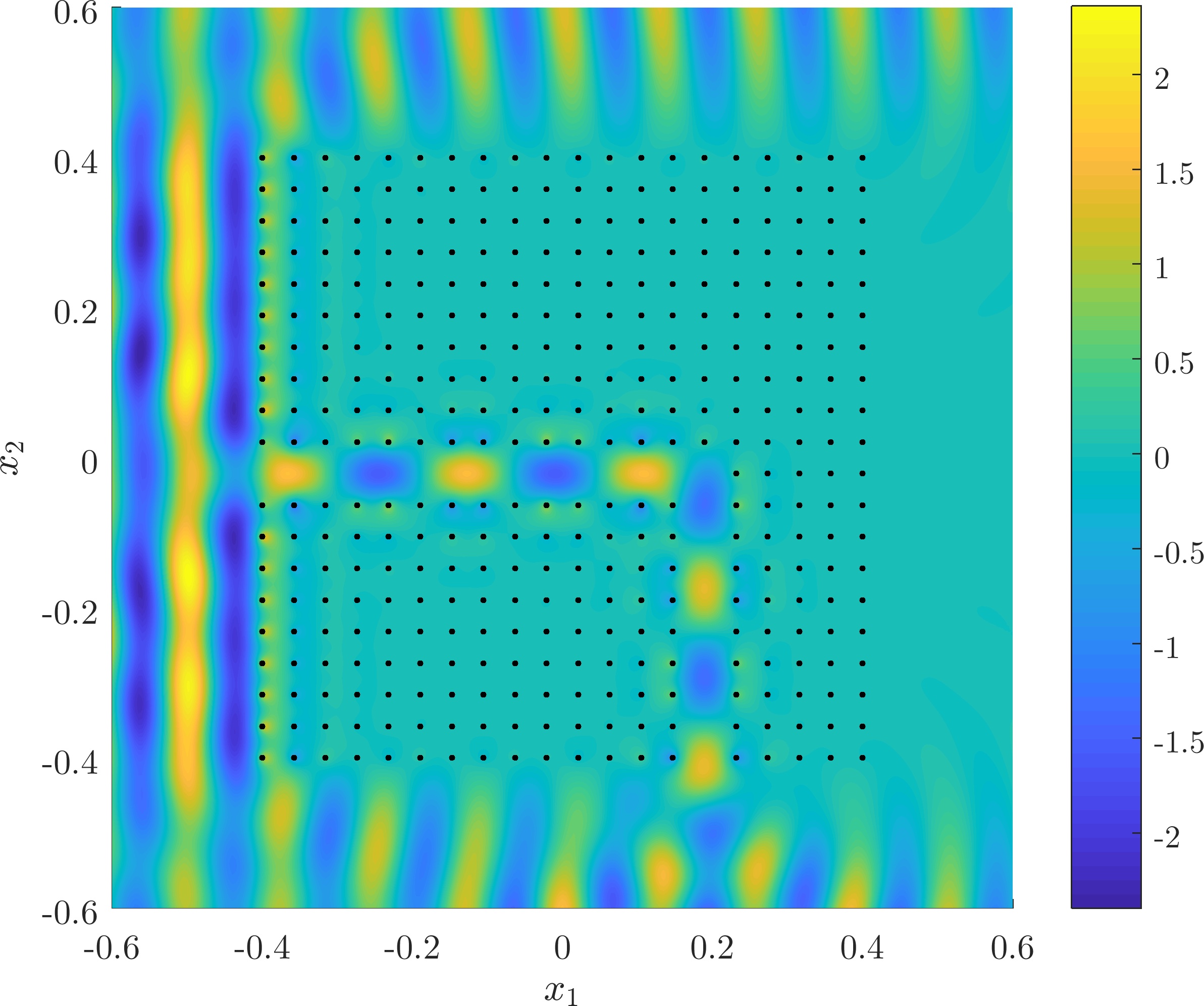}
        \caption{Real part of the total field. Centers of dielectric rods are shown with black dots. \label{fig:photonic_waveguide_solution}}
    \end{subfigure}
    \hfill
    \begin{subfigure}[t]{0.475\textwidth}
        \centering
        \includegraphics[width=\textwidth]{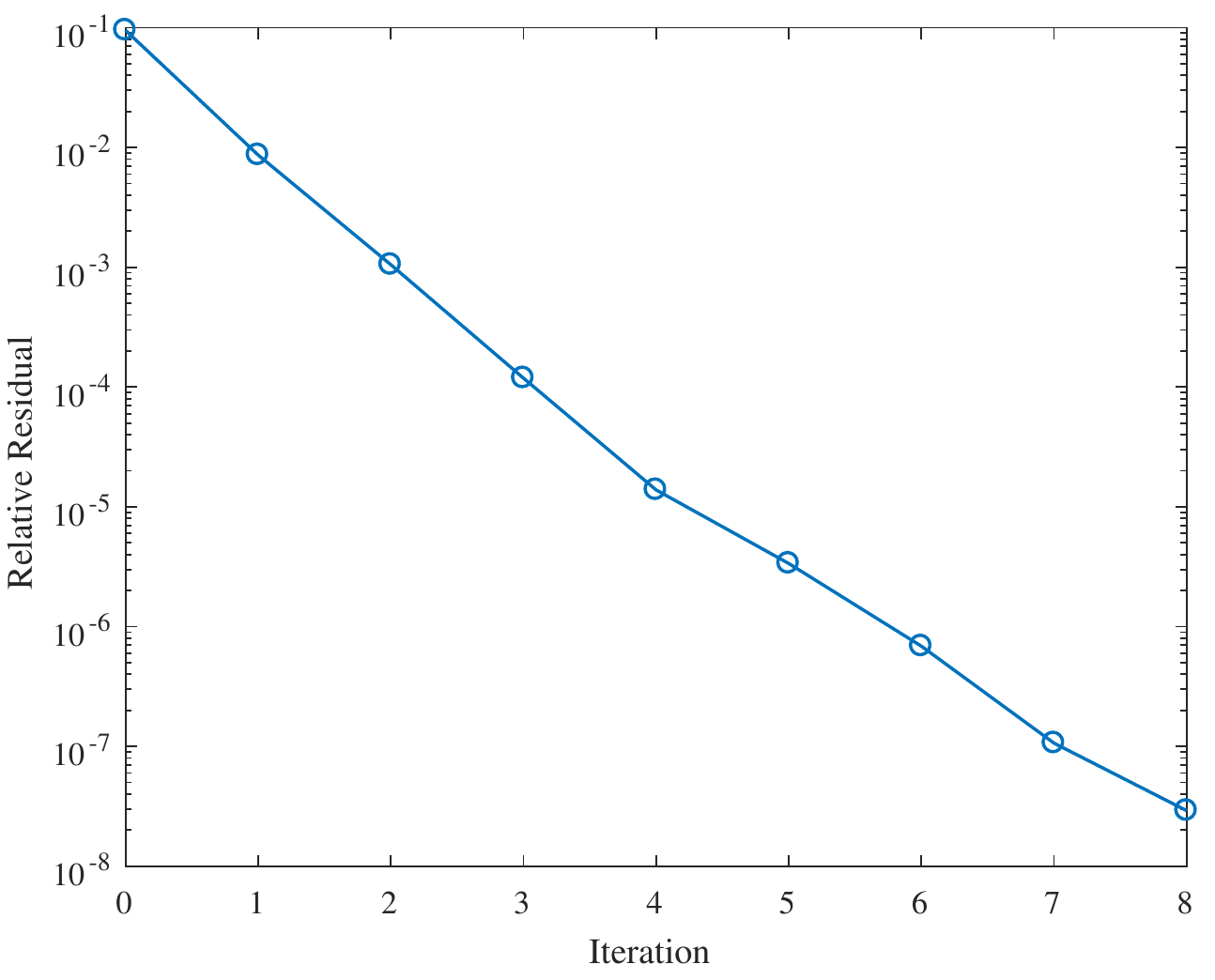}
        \caption{Preconditioned relative residual versus iteration number. \label{fig:photonic_waveguide_residual}}
    \end{subfigure}

    \caption{Results for the photonic crystal waveguide problem described in Section \ref{sec:photonic_waveguide}.}
\end{figure*}

The photonic crystal is comprised of a $20 \times 20$ array of dielectric cylinders, each of radius $a = 4/475$, enclosed in domain $\Omega = (-0.8,0.8)^2$. The centers of the rods are spaced uniformly in the region $[-0.4,0.4]^2$. A channel is removed along the 11th row and 15th column so as to form a type of waveguide bend (see Figure \ref{fig:photonic_waveguide_solution} for a schematic of the centers of the rods). The rods each have permittivity $\epsilon_{33} = 12.25$ (corresponding to silicon) and are immersed in free space $\epsilon_{33} = 1$. Again, we work with the $\textrm{TM}_{x_3}$ mode. The incident field is a plane wave with frequency $\omega = 50$ corresponding to a free space wavelength of $\lambda \approx 1/8$ so that the domain is approximately $13\lambda \times 13\lambda$ in size. This frequency is chosen to lie in the first bandgap of the photonic crystal.

The PML has thickness $w=0.2$ and decay rates $\sigma_i = 37$. The mesh has a maximum level of refinement of 9 and a minimum level of refinement of 4. All elements have polynomial degree 8 and all expansions are computed to a tolerance of $10^{-6}$. The iterative method is terminated with a relative residual reduction tolerance of $10^{-6}$. All other unspecified parameters are chosen as in Section \ref{sec:dielectric_cylinder}.

Figure \ref{fig:photonic_waveguide_mesh} illustrates the mesh used to resolve the photonic crystal waveguide problem. The mesh is very fine near dielectric cylinders and coarse away from them. Figure \ref{fig:photonic_waveguide_mesh} shows a detailed view of the mesh in the vicinity of one of the cylinders. Curvilinear elements are used to model the circular boundary. Similar mesh detail is used to model the boundaries of the remaining 375 cylinders.

The real part of the total field $E_i + E_s$ is shown in Figure \ref{fig:photonic_waveguide_solution} (the PML region is not shown). By virtue of choosing an incident field with frequency in the first bandgap of the photonic crystal, the total field is attenuated in the crystal, but propagates through the waveguide channel. Figure \ref{fig:photonic_waveguide_residual} shows the reduction in the relative residual for the domain decomposition method used to produce this solution. There are 3,662,658 unknowns in $\vect{\phi}$ with 324,387 rows in matrix $\vect{K}$. A weakness of the domain decomposition method is evident in this example. In particular, because we use a fixed number of constraints $l$ along each edge in the mesh to assemble the coarse problem, when the mesh must capture many fine features, the size of the coarse problem can grow rapidly. See Remark \ref{rem:remark_graded_domains} for a related discussion concerning mesh refinement near corners and a possible way to reduce the size of matrix $\vect{K}$.

\subsubsection{Perfect Electric Conducting Ogival Cylinder} \label{sec:ogive}

In our final two examples, we consider the $\textrm{TE}_{x_3}$ mode. In the first of these two examples, both $\vect{\mu}$ and $\vect{\epsilon}$ are diagonal. The third equation in \eqref{eq:curl_curl_H} becomes
\begin{equation} \label{eq:TE_z_equation}
	-\frac{\partial}{\partial x_1} \left( \frac{1}{\epsilon_{22}} \frac{\partial}{\partial x_1} H_{x_3} \right)
	-\frac{\partial}{\partial x_2} \left( \frac{1}{\epsilon_{11}} \frac{\partial}{\partial x_2} H_{x_3} \right) - \omega^2 \mu_{33} H_{x_3} = 0.
\end{equation}
Like with the $\textrm{TM}_{x_3}$ mode, we solve this equation by separating the total field $H_{x_3}$ into a known incident field $H_i$ and unknown scattered field $H_s$ such that $H_{x_3} = H_i + H_s$. Then \eqref{eq:TE_z_equation} becomes \eqref{eq:prototype_pde} with
\begin{equation}
	\vect{\alpha} =
	\begin{bmatrix}
    	\epsilon_{22}^{-1} & 0             \\
    	0             & \epsilon_{11}^{-1}
	\end{bmatrix}, \qquad
	\beta = -\omega^2 \mu_{33}, \qquad
	\phi = H_s, \qquad
	f = \nabla \cdot (\vect{\alpha} \nabla H_i) - \beta H_i,
\end{equation}
and we enforce zero Dirichlet boundary conditions on $\phi$ on an enclosing boundary beyond the PML.

In this section, we consider scattering from an infinitely long, perfectly conducting, ogival cylinder. We use this example to show that the method can be applied to geometries including corners and provide comparisons to the existing literature \cite{Jin2014}. The two boundary components of the ogive are circular arcs. Defining parameters $h_o = 2.07/2$, $w_o = 5/2$, and $\rho_o = (h_o^2 + w_o^2)/(2 h_o)$, both arcs have radius $\rho_o$. The upper arc is comprised of points satisfying $x_2 \ge 0$ on the circle with center $[0, h_o - \rho_o]^T$. The lower arc is comprised of points satisfying $x_2 \le 0$ on the circle with center $[0, -(h_o - \rho_o)]^T$. The ogive measures $2 w_o$ along the $x_1$ axis and $2 h_o$ along the $x_2$ axis.

We solve for the scattered field in domain $\Omega = (-3.5, 3.5)^2 \setminus \Omega_o$ where $\Omega_o$ is the interior of the ogive. We assume free space conditions $\vect{\epsilon} = \vect{I}$ and $\vect{\mu} = \vect{I}$ throughout the domain. We choose the plane wave incident field
\begin{equation} \label{eq:incident_plane_wave}
	H_i = e^{\jmath \omega \vect{k}^T \vect{x}}
\end{equation}
which means, together with the free space assumption, that the forcing function $f$ is zero. Note that the unit vector $\vect{k} = [\cos{(\varphi_i),\, \sin{(\varphi_i)}}]^T$ will be varied according to angle $\varphi_i$ but that we will fix frequency $\omega = 2\pi$, as in \cite{Jin2014}. This gives a free space wavelength $\lambda = 1$ so that the domain is $7\lambda \times 7\lambda$ in size.

We enforce the perfect electric conductor boundary condition \eqref{eq:pec_boundary_condition} on the ogive boundary. In our problem setting, \eqref{eq:pec_boundary_condition} implies
\begin{equation}
	\vect{n} \cdot (\vect{\alpha} \nabla H_{x_3}) = 0
\end{equation}
where $\vect{n}$ is the inward pointing normal to the perfect electric conductor. Thus to treat a scattering problem from a perfect electric conductor, we impose the inhomogeneous Neumann boundary condition
\begin{equation} \label{eq:pec_TEz}
	\vect{n} \cdot (\vect{\alpha} \nabla H_s) = -\vect{n} \cdot (\vect{\alpha} \nabla H_i)
\end{equation}
so that \eqref{eq:prototype_robin} takes parameters $\gamma = 0$ and $q = -\vect{n} \cdot (\vect{\alpha} \nabla H_i)$. The perfect electric boundary condition \eqref{eq:pec_TEz} on the ogive for the incident plane wave \eqref{eq:incident_plane_wave} requires
\begin{equation}
	q = -\jmath \omega \vect{n}^T \vect{\alpha} \vect{k} H_i 
\end{equation}
with inward pointing normal
\begin{equation}
	\vect{n}(\vect{x}) = \frac{-1}{\sqrt{x_1^2 + \big(x_2 \pm (h_o - \rho_o)\big)^2}}
	\begin{bmatrix}
    	x_1 \\
    	x_2 \pm (h_o - \rho_o)
	\end{bmatrix}
\end{equation}
(the sign changes depending on if on the upper or lower arc of the ogive).

\begin{figure*}[!tp]
    \centering
    \begin{subfigure}[t]{0.475\textwidth}
        \includegraphics[width=0.845\textwidth]{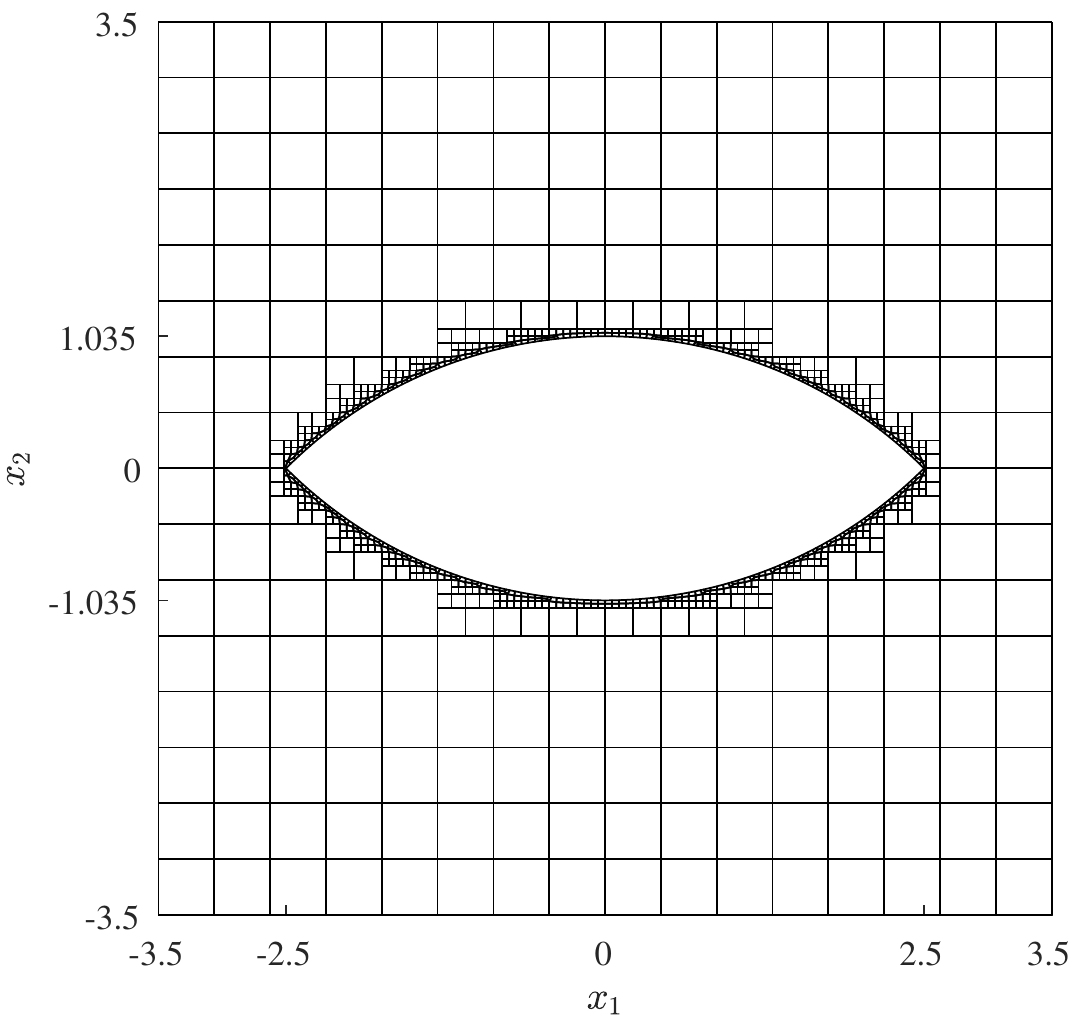}
        \caption{Mesh used to compute the scattered field. \label{fig:ogive_mesh}}
    \end{subfigure}
    \hfill
    \begin{subfigure}[t]{0.475\textwidth}
        \centering
        \includegraphics[width=\textwidth]{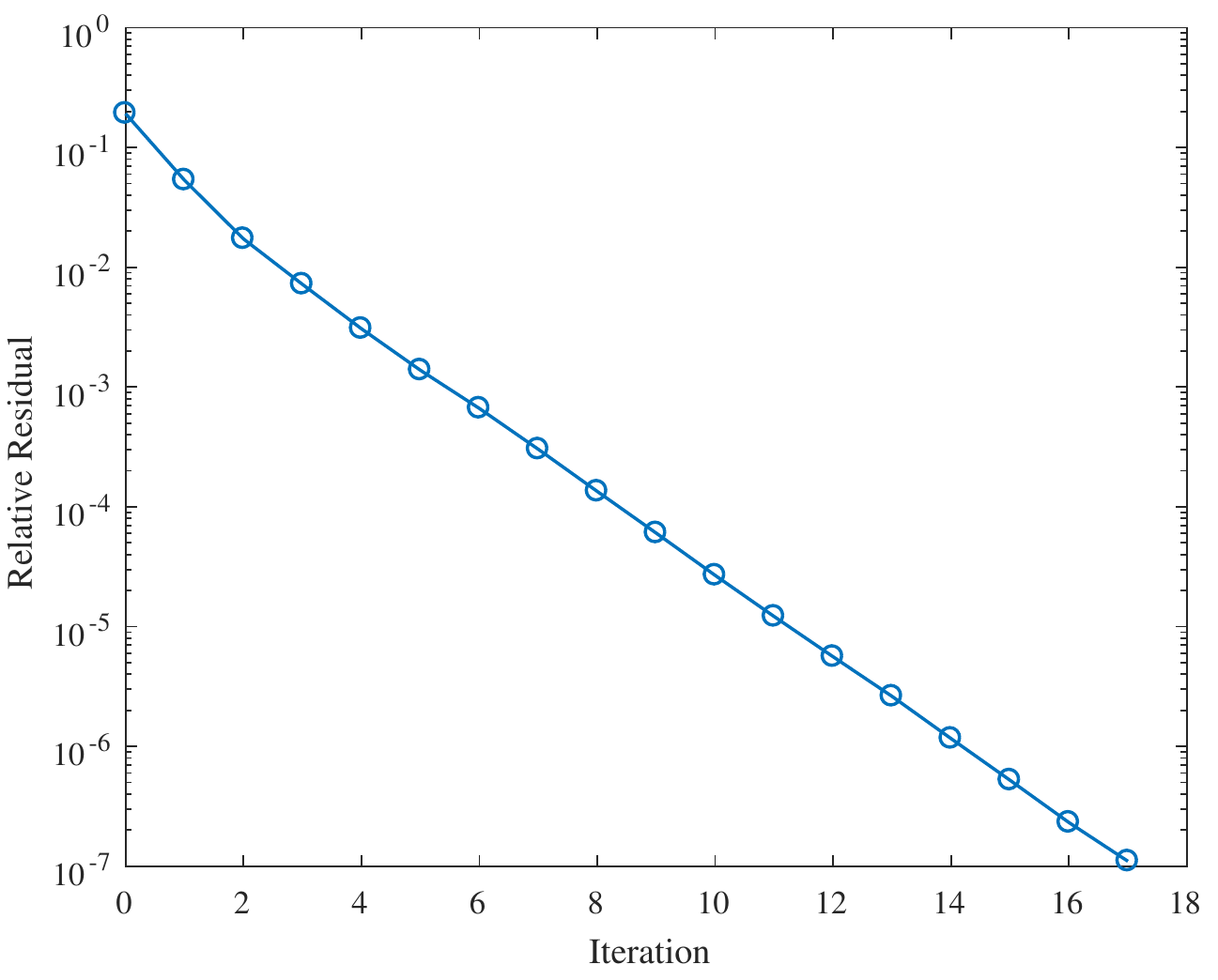}
        \caption{Preconditioned relative residual versus iteration number. \label{fig:ogive_residual}}
    \end{subfigure}
    
    \par\smallskip 
    
    \begin{subfigure}[t]{0.475\textwidth}
        \centering
        \includegraphics[width=0.925\textwidth]{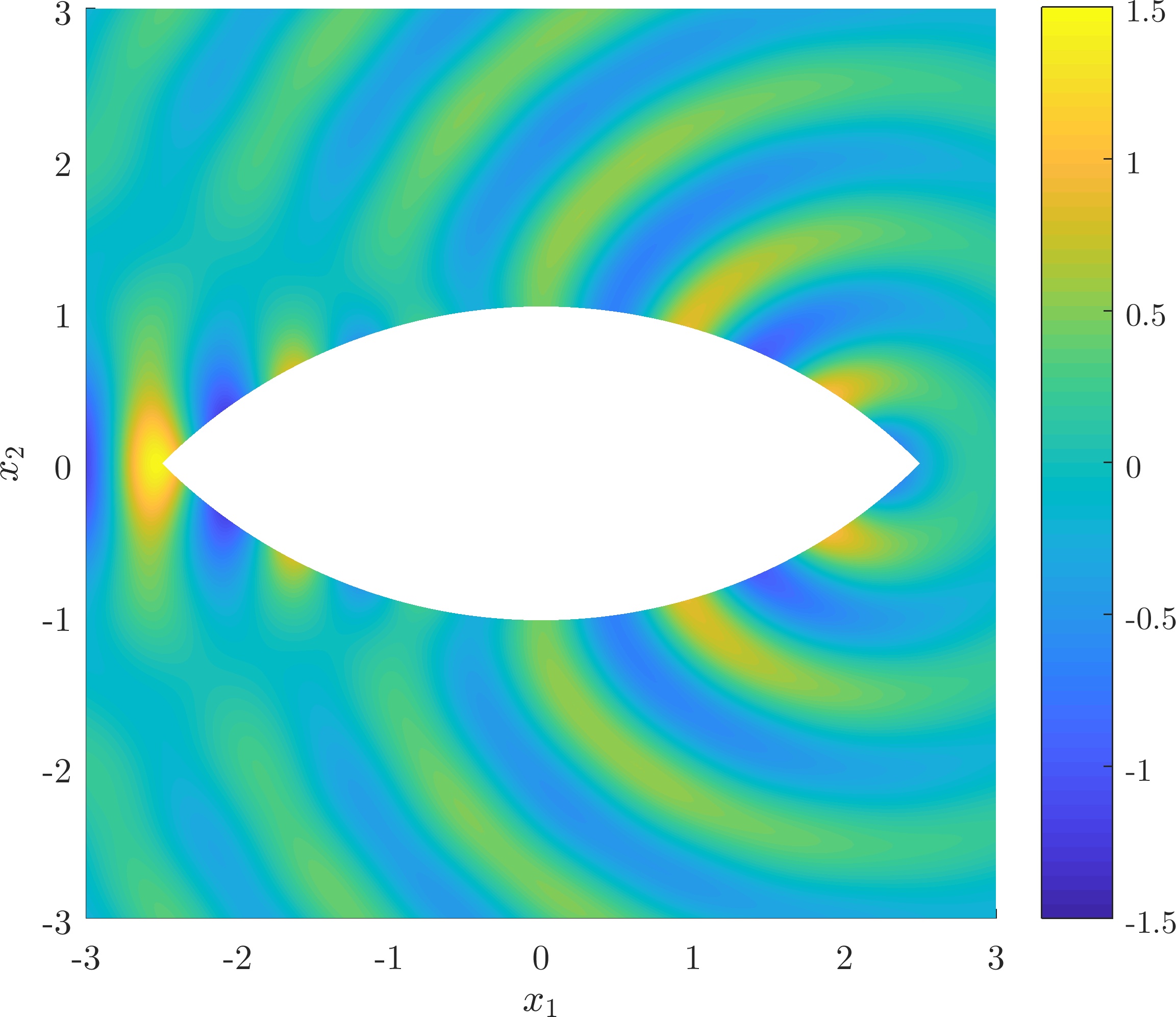}
        \caption{Real part of the scattered field. \label{fig:ogive_solution}}
    \end{subfigure}
    \hfill
    \begin{subfigure}[t]{0.475\textwidth}
        \centering
        \includegraphics[width=\textwidth]{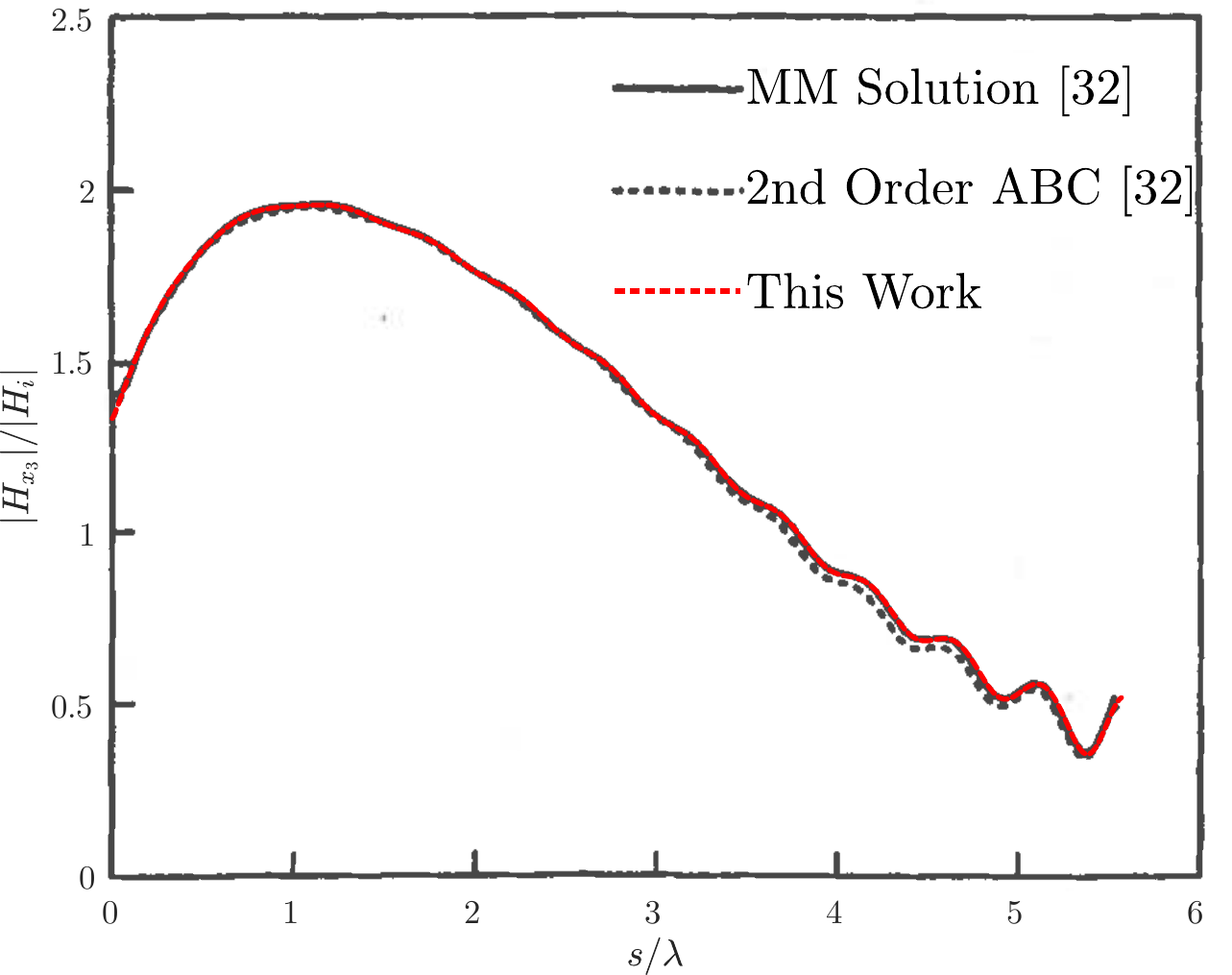}
        \caption{Magnetic field observed at the surface of the ogive. \label{fig:ogive_surface_computation}}
    \end{subfigure}
    
    \par\medskip 
    
    \begin{subfigure}[t]{0.475\textwidth}
        \centering
        \includegraphics[width=\textwidth]{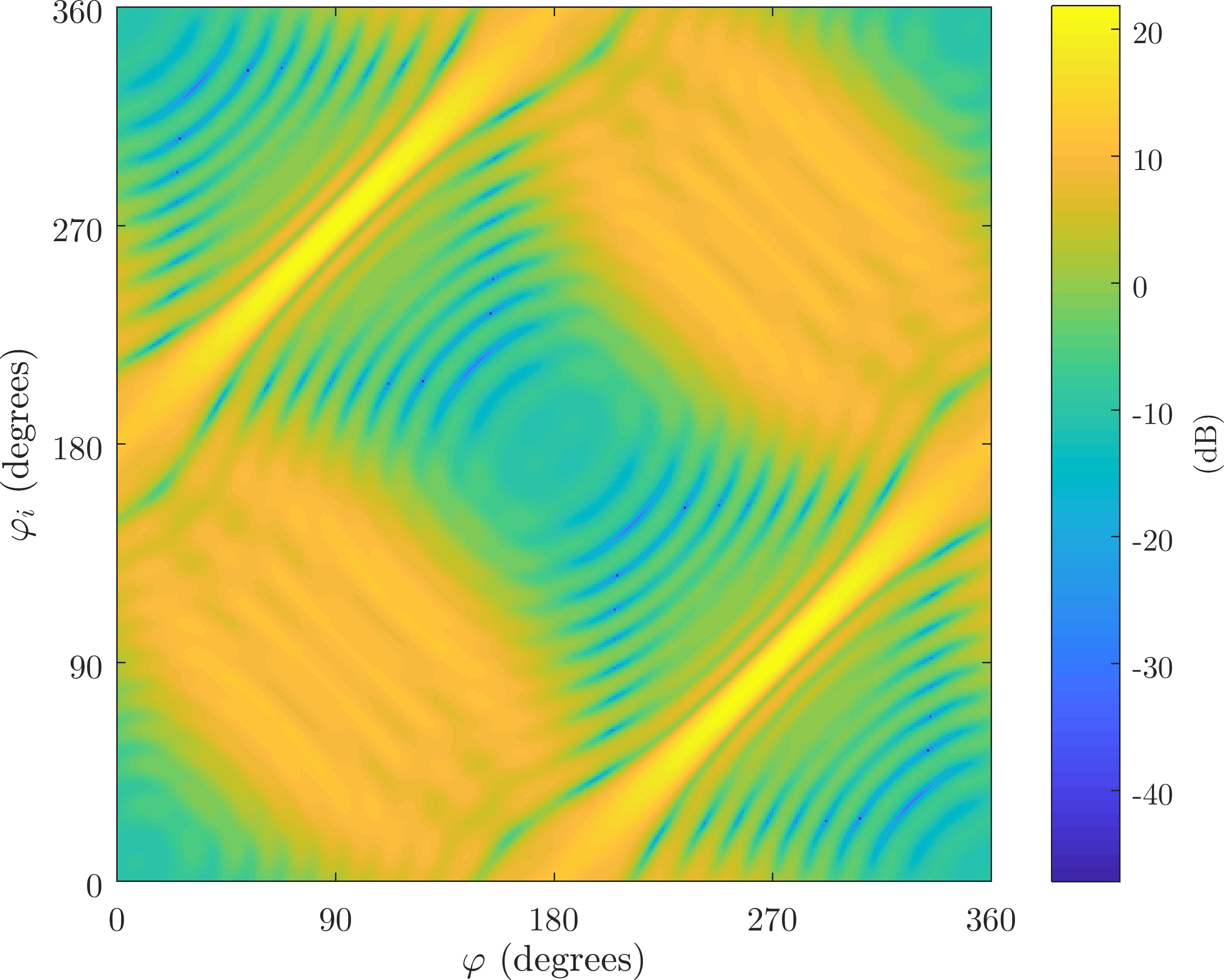}
        \caption{Bistatic scattering width (RCS). \label{fig:ogive_bistatic_RCS}}
    \end{subfigure}
    \hfill
    \begin{subfigure}[t]{0.475\textwidth}
        \centering
        \includegraphics[width=\textwidth]{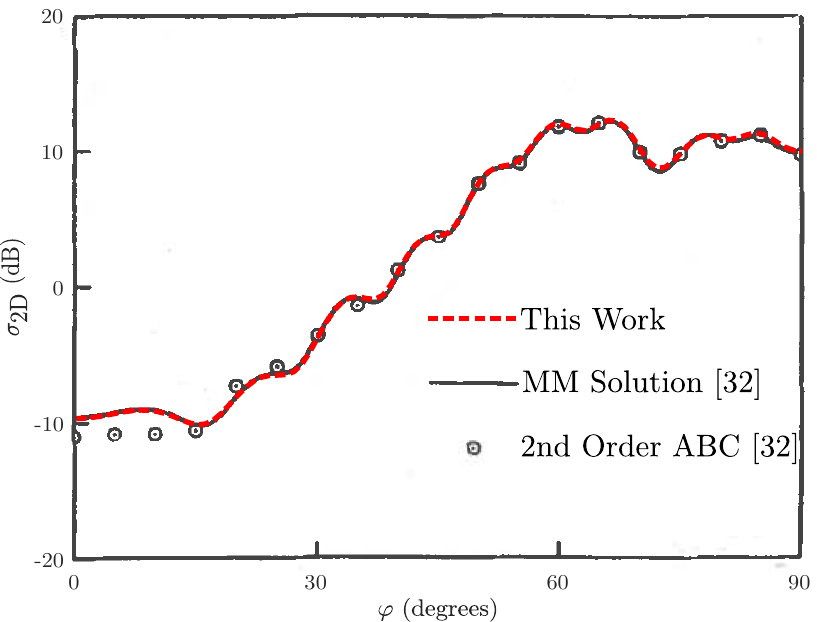}
        \caption{Monostatic scattering width (RCS). \label{fig:ogive_monostatic_RCS}}
    \end{subfigure}

    \caption{Results for the ogive problem described in Section \ref{sec:ogive}.}
\end{figure*}

A PML of thickness $w = 0.4375$ with decay rates $\sigma_i = 16$ is chosen to attenuate the scattered field. The quadtree mesh is refined to a maximum level of 7 and minimum level of 4 with points $[\pm w_o, 0]^T$ fixed in the mesh to preserve the corners of the ogive (see Figure \ref{fig:ogive_mesh} for an illustration of this mesh). In all results, the solution is expanded with degree 16 polynomials on all elements, and coefficient expansions and boundary representations are computed to a tolerance of $10^{-6}$. The domain decomposition algorithm is run until the preconditioned relative residual is reduced by a factor of $10^{-6}$.

First, we show results for a single angle of incidence $\varphi_i = 0$. Figure \ref{fig:ogive_residual} shows the preconditioned relative residual as a function of number of iterations associated with the computation of the real part of the scattered field, illustrated in Figure \ref{fig:ogive_solution} (the PML region is not shown). For comparison purposes, we reproduce Figure 4.23(b) in \cite{Jin2014} which demonstrates the ratio of the magnitude of the total field $\abs{H_{x_3}}$ and the incident field $\abs{H_i}$ observed at the surface of the ogive. Figure \ref{fig:ogive_surface_computation} reproduces Figure 4.23(b) in \cite{Jin2014}, where $s$ is the distance along the upper arc of the ogive (or lower arc due to symmetry of the solution) measured from its leading edge (the rightmost corner of the ogive). There are 269,926 unknowns in $\vect{\phi}$ with 6,158 rows in matrix $\vect{K}$. Our solution matches the moment method integral equation reference solution (labelled MM Solution on the figure) and improves upon the reported finite element solution with second order absorbing boundary conditions (labelled 2nd Order ABC on the figure).

Second, we compute the RCS of the ogive. We do so for 360 equally spaced incident angles $\varphi_i$ and 360 observation angles $\varphi$ in $[0,2\pi)$. The RCS is computed in the same way as \eqref{eq:rcs_def} and \eqref{eq:far_field} replacing $E_s$ with $H_s$ and $E_{\textrm{far}}$ with $H_{\textrm{far}}$. We choose the boundary of the ogive as $\Gamma$ when computing the far field integral. Since the ogive is not invariant under rotations about the $x_3$ axis, we compute the bistatic RCS illustrated in Figure \ref{fig:ogive_bistatic_RCS}. We do so by computing $H_s$ for a given incident angle $\varphi_i$ using the domain decomposition method, then compute \eqref{eq:far_field} and \eqref{eq:rcs_def} at the 360 observation angles $\varphi$, repeating this process for every new incident angle. We also include the monostatic RCS (where incident and observation angles are equal) for angles between 0 and 90 degrees in Figure \ref{fig:ogive_monostatic_RCS} for direct comparison with the result presented in Figure 4.25(b) in \cite{Jin2014}. This corresponds to sampling the bistatic RCS along the line $\varphi_i = \varphi$ (only angles 0 to 90 degrees are required to characterize the monostatic RCS due to symmetries of the ogive). That being said, we have not exploited any symmetries when computing the bistatic RCS of Figure \ref{fig:ogive_bistatic_RCS} which shows that such symmetries exist. Our monostatic RCS matches the moment method integral equation reference monostatic RCS (labelled MM Solution on the figure) and improves upon the reported finite element monostatic RCS computed with second order absorbing boundary conditions (labelled 2nd Order ABC on the figure).

\subsubsection{Electromagnetic Cloak} \label{sec:electromagnetic_cloak}

As a final example, we consider the electromagnetic cloak in \cite{Cummer2006}. We use this example to show how the method behaves with anisotropic, spatially varying permittivity and permeability and again demonstrate the accuracy of the method. The cloak is meant to eliminate scattering from objects inside the radius $R_1$ infinite cylinder coaxial with the $x_3$ axis. It is comprised of a layer of anisotropic and spatially varying permittivity and permeability that surrounds the cylinder. In practice, anything can be placed inside the layer, but we will assume that the cylinder is a perfect electric conductor. The layer has outer radius $R_2$ and its permittivity has components
\begin{align}
	\epsilon_{\rho} &= \frac{\rho - R_1}{\rho}, \\
	\epsilon_{\varphi} &= \frac{1}{\epsilon_{\rho}}, \\
	\epsilon_{33} &= \left( \frac{R_2}{R_2 - R_1} \right)^2 \epsilon_{\rho},
\end{align}
where $\rho = \norm{\vect{x}}_2$ and $\varphi = \textrm{atan2}(x_2,x_1)$ are cylindrical coordinates. Conversion to Cartesian coordinates yields
\begin{align}
	\epsilon_{11} &= \epsilon_{\rho} \cos^2 \varphi + \epsilon_{\varphi} \sin^2 \varphi, \\
	\epsilon_{12} &= (\epsilon_{\rho} - \epsilon_{\varphi}) \sin \varphi \cos \varphi, \\
	\epsilon_{22} &= \epsilon_{\rho} \sin^2 \varphi + \epsilon_{\varphi} \cos^2 \varphi.
\end{align}
The cloak also has $\vect{\mu} = \vect{\epsilon}$ so that both are of the form \eqref{eq:anisotropic_dielectric}. Both the permeability and permittivity hold for $R_1 \le \rho \le R_2$. Note that $\epsilon_{\varphi}$ is singular at radius $R_1$ which can cause problems in our solver when computing Legendre expansions. We regularize $\epsilon_{\varphi}$ by adding a small quantity $\delta = 10^{-8}$ to its denominator. Outside the cloak layer, we have free space with $\vect{\epsilon} = \vect{\mu = \vect{I}}$.

Working with the $\textrm{TE}_{x_3}$ mode and using the decomposition $H_{x_3} = H_i + H_s$, equation \eqref{eq:curl_curl_H} becomes \eqref{eq:prototype_pde} with
\begin{equation}
	\vect{\alpha} =
	\begin{bmatrix}
    	\epsilon_{11} & \epsilon_{12} \\
    	\epsilon_{12} & \epsilon_{22}
	\end{bmatrix}, \qquad
	\beta = -\omega^2 \mu_{33}, \qquad
	\phi = H_s, \qquad
	f = \nabla \cdot (\vect{\alpha} \nabla H_i) - \beta H_i.
\end{equation}
The simplicity of $\vect{\alpha}$ is due to the fact that $\epsilon_{11}\epsilon_{22} - \epsilon_{12}^2 = 1$.

\begin{remark} \label{rem:remark_boundary_correction}
	This formulation is insufficient because it does not enforce tangential continuity \eqref{eq:tangential_continuity} at the interface between the cloaking layer and free space. This is a problem in this example but not in other examples presented in Section \ref{sec:results} because of the discontinuity in $\vect{\epsilon}$ at an interface for the $\textrm{TE}_{x_3}$ mode (a discontinuity in $\vect{\mu}$ would have the same effect in $\textrm{TM}_{x_3}$ mode examples).
	To enforce tangential continuity, we add a boundary term to elements in free space adjacent to the interface of the form
	\begin{equation}
		\int_{\Gamma} \psi \vect{n} \cdot (\vect{\alpha}_2 -  \vect{\alpha}_1) \nabla H_i \, d\Gamma
	\end{equation}
	where $\vect{\alpha}_1$ denotes $\vect{\alpha}$ outside the cloak and $\vect{\alpha}_2$ denotes $\vect{\alpha}$ inside the cloak with $\vect{n}$ the unit normal into the cloak. This is needed because \eqref{eq:tangential_continuity} implies 
	\begin{equation}
		\vect{n} \cdot \Big(\vect{\alpha}_2 \nabla (H_{x_3})_2 - \vect{\alpha}_1 \nabla (H_{x_3})_1 \Big) = 0
	\end{equation}
	at interfaces but the weak form without the added boundary term only implies such a constraint on the scattered field $H_s$ rather than the total field $H_{x_3}$. Adding the boundary forcing term corrects the condition by including the effect of the incident field $H_i$ while only changing the interpretation of the Lagrange multipliers $\vect{\nu}$. Under this change, rather than represent the normal flux of the scattered field, the Lagrange multipliers represent the normal flux of the total field. \remarkend
\end{remark}

In our example, we choose a plane wave incident field
\begin{equation}
	H_i = e^{-\jmath \omega \vect{k}^T \vect{x}}
\end{equation}
with unit vector $\vect{k} = \vect{e}_1$ and frequency $\omega = 20 \cdot 2\pi$ corresponding to a free space wavelength $\lambda = 1/20$. This choice leads to the imposition of Robin boundary conditions \eqref{eq:prototype_robin} with parameters
\begin{equation}
	\gamma = 0, \qquad
	q = -\jmath \omega \vect{n}^T (\vect{\alpha}_2 - \vect{\alpha}_1)\vect{k} H_i,
\end{equation}
on the boundary between the cloak and free space (see Remark \ref{rem:remark_boundary_correction} for the orientation of the normal and the meaning of subscripts 1 and 2) and Robin boundary conditions \eqref{eq:prototype_robin} with parameters
\begin{equation}
	\gamma = 0, \qquad
	q = \jmath \omega \vect{n}^T \vect{\alpha} \vect{k} H_i,
\end{equation}
on the perfect electric conductor cylinder (see \eqref{eq:pec_TEz} from Section \ref{sec:ogive}). Computation of the forcing function inside the cloak (using the chain rule and product rule) gives
\begin{equation}
	f = \left[ \omega^2 (\mu_{33} - \epsilon_{11}) -\jmath \omega \left( \frac{\partial \epsilon_{11}}{\partial x_1} + \frac{\partial \epsilon_{12}}{\partial x_2} \right) \right] H_i 
\end{equation}
with
\begin{align}
	\frac{\partial \epsilon_{11}}{\partial x_1} &= \frac{x_1}{\rho} \left(\frac{\partial \epsilon_{\rho}}{\partial \rho} \cos^2 \varphi + \frac{\partial \epsilon_{\varphi}}{\partial \rho} \sin^2 \varphi  \right) + 2\frac{x_2}{\rho^2} (\epsilon_{\rho} - \epsilon_{\varphi}) \sin \varphi \cos \varphi, \\
	\frac{\partial \epsilon_{12}}{\partial x_2} &= \frac{x_2}{\rho} \left( \frac{\partial \epsilon_{\rho}}{\partial \rho} - \frac{\partial \epsilon_{\varphi}}{\partial \rho} \right) \sin \varphi \cos \varphi + \frac{x_1}{\rho^2} (\epsilon_{\rho} - \epsilon_{\varphi}) (\cos^2 \varphi - \sin^2 \varphi),
\end{align}
and
\begin{align}
	\frac{\partial \epsilon_{\rho}}{\partial \rho} &= \frac{R_1}{\rho^2}, \\
	\frac{\partial \epsilon_{\varphi}}{\partial \rho} &= -\frac{R_1}{(\rho - R_1)^2}.
\end{align}
In the presence of regularization, $\epsilon_{\varphi} = \rho / (\rho - R_1 + \delta)$ so that this last derivative is $(-R_1 + \delta)/(\rho - R_1 + \delta)^2$.

We choose the radius of the perfectly conducting cylinder as $R_1 = 0.1$ and the outer radius of the cloak as $R_2 = 2R_1$. We solve for the scattered field in domain $\Omega = (-0.8,0.8)^2 \setminus \Omega_c$ where $\Omega_c$ is the interior of the perfect conducting cylinder of radius $R_1$. This domain is $32\lambda \times 32 \lambda$ in size. A PML with thickness $w=0.2$ and decay rates $\sigma_i = 70$ is chosen to attenuate the scattered field. The quadtree mesh is refined to a maximum of 6 levels and minimum of 4. We use degree 10 polynomials outside the cloak and increase the degree of polynomials to 20 in a graded fashion approaching the perfect conducting cylinder (the grading is based on how complicated the forcing function $f$ is to resolve using Legendre polynomials). All coefficient expansions and boundary representations are computed to a tolerance of $10^{-12}$. We continue to choose the parameter $l$ as in Section \ref{sec:dielectric_cylinder} and the domain decomposition method is run until the preconditioned relative residual is reduced by a factor of $10^{-10}$.

\begin{figure*}[!t]
    \centering
    \begin{subfigure}[t]{0.475\textwidth}
        \centering
        \includegraphics[width=0.94\textwidth]{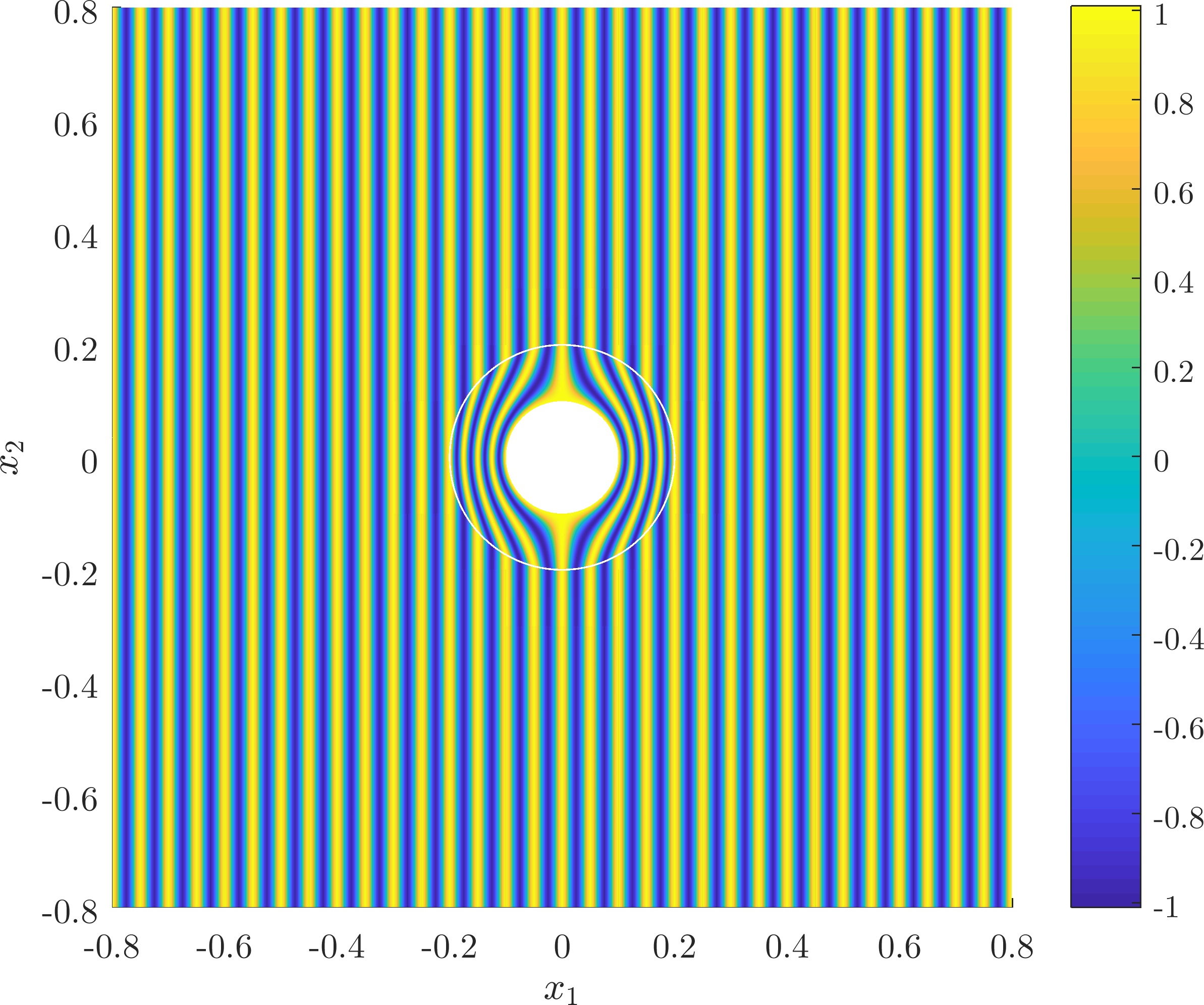}
        \caption{Real part of the total field. \label{fig:cloak_solution}}
    \end{subfigure}
    \hfill
    \begin{subfigure}[t]{0.475\textwidth}
        \centering
        \includegraphics[width=\textwidth]{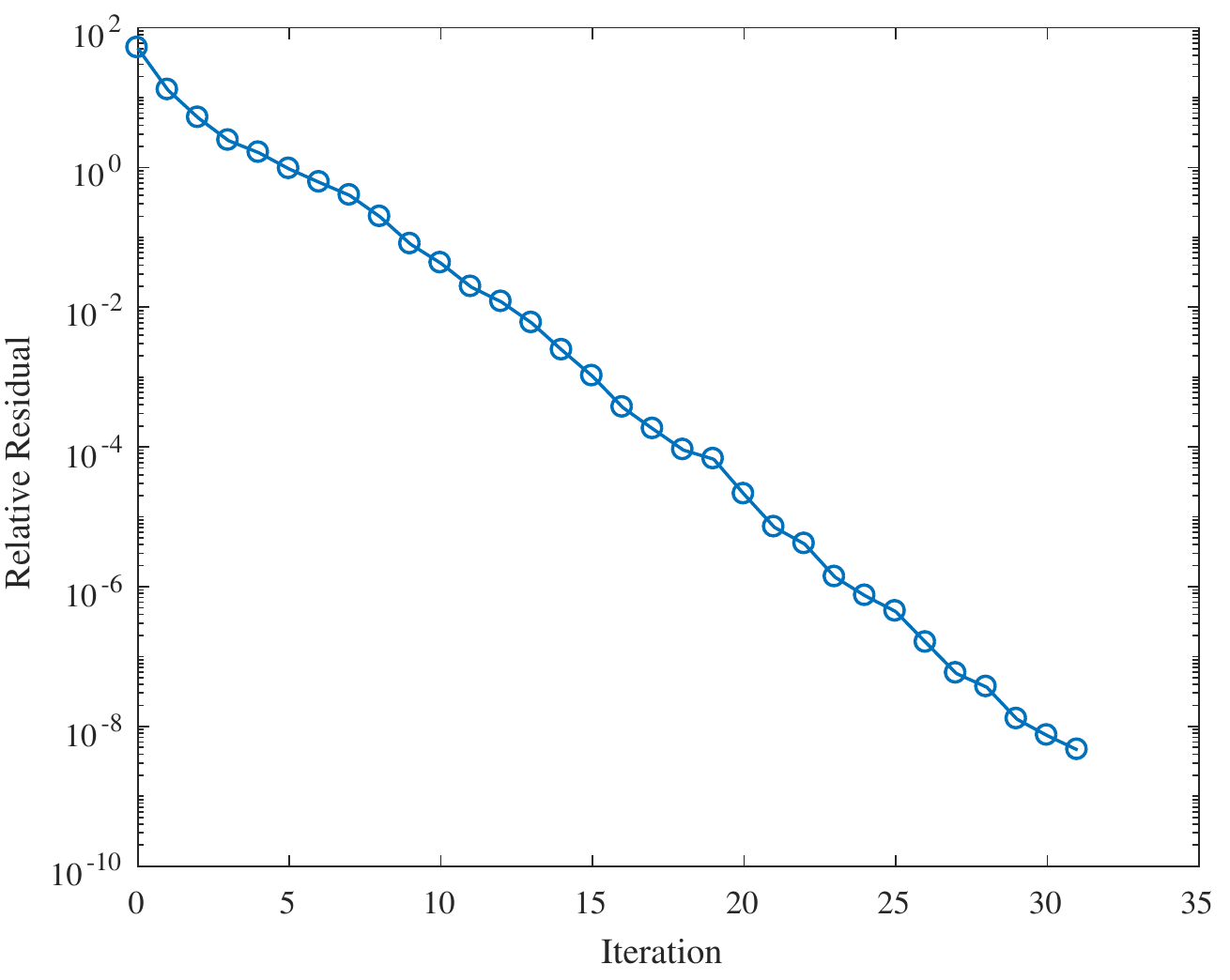}
        \caption{Preconditioned relative residual versus iteration number. \label{fig:cloak_residual}}
    \end{subfigure}
    
    \vskip \baselineskip 
    
    \begin{subfigure}[t]{0.475\textwidth}
        \centering
        \includegraphics[width=0.94\textwidth]{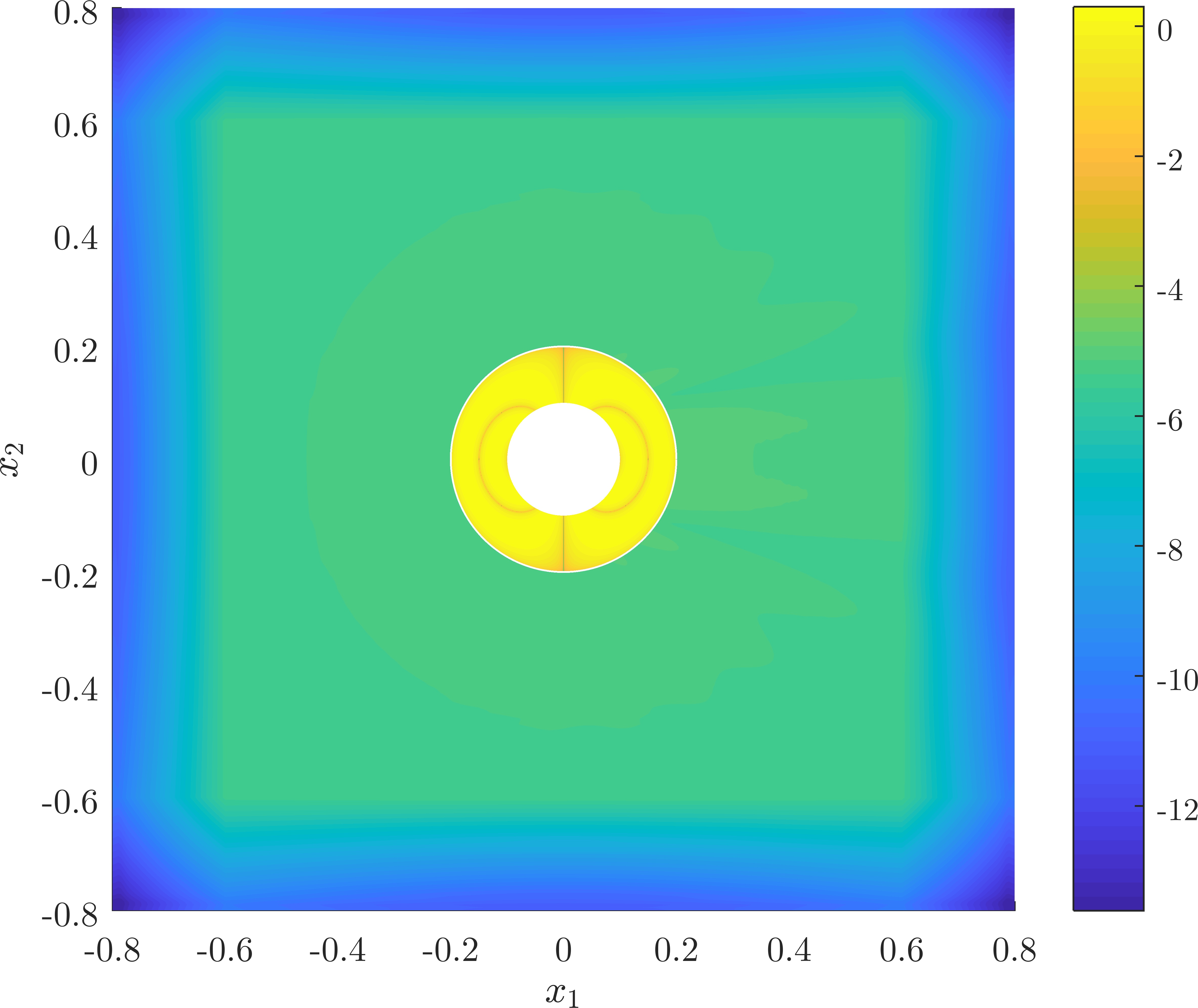}
        \caption{Logarithm of the absolute value of the scattered field ($\log_{10}\abs{H_s}$). \label{fig:cloak_error}}
    \end{subfigure}
    \hfill
    \begin{subfigure}[t]{0.475\textwidth}
        \centering
        \includegraphics[width=\textwidth]{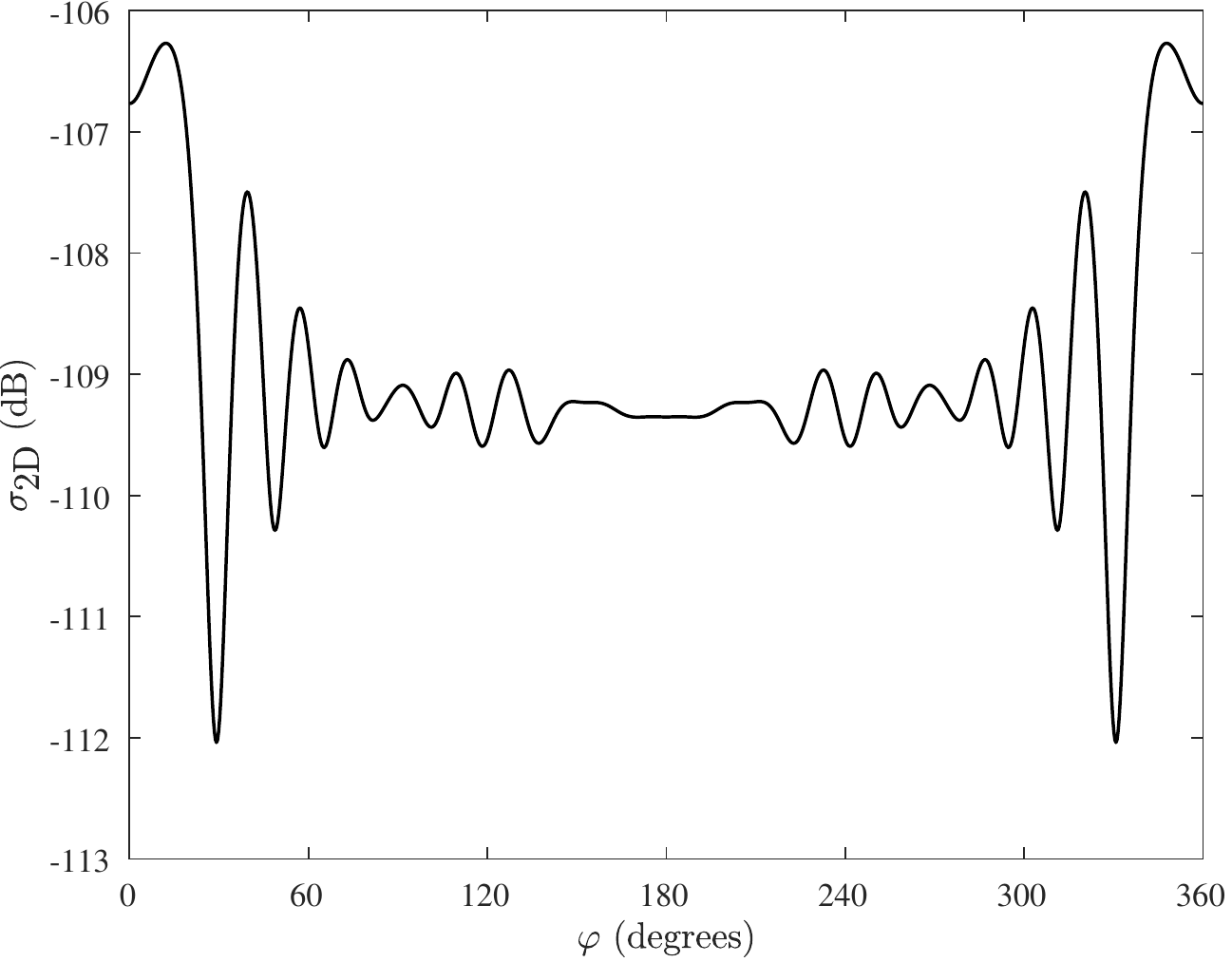}
        \caption{Scattering width (RCS). \label{fig:cloak_rcs}}
    \end{subfigure}

    \caption{Results for the electromagnetic cloak problem described in Section \ref{sec:electromagnetic_cloak}.}
\end{figure*}

Figure \ref{fig:cloak_solution} illustrates the real part of the computed total field $H_{x_3}$ for the electromagnetic cloak problem. Outside the cloak, the total field appears to be identical to the incident field, as though no scatterer were present. Figure\ref{fig:cloak_residual} shows the preconditioned relative residual as a function of number of iterations. There are a total of 117,791 unknowns in $\vect{\phi}$ and 9,676 rows in matrix $\vect{K}$. Figure \ref{fig:cloak_error} shows the base ten logarithm of the pointwise absolute error in the total field (this is the same as measuring the scattered field since we take the incident field as exact solution). Note that, as expected, the error in the cloak is large (on the order of 1 depicted in yellow) but that, immediately outside the cloak, the error is on the order of $10^{-5}$ (depicted in green) and that the error decays to zero in the PML (depicted in blue). The fact that we only recover 5 digits of accuracy is a consequence of the choice of regularization parameter $\delta$ and not our discretization. Decreasing $\delta$ decreases the error, but makes computing the Legendre expansions near the cylinder increasingly costly. The error is small in the PML because the PML causes the scattered wave to decay to zero which happens to be the desired solution for this example (in almost all other scattering problems, this decay would cause the error to be large in the PML; see Section \ref{sec:dielectric_cylinder}, for example). Figure \ref{fig:cloak_rcs} shows the RCS of the cloak, computed in the same way as in Sections \ref{sec:dielectric_cylinder} and \ref{sec:ogive}. We choose the surface for the far field integral to be the interface between the cloak and free space. The RCS confirms that the cloak is effective at reducing reflected waves, as expected.

\section{Conclusions}

In this paper, we have described an efficient iterative method suitable for obtaining high accuracy solutions to high frequency time-harmonic scattering problems and have shown how this method can be used to solve certain challenging electromagnetic problems in two dimensions. The method is a generalization of FETI-DP and can be viewed as a method of FETI-DPH type with a coarse space that can be grown so that the number of iterations of PGMRES depends only weakly on the frequency $k$ for Helmholtz problems. As with all domain decomposition methods, the size of the associated coarse problem remains a computational bottleneck and can be particularly large for problems with fine mesh features (for example, see Section \ref{sec:photonic_waveguide}) or for high frequency problems (for example, see Section \ref{sec:eaton_lens}).

For these reasons, future work is aimed at reducing the size of the coarse problem. Since we intend to apply the method in an $hp$-adaptive framework, relaxing the constraint that each element belong to its own subdomain is important. Aggregating small elements of low degree near singularities or fine features into subdomains is one way to reduce the size of the coarse space (as in standard low polynomial degree domain decomposition methods). Another approach to reducing the size of the coarse problem involves choosing the number of constraints per edge $l$ in a non-uniform manner reflecting the non-uniform refinement of the mesh. Both of these changes may extend the range of problems for which the method is efficient.

While our mesh generation procedure has been applied to force many elements to be affine maps of the canonical domain $(-1,1)^2$, this can come at the expense of relatively poorly shaped elements near boundaries. These elements can require Legendre expansions of high degree to capture the variation of the map with sufficient accuracy. Future work includes improving the shape quality of elements near boundaries. One possible approach is to allow a larger number of elements with maps that are not affine, but whose maps, on average, are closer to affine than those in the current method. When maps are nearly affine, low order terms in the Legendre expansions representing effective parameters on elements allow us to use the local fast solver as a preconditioner.

Finally, all computations presented in this paper were performed using MATLAB on a workstation with a four core Intel Xeon E5-1603 processor and 32 GB of RAM. Future work includes implementation of the algorithm in C/C++ using MPI so as to test parallel scalability of the method. This is particularly crucial when implementing the method in three dimensions with local polynomial degree higher than the choices made in this paper. Extension to three-dimensional electromagnetic scattering problems via discretization of the curl-curl equation is planned.


\section*{Acknowledgments}
	We acknowledge the support of the Natural Sciences and Engineering Research Council of Canada (NSERC), [funding reference number 121777].

	Cette recherche a \'et\'e financ\'ee par le Conseil de recherches en sciences naturelles et en g\'enie du Canada (CRSNG), [num\'ero de r\'ef\'erence 121777].

\interlinepenalty=10000 
{\footnotesize
\bibliography{jcp_review}}

\end{document}